\documentclass[12pt]{article}

\usepackage{float}
\usepackage[makeroom]{cancel}
\usepackage{color}
\usepackage{graphicx}
\usepackage{amsmath}
\usepackage{amssymb}
\usepackage{xspace}
\usepackage[small]{subfigure}
\usepackage[numbers,compress]{natbib}
\usepackage[hyperfootnotes=false]{hyperref}
\usepackage{tcolorbox}

\usepackage{framed}
\usepackage{physics}
\usepackage{tensor}

\newlength{\fighskip} \fighskip=2pt
\newlength{\figvskip} \figvskip=3pt

\newcommand*{\figbox}[2]{{
  \def\figscale{#1}
  \def\arraystretch{0.8}
  \arraycolsep=0pt
  \begin{array}{c}
    \vbox{\vskip\figscale\figvskip
      \hbox{\hskip\figscale\fighskip
        \includegraphics[scale=\figscale]{#2}}}
  \end{array}}}

\usepackage{mciteplus} 
\usepackage{dcolumn}
\usepackage{bm}
\usepackage{verbatim}
\usepackage{amscd}
\usepackage{amsfonts}
\usepackage{setspace}
\usepackage{amsthm}
\usepackage{enumerate}
\usepackage{mathtools}

\theoremstyle{plain}
\newtheorem{claim}{Claim}
\theoremstyle{plain}

\theoremstyle{plain}
\newtheorem{proposal}{Proposal}
\theoremstyle{plain}
\newtheorem{theorem}{Theorem}
\theoremstyle{plain}

\theoremstyle{plain}
 
\theoremstyle{plain}

\theoremstyle{remark}

\theoremstyle{conjecture}

\theoremstyle{observation}

\theoremstyle{definition}

\theoremstyle{corollary}
\newtheorem{corollary}{Corollary}
\theoremstyle{definition}

\theoremstyle{definition}

\theoremstyle{result}

\theoremstyle{assumption}

\theoremstyle{definition}

\theoremstyle{problem}

\theoremstyle{fact}

\usepackage{authblk}

\title{
\vspace{-80pt}
\hfill
{\normalsize YITP-24-149}\\
\vspace{40pt}
\bf 
Does connected wedge imply distillable entanglement? 
}

\author[1,2]{Takato Mori
}
\author[1]{Beni Yoshida
}
\affil[1]{\em \small Perimeter Institute for Theoretical Physics, Waterloo, Ontario N2L 2Y5, Canada}
\affil[2]{\em \small Center for Gravitational Physics and Quantum Information, 
Yukawa Institute for Theoretical Physics, Kyoto University, 
Kitashirakawa Oiwakecho, Sakyo-ku, Kyoto 606-8502, Japan}
\affil[ ]{\textit {\href{mailto:takato.mori@yukawa.kyoto-u.ac.jp}{takato.mori@yukawa.kyoto-u.ac.jp}, \hspace{3pt} \href{mailto:byoshida@perimeterinstitute.ca}{byoshida@perimeterinstitute.ca}}}

\date{}

\begin{document}
\maketitle

\begin{abstract}
The Ryu-Takayanagi formula predicts that two boundary subsystems $A$ and $C$ can exhibit large mutual information $I(A:C)$ even when they are spatially disconnected on the boundary and separated by a buffer subsystem $B$, as long as $A$ and $C$ have connected entanglement wedge in the bulk. 
However, whether the reduced state $\rho_{AC}$ contains distillable EPR pairs has remained a longstanding open problem. 
In this work, we resolve this problem by showing that: i) there is no LO-distillable entanglement at leading order in $G_N$, suggesting the absence of bipartite entanglement in a holographic mixed state $\rho_{AC}$, and 
ii) one-shot, one-way LOCC-distillable entanglement is given at leading order by locally accessible information $J^W(A|C)$, which is related to the entanglement wedge cross section $E^W$ involving the (third) purifying system $B$ via $J^W(A|C) = S_A - E^W(A:B)$.   
Namely, we demonstrate that a connected entanglement wedge does not necessarily imply nonzero distillable entanglement in one-shot, one-way LOCC. 
We also show that entanglement of formation $E_{F}(A:C)$ is given by $E^W(A:C)$ at leading order in holography.
\end{abstract}


\tableofcontents

\newpage

\section{Introduction}\label{sec:introduction}

\begin{figure}[H]
\centering
\includegraphics[width=0.22\textwidth]{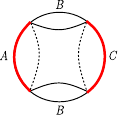}
\caption{Connected entanglement wedge. How are two subsystems $A$ and $C$ entangled?
}
\label{fig_connected}
\end{figure}

In the AdS/CFT correspondence, for static geometries, entanglement entropy $S_{A}$ of a boundary subsystem $A$ can be computed by the Ryu-Takayanagi (RT) formula 
\begin{align}
S_{A} = \frac{1}{4G_{N}} \min_{\gamma_A} \text{Area}(\gamma_A) + \cdots 
\end{align}
at leading order in $1/G_{N}$.
This remarkable formula predicts that two boundary subsystems $A$ and $C$ can have large mutual information 
\begin{align}
I(A:C) \equiv S_{A} + S_{C}-S_{AC} = O(1/G_{N})
\end{align}
even when they are spatially disconnected on the boundary with a buffer subsystem $B$, provided that $A$ and $C$ have \emph{connected entanglement wedge} in the bulk. A prototypical example is illustrated in Fig.~\ref{fig_connected} for the AdS${_3}$/CFT$_{2}$. 

However, the precise nature of entanglement in $\rho_{AC}$ remains elusive.
For one thing, the mutual information is sensitive to classical correlations such as those in the GHZ state. 
Fortunately, several evidences from quantum gravity thought experiments and toy models suggest that correlations in $\rho_{AC}$ in holography are not of classical nature at leading order in $1/G_{N}$~\cite{Susskind:2014yaa, Nezami:2016zni, Dong:2021clv}. 
However, even in the absence of classical correlation, mutual information cannot distinguish bipartite from tripartite entanglement. 
For instance, one could achieve large mutual information by simply distributing $\frac{1}{2}I(A:C)$ copies of EPR pairs between $A$ and $C$. 
Recent studies have, however, shown that correlations in $\rho_{AC}$ in holography contain genuinely tripartite entanglement in $|\psi_{ABC}\rangle$~\cite{Akers:2019gcv,Mori:2025gqe}.

\subsection{Entanglement measures for mixed state}

Then, how should we characterize the quantum entanglement in $\rho_{AC}$? Fortunately (or unfortunately), there are a plethora of entanglement measures for mixed states, each with distinct operational meanings. Several examples are listed below:

\begin{enumerate}[a)]

\item Entanglement of purification, $E_P$~\cite{Terhal:2002riz}: Considering all possible purifications $|\psi_{AA'CC'}\rangle$ of $\rho_{AC}$, it is the minimal entanglement entropy between $AA'$ and $CC'$:
\begin{align}
E_{P}(A:C) = \min_{\Tr_{A'C'}(|\psi\rangle\langle \psi|)=\rho_{AC}} S_{AA'}.
\end{align}

\item Entanglement of formation, $E_{F}$~\cite{Bennett:1996gf}: Considering all possible convex decompositions of $\rho_{AC}$ by pure states, $\rho_{AC}= \sum_{i} p_i |\psi_i\rangle\langle \psi_i|$, it is the minimal average entanglement entropy of $A$:
\begin{align}
E_{F}(A:C) = \inf_{ \rho_{AC}= \sum_{i} p_i |\psi_i\rangle\langle \psi_i| } \sum_{i} p_i S( \rho_{A}^i ), \qquad \rho_{A}^i = \Tr_{B}(|\psi_i\rangle\langle \psi_i|).
\end{align}

\item Entanglement cost, $E_{C}$~\cite{Hayden_2001}: Defined as the asymptotic limit of entanglement of formation $E_{F}$:
\begin{align}
E_{C}(A:C) = \lim_{m\rightarrow \infty}\frac{E_{F}({\rho_{AC}}^{\otimes m})}{m}.
\end{align}
This measures the number of EPR pairs per copy required to create $\rho_{AC}$ with vanishing error at the asymptotic limit $m\rightarrow \infty$. 

\item Squashed entanglement, $E_{sq}$~\cite{Christandl_2004}: Considering all possible extensions $\rho_{ACE}$ of $\rho_{AC}$, squashed entanglement is given by a half of the minimum conditional mutual information:
\begin{align}
E_{sq}(A:C) =\frac{1}{2} \inf_{\Tr_{E}(\rho_{ACE})=\rho_{AC}} I(A:C|E),
\end{align}
where $I(A:C|E)\equiv I(A:CE)-I(A:E)$.

\item Distillable entanglement, $E_{D}$~\cite{Bennett:1996gf,Rains:1998gp}: This measures the number of EPR pairs that can be extracted from $\rho_{AC}$ via Local Operations and Classical Communication (LOCC). It is usually defined in the asymptotic setting:
\begin{align}
E_{D}(A:C) = \sup_{r} \left\{  r \Big| \lim_{m\rightarrow \infty} \Big[ \inf_{\Lambda \in \text{LOCC}} D\big( \Lambda(\rho_{AC}^{\otimes m}) , \Phi_{\text{EPR}, 2^{rm}}   \big)  \Big] = 0 \right\},
\end{align}
where $D$ represents the trace distance, and $\Phi_{\text{EPR},2^n}$ represents $n$ EPR pairs, $\qty(\frac{\ket{00}+\ket{11}}{\sqrt{2}})^{\otimes n}$.

\end{enumerate}  

In quantum information, these measures are known to obey the following chain of inequalities:
\begin{align}
\text{hash}(A:C) \leq  E_{D} \leq E_{sq} \leq E_{C} \leq E_{F} \leq E_{P}\leq \min(S_{A}, S_{C}) 
\label{eq:hier}
\end{align}
where the \emph{hashing bound}~\cite{Bennett:1996gf, Devetak_2005,PhysRevLett.85.433}, $\text{hash}(A:C) \leq  E_{D}(A:C)$, is given by
\begin{align}
\text{hash}(A:C)\equiv \max(S_{A} - S_{AC}, S_{C} - S_{AC}, 0). 
\end{align}
Here, $I_{coh}(A|C)\equiv - S(A|C)= S_{A} - S_{AC}$ is known as the coherent information.

\subsection{Holographic entanglement measures}

Computing these entanglement measures is generally challenging, as it involves optimization over all quantum states. 
Fortunately, in the AdS/CFT correspondence, there are promising proposals for $E_{P}$ and $E_{sq}$ at leading order in $1/G_{N}$, summarized below~\cite{Hayden:2011ag, Nguyen:2017yqw, Takayanagi:2017knl, Cheng:2019aqf}: 
\begin{align}
E_{sq} \approx \frac{1}{2}I(A:C), \qquad E_{P} \approx E^W(A:C)
\end{align}
where $E^W(A:C)$ is the entanglement wedge cross section~\cite{Takayanagi:2017knl}, whose definition we recall shortly.\footnote{
We use ``$f \approx g$'' to indicate agreement at leading order, i.e., $|f - g| \sim o(n)$ or $o(1/G_N)$. 
For Haar random states, leading order refers to $O(n)$ scaling with system size $n$, while in holography, it corresponds to $O(1/G_N)$ as $G_N \to 0$. 
Subleading terms grow more slowly, and denoted by $o(n)$ or $o(1/G_N)$. 
In holography, subleading terms often scale polynomially, e.g., $O(1/G_N^{1-a})$ with $0 < a < 1$, while in Haar random states and tensor networks, they are exponentially suppressed, and thus $o(1)$.
} 
These proposals have been supported under physically reasonable assumptions and have passed a range of non-trivial consistency checks.

\begin{figure}
\centering
\raisebox{\height}{a)\hspace{10pt}}\raisebox{-0.85\height}{\includegraphics[width=0.15\textwidth]{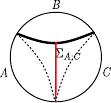}}
\hspace{10pt}
\raisebox{\height}{b)\hspace{10pt}}\raisebox{-0.85\height}{\includegraphics[width=0.33\textwidth]{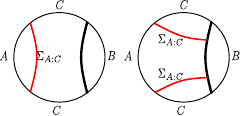}}\\
\raisebox{\height}{c)\hspace{10pt}}\raisebox{-0.85\height}{\includegraphics[width=0.18\textwidth]{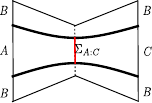}}
\hspace{10pt}
\raisebox{\height}{d)\hspace{10pt}}\raisebox{-0.85\height}{\includegraphics[width=0.36\textwidth]{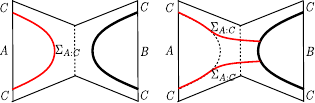}}
\caption{Examples of the minimal cross section $\Sigma_{A:C}$ (shown in red lines). 
Thick lines represent the boundaries of the entanglement wedge $\mathcal{E}_{AC}$. 
a) Pure AdS$_3$ divided into three segments $A,B,C$. 
b) Pure AdS$_3$ where $C$ (instead of $B$) is a buffer region.
c) Two-sided BTZ black hole where $A,C$ are placed on opposite sides. 
d) Another example for two-sided BTZ black hole.
Note that b) and d) have two candidate cross sections.
}
\label{fig_EW_example}
\end{figure}

The entanglement wedge cross section $E^{W}(A:C)$ is defined as follows. 
When $A,C$ have connected wedge, $E^{W}$ is defined by
\begin{align}
E^{W}(A:C) \equiv \min_{\Sigma_{A:C}} \frac{\text{Area}(\Sigma_{A:C})}{4G_{N}} = \frac{1}{4G_{N}} \ \figbox{1.5}{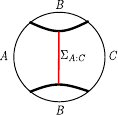}
\label{eq:def-ewcs}
\end{align}
where the minimization is taken over all possible surfaces (cross sections) $\Sigma_{A:C}$ that divide the entanglement wedge into two parts, one containing $A$ and the other $C$. 
Here, we have schematically illustrated the minimal cross section for pure AdS$_3$. 
In contrast, when $A,C$ do not have a connected wedge, we set $E^{W}(A:C)=0$.
Since $E^W(A:C)$ plays a central role throughout this work, we present further examples in Fig.~\ref{fig_EW_example}, for both pure AdS$_3$/CFT$_{2}$ and two-sided BTZ black holes.

The proposal $E_{P}\approx E^W$ suggests that the optimal purification for computing $E_{P}$ is obtained by choosing the purifying system $A'C'$ to lie along the minimal surface of $AC$, divided by the minimal cross section $\Sigma_{A:C}$, as illustrated schematically:
\begin{align}
|\Psi_{AA'CC'}\rangle = \figbox{1.7}{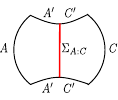}.
\label{eq:EWCS}
\end{align}
Here, the auxiliary purifying system $A'C'$ does not necessarily lie on asymptotic AdS boundaries; instead, it can be thought of as degrees of freedom (DOFs) on the bulk.
One may interpret these DOFs via the state-surface correspondence~\cite{Miyaji:2015yva} or justify them using the tensor network picture~\cite{Pastawski:2015qua, Hayden:2016cfa}. 
In the boundary, these may be viewed as coarse-grained DOFs for the complementary subsystem $B$. 

We now propose an additional entry to the holographic dictionary. 
Namely, a similar line of observation prompts us to propose the following holographic relation for $E_{F}$ at leading order:
\begin{align}
E_{F} \approx E^W(A:C).
\end{align}
Specifically, we propose that the optimal decomposition for computing $E_{F}$ involves placing disentangled basis states along the minimal surface of $AC$, as illustrated schematically:
\begin{align}
\figbox{1.9}{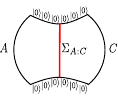}\ .
\end{align}
Again, these basis states may be justified via the state-surface correspondence and the tensor network picture.
In the geometric picture, they correspond to configurations where End-of-the-World (EoW) brane-like objects are placed along the minimal surface of $AC$, as we will discuss later. 

To the best of our knowledge, the proposal of $E_{F} \approx E^W(A:C)$ has not previously appeared in the literature. 
In fact, a prior work has presented a particular no-go argument that appears to suggest $E_{F} \not= E^W(A:C)$ in the holographic context~\cite{Umemoto:2019jlz}. 
Later in this paper, we show how our proposal circumvents this argument. 
(In short, $E_{F} \approx E^W(A:C)$ holds at leading order in $1/G_{N}$, but may deviate at subleading order.)
Given the novelty and central importance of this proposal, we will present additional evidences for the $E_{F} \approx E^W(A:C)$ proposal later in this paper. 
We will also demonstrate that this relation holds for a tripartite Haar random state $|\psi_{ABC}\rangle$ when regarded as a toy model of holography. 

In summary, we arrive at the following hierarchy for holographic entanglement measures:
\begin{align}
\text{hash}(A:C)\leq E_{D} \leq \frac{1}{2}I(A:C) \leq E_{F}, E_{P} \approx E^{W}(A:C)
 \label{eq:holo_ineq}
\end{align}
at leading order.\footnote{
It is known that $E_{F}$ can violate the additivity by $O(1)$ amount~\cite{Hastings:2009ybd}, i.e. $E_{C}\not=E_{F}$ in general. 
A possibility of an extensive violation has been suggested in holographic contexts~\cite{Hayden:2020vyo}, but this argument can be avoided under the assumption that disentangled basis states exist on the black hole horizon.
}
The most significant open question lies in determining the distillable entanglement $E_{D}$ in holography, which is the major focus of the present work. 
We will also address the $E_{F} \approx E^W(A:C)$ proposal extensively in the rest of the paper.

\subsection{Locally accessible information}

Another important quantity is the \emph{locally accessible information} $J(A|C)$~\cite{Zurek_2003,Henderson:2001wrr,Badzia_g_2003}. 

\begin{enumerate}[f)]
\item Locally accessible information, $J(A|C)$: Considering all possible measurements described by positive operator-valued measures (POVMs) $\{ \Pi^{i}_C \}_i$ acting on $C$, and the resulting marginal states $\{p_j, \rho_A^{j}  \}_j$, it quantifies the maximal average entropy reduction in $A$:\footnote{It is also useful to write $J(A|C) = \max_{\{ \Pi^{i}_C \}_i} \sum_{j}p_j S(\rho_A^j||\rho_A)$ where $S(\rho||\sigma)=\Tr(\rho\log\rho-\rho\log\sigma)$ is the quantum relative entropy. 
Hence, $J(A|C)$ remains UV finite in holography due to cancellation of divergences.
}
\begin{align}
J(A|C) &\equiv S_{A} - \min_{\{ \Pi^{i}_C \}_i} \sum_{j}p_j S_{A}(\rho_A^j), \qquad 
p_j = \Tr (\Pi^j_C \rho_C).
\end{align}
\end{enumerate}

When the global state is a pure tripartite state $|\psi_{ABC}\rangle$, $J(A|C)$ is closely related to the entanglement of formation $E_{F}(A:B)$ via the Koashi-Winter relation~\cite{Koashi:2004fvj}:
\begin{align}
J(A|C) = S_{A} - E_{F}(A:B).
\end{align}
This is an information theoretic identity and is not restricted to holographic context.
This relation can be viewed as a manifestation of entanglement monogamy, indicating that $J(A|C)$ and $E_{F}(A:B)$ cannot both be large simultaneously. 

Assuming the holographic correspondence $E_{F}\approx E^W$, we are naturally led to the following dual expression for $J(A|C)$:
\begin{align}
J(A|C)  \approx J^W(A|C)\qquad \mbox{where}\quad
J^{W}(A|C) \equiv S_{A} - E^W(A:B)  \geq 0.
\label{eq:JWdef}
\end{align}
It is helpful to visualize $J^W(A|C)$ in pure AdS$_3$ geometry:
\begin{align}
J^{W}(A|C) = \frac{1}{4G_{N}} \max\qty(  \figbox{1.5}{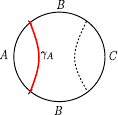} -\figbox{1.5}{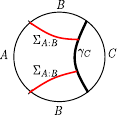}\, ,\quad 0 )\ .
\end{align}

This proposal implies that the maximal entropy drop achievable via local measurements on $C$ corresponds to projecting onto disentangled basis states along the minimal surface of $C$, yielding:
\begin{align}
\figbox{1.7}{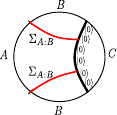} \ .
\end{align}
In such a measurement process, the entropy change is given by $\Delta S_{A} = S_{A}^{\text{before}} - S_{A}^{\text{after}}$ with
\begin{align}
S_{A}^{\text{before}} = \frac{1}{4G_{N}} \figbox{1.5}{fig_SA}\,  , \qquad S_{A}^{\text{after}} = \frac{1}{4G_{N}} \figbox{1.5}{fig_EWAB}.\label{eq:post-measurement}
\end{align}
Thus, at leading order, we identify $S_A^{\text{after}} = E^W(A:B)$.
Here, $S_{A}^{\text{after}}$ is evaluated in the presence of EoW brane-like objects. Note that the geometry of post-measurement states remains nearly identical across different measurement outcomes.

We are not aware of prior work analyzing the locally accessible information $J(A|C)$ in the holographic context. 
Later in this paper, we will present additional evidences for the $J(A|C) \approx J^W(A|C) \equiv S_{A} - E^W(A:B)$ proposal (which also support the $E_{F}\approx E^W$ proposal via the Koashi-Winter relation). 
We will also demonstrate that the distillable entanglement in holography is governed by $J(A|C)$.

\section{Main results: Entanglement distillation in holography}

Entanglement distillation is traditionally studied in the asymptotic setting, where $m$ copies of $\rho_{AC}$ are given with LOCCs applied jointly on ${\rho_{AC}}^{\otimes m}$, and the distillation rate per copy is considered at the $m\rightarrow \infty$ limit.
However it is unclear whether multiple identical copies of a holographic system can be meaningfully prepared or manipulated with joint operations.\footnote{
Many-copy systems can serve as a useful computational tool in replica calculations of single-copy quantities.
}
For this reason, we instead focus on \emph{one-shot} distillable entanglement $E_{D}^{(1)}(A:C)$ which corresponds to the number of (approximate) EPR pairs that can be prepared from a single copy of $\rho_{AC}$ via LOCCs.

Formally, $E_{D}^{(1)}(A:C)$ needs to be defined with some small tolerance $\epsilon$ as follows
\begin{align}
E_{D}^{(1)}(A:C) \equiv \sup_{r} \left\{  r \Big|  \inf_{\Lambda \in \text{LOCC}} D\big( \Lambda(\rho_{AC}) , \Phi_{\text{EPR}, 2^{r}}   \big)  \leq \epsilon \right\}. \label{eq:ED_def}
\end{align}
In this paper, we will require that 
\begin{align}
\epsilon \rightarrow 0 \qquad \mbox{for}\quad 1/G_{N} \rightarrow \infty \label{eq:ED_def2}
\end{align}
or more specifically, $\epsilon \lesssim \text{Poly}(G_{N})$.
Although this may appear stricter than necessary, it enables us to obtain rigorous results. 
For Haar random states, this condition can be further relaxed, see an accompanying work~\cite{Mori:2025mcd}.

We now clarify the types of LOCC protocols considered in this work. 
In general LOCC, the two parties may engage in multiple rounds of classical communication (CC), with each party choosing their measurements based on outcomes shared by the other.
This is often called two-way (2WAY) LOCC.
In contrast, we focus primarily on one-way (1WAY) LOCC, where only one party sends measurement outcomes to the other.
In some holographic scenarios (such as traversable wormholes and holographic scattering), only one-way or one-round communication is naturally feasible. 

Another crucial issue in holography is the role of classical communication itself.
Although CC is conventionally assumed to be freely available in traditional studies in quantum information theory, its applicability to spacelike-separated regions in quantum gravity requires careful justification. 
Hence, we also consider LO-distillable entanglement where LO (local operation) refers to quantum channels that act locally on $A$ and $C$ without sharing CCs. 
Note that a quantum channel can be also thought of as a unitary operator acting on the system and ancilla qubits with trace operations. 

To summarize, our main objects of study are:
\begin{equation}
\begin{split}
&E_{D}^{[\text{1WAY LOCC}]}: \text{one-shot 1WAY LOCC distillable EPR pairs} \\
&E_{D}^{[\text{LO}]} : \text{one-shot LO distillable EPR pairs}
\end{split}
\end{equation}
where we suppress the superscript $(1)$ in $E_{D}^{(1)}$ for brevity, with the understanding that we work in the one-shot setting unless stated otherwise.
The central goal of this paper is to study $E_{D}^{[\text{LOCC}]}(A:C)$ and $E_{D}^{[\text{LO}]}(A:C)$ for a holographic mixed state $\rho_{AC}$.

Previously, $E_D^{[\text{LOCC}]}(A:C)$ has been studied in the literature when $|\psi_{AC}\rangle$ is a pure state (i.e. $B=\emptyset$)~\cite{Bao:2018pvs}. 
In the asymptotic setting, it is well known that
\begin{align}
E_D^{[\text{LOCC(asympt)}]}(A:C) = S_{A}.
\end{align}
Moreover, for pure holographic states in the one-shot setting, it has been shown that
\begin{align}
E_D^{[\text{LOCC($1$)}]}(A:C) = S_{A} + O\qty(\frac{1}{\sqrt{G_N}}) \label{eq:Holo_distill}
\end{align}
i.e., it matches the entanglement entropy $S_A$ at leading order in $1/G_N$ while the subleading correction arises from the variance in the area operator around the saddle point.
This relation is a special property of holographic states and does not hold for generic pure states.

\subsection{LO-distillable entanglement}

\subsubsection*{--- Holographic proposal for $E_{D}^{[\mathrm{LO}]}$} 

\begin{figure}
\centering
\includegraphics[width=0.21\textwidth]{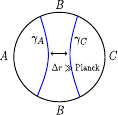}
\caption{EPR pairs cannot be LO-distilled when minimal surfaces are separated in the bulk.
}
\label{fig_ent_distillation}
\end{figure}

We now present our central proposal for one-shot LO-distillable entanglement:
\begin{align}
E_{D}^{[\text{LO}]}(A:C) \approx 0 \qquad \text{if $\gamma_A, \gamma_C$ are separated in the bulk}
\end{align}
at leading order in $1/G_{N}$ where $\gamma_A, \gamma_C$ denote the minimal surfaces of $A$ and $C$ respectively. 
By ``separated in the bulk'', we mean that the minimal separation between $\gamma_A$ and $\gamma_C$, measured in the proper length, is much larger than the Planck length (Fig.~\ref{fig_ent_distillation}).
This applies even when $A$ and $C$ have a connected wedge with $I(A:C)\sim O(1/G_{N})$, as in Fig.~\ref{fig_connected}. 

This proposal stands in contrast to the so-called \emph{mostly bipartite} conjecture~\cite{Cui:2018dyq, Freedman:2016zud}, which suggested that $\rho_{AC}$ effectively contains $\approx \frac{1}{2} I(A:C)$ copies of unitarily rotated EPR pairs.
However, this conjecture has been challenged by arguments showing that $\rho_{AC}$ contains genuine multipartite entanglement~\cite{Akers:2019gcv}.
Our proposal strengthens this perspective, suggesting that the entanglement in $\rho_{AC}$ is not just multipartite but \emph{mostly non-bipartite}.

On the other hand, we propose that LO distillation becomes possible when $\gamma_A$ and $\gamma_C$ are separated only at the Planck scale.
In such cases, we propose
\begin{align}
E_{D}^{[\text{LO}]}(A:C) \approx \frac{1}{4G_N} \text{Area}(\gamma_A \cap \gamma_C) 
, 
\label{eq:recovery}
\end{align}
where $\gamma_A \cap \gamma_C$ denotes the portion where the two minimal surfaces are Planck-scale close.\footnote{
In random tensor networks, the Petz recovery map distills EPR pairs along the overlapping portion $\gamma_A \cap \gamma_C$.
}
However we emphasize that Eq.~\eqref{eq:recovery} is heuristic, due to ambiguities in defining the overlap.

\begin{figure}
\centering
\raisebox{\height}{a)\hspace{10pt}}\raisebox{-0.85\height}{\includegraphics[width=0.2\textwidth]{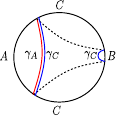}}
\hspace{10pt}
\raisebox{\height}{b)\hspace{10pt}}\raisebox{-0.85\height}{\includegraphics[width=0.24\textwidth]{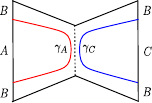}}
\caption{Examples of (nearly) overlapping minimal surfaces. 
}
\label{fig_overlap}
\end{figure}

The simplest case illustrating this is when $B$ is empty, so that $\rho_{AC}$ is pure and $E_{D}^{[\text{LO}]}(A:C) \approx S_A = S_C$.
A more nontrivial example appears when $A$ and $B$ have a disconnected wedge, implying $\gamma_A \subset \gamma_C$, as shown in Fig.~\ref{fig_overlap}(a).
In this case, $A$ is effectively decoupled from $B$ and entangled only with $C$, leading to $E_{D}^{[\text{LO}]}(A:C) \approx S_A$.
Another illustrative case is the two-sided AdS black hole at the $t=0$ slice, where $A$ and $C$ reside on opposite boundaries (Fig.~\ref{fig_overlap}(b)).
When $A$ and $C$ are large enough, their minimal surfaces $\gamma_A$ and $\gamma_C$ approach each other at the Planck scale.

\subsubsection*{--- Haar random state}

We demonstrate that this proposal holds for Haar random states viewed as a toy model of holography.
Consider a tripartite Haar random state $|\psi_{ABC}\rangle$ on $n$ qubits, with each subsystem $A,B,C$ satisfying $n_{R} < \frac{n}{2}$ for $R=A,B,C$. 
This mimics a connected wedge geometry without overlapping minimal surfaces, as schematically illustrated below:
\begin{align}
S_{A} \approx \figbox{1.8}{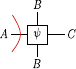}\qquad
S_{C} \approx \figbox{1.8}{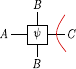}\qquad
S_{AC} \approx \figbox{1.8}{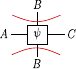} 
\end{align}
where $B$ is split into two subsystems for a clear analogy. 
In this regime, we find that
\begin{align}
E_{D}^{[\mathrm{LO}]}(A:C) = 0,
\end{align}
almost surely, providing strong support for the holographic proposal. 
We will provide a heuristic counting argument while a rigorous proof is presented in an accompanying work~\cite{Mori:2025mcd}. 

\subsubsection*{--- Pretty good bound for holographic mixed state}

For generic holographic states where a direct counting argument fails, we establish an alternative method of bounding $E_{D}^{[\mathrm{LO}]}$.
Specifically, we show that if EPR pairs are LO-distillable, then the Petz recovery map can approximately distill them. 
Viewing a holographic mixed state $\rho_{AC}$ as a quantum channel from $A$ to $C$, the Petz map prepares a \emph{double-copy state} (while ignoring the subleading fluctuations of area operators):
\begin{align}
\sigma_{A,A'} = \figbox{1.7}{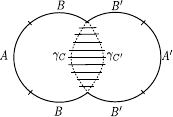}
\end{align}
formed by gluing two copies along minimal surfaces $\gamma_C$ and $\gamma_{C'}$. 
This leads to the bound
\begin{align}
E_{D}^{[\mathrm{LO}]}(A:C) \lesssim \frac{1}{2}I(A:A') = \frac{1}{2}(S_{A} + S_{A'} - S_{AA'}) 
\end{align}
where all quantities are evaluated for $\sigma_{A,A'}$. 

Here, $S_{AA'}$ corresponds to the \emph{reflected entropy} of $\rho_{AB}$~\cite{Dutta:2019gen}, and thus yielding
\begin{align}
S_{AA'} \approx 2E^W(A:B), \qquad   \frac{1}{2}I(A:A') \approx J^{W}(A|C) \equiv S_A - E^W(A:B). 
\end{align}
This gives the bound
\begin{align}
E_{D}^{[\mathrm{LO}]}(A:C) \lesssim \min(J^{W}(A|C), J^{W}(C|A)) .
\end{align}
While the bound is looser than our proposal ($E_{D}^{[\text{LO}]}\approx 0$), this suffices to prove the existence of a regime where $E_{D}^{[\text{LO}]}(A:C)\approx 0$, but $I(A:C)=O(n)$. 

\subsubsection*{--- Entanglement wedge reconstruction}

Our findings also have implications for bulk reconstruction.
The entanglement wedge reconstruction asserts:
\begin{align}
\text{$\phi$ can be reconstructed on $A$} \ \Leftarrow \ \text{$\phi \in \mathcal{E}_A$}
\end{align}
but whether the converse holds (i.e., whether reconstructability implies containment in $\mathcal{E}_A$) remains unclear:
\begin{align}
\text{$\phi$ can be reconstructed on $A$} \ \overset{?}{\Rightarrow} \ \text{$\phi \in \mathcal{E}_A$}.
\end{align}
Our proposal suggests that the converse holds even when the bulk contains $O(1/G_N)$ entropy.

Specifically, we can interpret reconstruction as LO-entanglement distillation, as illustrated for a Haar random state $|\psi\rangle$:
\begin{align}
\figbox{1.8}{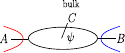}
\end{align}
where $C$ is viewed as bulk DOFs that are holographically encoded into boundary DOFs $A,B$ via an (approximate) isometry $\psi : \mathcal{H}_C\rightarrow \mathcal{H}_{AB}$.
Here, $C$ lies outside both $\mathcal{E}_A$ and $\mathcal{E}_B$, and $E_D^{[\text{LO}]}(A:C), E_D^{[\text{LO}]}(B:C) \approx 0$, implying that unitary operators on $C$ are not reconstructible from either $A$ or $B$.
This predicts the existence of extensive bulk regions whose unitary operators are not recoverable from either side in a bipartition.

\subsection{LOCC-distillable entanglement}

\subsubsection*{--- Holographic proposal for $E_{D}^{[\mathrm{LOCC}]}$}

Next, we present our central proposal for $E_{D}^{[\mathrm{LOCC}]}$:
\begin{equation}
\begin{split}
&E_{D}^{[\text{1WAY LOCC}]}(A \leftarrow C) \approx J^W(A|C) \equiv S_{A} - E^{W}(A:B) \\
&E_{D}^{[\text{1WAY LOCC}]}(A:C) \approx J^W(A:C) = \max \big(J^W(A|C),J^W(C|A) \big)
\end{split}
\end{equation}
at leading order in $1/G_{N}$. 
These proposals follow from the holographic relation for $E_{F}$: 
\begin{align}
E_{F}(A:C) \approx E^W(A:C).
\end{align}
Furthermore, we obtain the Koashi-Winter entanglement monogamy relation
\begin{align}
E_{D}^{[\text{1WAY LOCC}]}(A \leftarrow C) + E_{F}(A:B) \approx S_{A}
\end{align}
in the holographic context. 

\subsubsection*{--- Haar random state}

We demonstrate that this proposal holds for tripartite Haar random state $|\psi_{ABC}\rangle$.
Namely, we present a 1WAY LOCC protocol that distills $\approx \max (0, n_A -n_{B} )$ EPR pairs by performing projective measurements on  part of $C$ and sending the outcomes to $A$.  
We then prove that this protocol is optimal at leading order, suggesting 
\begin{align}
E_{D}^{[\text{1WAY LOCC}]}(A \leftarrow C)\approx \max (0, n_A -n_{B} ).
\end{align}
For Haar random states, entanglement cross section $E^W(A:B)$ can be identified as
\begin{align}
E^W(A:B) = \min \left( \figbox{2.0}{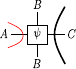} \ \figbox{2.0}{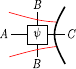} \right) \ = \min (n_{A}, n_{B} )
\end{align}
which yields 
\begin{align}
J^W(A|C) \approx \max (0, n_A -n_{B} ), 
\end{align}
and thus 
\begin{align}
E_{D}^{[\text{1WAY LOCC}]}(A\leftarrow C)\approx J^W(A|C).
\end{align}
Moreover, recalling that~\cite{Hayden_2006} 
\begin{align}
E_{F}(A:B) \approx \min (n_{A}, n_{B} ) = E^W(A:B),
\end{align}
we find the entanglement monogamy relation $E_{D}^{[\text{1WAY LOCC}]}(A \leftarrow C) + E^W(A:B)\approx n_{A}$.

\subsubsection*{--- Holographic state}

We then conjecture that the proposal holds for generic holographic states by presenting analytical bounds for fixed-area states. 
We construct a 1WAY LOCC protocol based on holographic measurements (see Section~\ref{sec:hol-meas} for their definition and assumptions), in which EoW brane-like objects are introduced along part of $\gamma_C$ via projective measurements.
This protocol distills $\approx J^W(A|C)$ EPR pairs, hence establishing 
\begin{align}
E_{D}^{[\text{1WAY LOCC}]}(A\leftarrow C) \gtrsim J^W(A|C).
\end{align}
On the other hand, the following inequality can be shown to hold generally:
\begin{align}
E_{D}^{[\text{1WAY LOCC}]}(A\leftarrow C) \lesssim J(A|C). 
\end{align}
The optimality then follows from the holographic proposal $E_{F}\approx E^W$ (or its dual proposal $J(A|C)\approx J^W(A|C)$): 
\begin{align}
E_{D}^{[\text{1WAY LOCC}]}(A\leftarrow C) \approx J^W(A|C).
\end{align}

\subsubsection*{--- On the $E_{F}\approx E^W$ proposal}
Since our proposal for $E_{D}^{[\text{1WAY LOCC}]}$ is closely related to the $E_F(A:B) \approx E^W(A:B)$ proposal, it is important to critically examine its validity.
We present several pieces of evidence supporting both $E_F(A:B) \approx E^W(A:B)$ and its dual formulation $J(A|C) \approx J^W(A|C)$.

\begin{enumerate}[i)]
\item We revisit a no-go argument previously raised by Umemoto, which claimed that $E_F \neq E^W$ in certain holographic regimes~\cite{Umemoto:2019jlz}.
We show that this apparent contradiction can be resolved by interpreting the proposal as holding only at leading order in $1/G_N$.

\item Additional support for $J(A|C) \approx J^W(A|C)$ arises from the generalized RT formula, which includes the matter contribution. Namely, this implies that disentangled projective measurements across the entanglement wedge cross section achieve the maximal entropy reduction on $A$.

\item Considerations from bulk causality provide an independent argument: since region $C$ is spacelike separated from the entanglement wedge $\mathcal{E}_{AB}$, any local operation or measurement on $C$ should not cause a leading-order backreaction in $\mathcal{E}_{AB}$.
This provides another support for $J(A|C) \approx J^W(A|C)$. 
\end{enumerate}

\subsubsection*{--- Holographic bound entanglement}

Our main results can be summarized as the following hierarchy of holographic entanglement measures:
\begin{align}
\boxed{ \ 
\ \underbrace{E_{D}^{[\text{LO}]}}_{\approx 0} \leq \text{hash}(A:C)\leq \underbrace{E_{D}^{[\text{1WAY LOCC}]}}_{\approx J^W(A:C)} \leq \underbrace{E_{sq}}_{\approx \frac{1}{2}I(A:C)} \leq \underbrace{E_{F}, E_{P}}_{\approx E^{W}(A:C)} \leq \min(S_A,S_C) \
\ }
\end{align}
at leading order when minimal surfaces $\gamma_A$ and $\gamma_C$ do not overlap.\footnote{The logarithmic negativity satisfies $E_{N}\approx \frac{1}{2}I(A:C)$ for random tensor networks and fixed area states~\cite{Dong:2021clv}. For generic holographic states, there may be deviations from $\frac{1}{2}I(A:C)$, see~\cite{Dong:2024gud} for a recent discussion.}
These entanglement measures can have $O(1/G_N)$ separations. 

\begin{figure}
\centering
\raisebox{\height}{\hspace{5pt}}\raisebox{-0.95\height}{\includegraphics[width=0.9\textwidth]{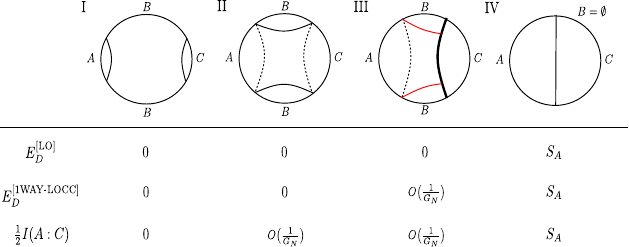}}
\caption{Distillable entanglement and connected wedge as sizes of $A,C$ increase from left to right. 
The transition point between II and III is obtained in Appendix \ref{app:hol-calc}.
}
\label{fig_claim}
\end{figure}

One important implication is that the presence of a connected entanglement wedge does not guarantee distillable entanglement under 1WAY LOCC.
Consider the pure AdS$_3$ setup where $A$ and $C$ are symmetric and their sizes vary (see Fig.~\ref{fig_claim}):

\begin{enumerate}[I)]
\item When $A$ and $C$ occupy less than quarters of the whole system, the entanglement wedge of $AC$ is disconnected with $I(A:C) \approx 0$, and we have $,E_{D}^{[\text{1WAY LOCC}]} \approx  0$ at leading order.

\item When $A$ and $C$ occupy slightly more than quarters, the entanglement wedge of $AC$ will be connected with $I(A:C) \sim O(1/G_{N})$, but $J^W(A|C), J^W(C|A) \approx 0$, and thus $,E_{D}^{[\text{1WAY LOCC}]} \approx 0$ at leading order.

\item When $A$ and $C$ occupy much more than quarters, $J^W(A:C) \sim O(1/G_N)$, and one can distill $J^W(A|C)$ EPR pairs at leading order.
However, $E_{D}^{[\text{1WAY LOCC}]}(A:C)$ remains smaller than $\frac{1}{2}I(A:C)$.

\item When $B$ becomes empty and $\gamma_A,\gamma_C$ overlap, we have 
$E_{D}^{[\text{LO}]},E_{D}^{[\text{1WAY LOCC}]}, \frac{1}{2}I(A:C) \approx S_{A}$ at leading order.
\end{enumerate}

In quantum information theory, entangled states that are not distillable are called bound entangled states~\cite{Horodecki:1998kf}. 
Our results suggest that holographic states with connected entanglement wedge, but with $J^W(A:C)=0$, may be viewed as examples of bound entangled states in one-shot 1WAY settings at the limit of $G_{N}\rightarrow 0$.

\subsection{On subleading contribution}

\subsubsection*{--- Subleading effects}

We discuss possible subleading contributions to $E_{D}^{[\text{LO}]}$ and $E_{D}^{[\text{LOCC}]}$. 
In particular, we identify three potential physical mechanisms that may lead to nonzero subleading distillable entanglement:

\begin{enumerate}[i)]
\item \emph{Traversable wormholes}: One-way LOCC versions of the traversable wormhole protocol may allow for entanglement distillation in two-sided BTZ black hole geometries.
\item \emph{Holographic scattering}: The requirement of a connected entanglement wedge for bulk scattering processes hints at the possibility of subleading LOCC-distillable entanglement.
\item \emph{Planck-scale proximity}: When the minimal surfaces $\gamma_A$ and $\gamma_C$ approach within the Planck scale, significant corrections to $E_{D}^{[\text{LO}]}(A:C)$ are expected.
\end{enumerate}

\subsubsection*{--- Instability of entanglement measures}

Certain entanglement measures are highly sensitive to small perturbations. 
We will highlight this subtlety by studying the entanglement properties of an isotropic state:
\begin{align}
\rho_{AA'} = \frac{1-F}{{d_A}^2-1} (I - |\text{EPR}\rangle \langle \text{EPR} |)  +  F |\text{EPR}\rangle \langle \text{EPR} |,\quad 0\le F\le 1
\end{align}
where $d_A = 2^{n_A}$. When $F$ is exponentially small in $n_A$, $\rho_{AA'}$ is close to the maximally mixed state, and thus is nearly disentangled.
Despite that, some entanglement measures score very high for $\rho_{AA'}$: 

\begin{enumerate}[i)]

\item \emph{Logarithmic negativity $E_{N}$}: We find $E_N \approx \max(n_A + \log_2 F, 0)$,  suggesting that $E_{N}$ can be extensive even when $\rho_{AA'}$ is almost separable. 

\item \emph{Exact entanglement cost $E_C^{[\mathrm{exact}]}$}: 
It quantifies the number of EPR pairs per copy required to \emph{exactly} create the target state before taking the asymptotic limit. We find $E_C^{[\mathrm{exact}]}\approx E_{N}$ while the commonly-used entanglement cost satisfies $E_C\approx 0$.
\end{enumerate}

\subsection*{Plan of the paper}

This paper is organized into three parts. 
\begin{enumerate}
\item[] Part I: LO-distillable entanglement\\
Part II: LOCC-distillable entanglement\\
Part III: Relevant topics (subleading contribution and outlook)
\end{enumerate}
This paper contains the following appendices.
\begin{enumerate}
\item[] Appendix~\ref{app:hol-calc}~\ref{app:metric}: Transition point for $J^W(A|C)>0$ in pure AdS$_3$ and two-sided BTZ\\
Appendix~\ref{app:Haar}~\ref{app:TN}: Entanglement structure of double-copy states for Haar and holography
\end{enumerate}

\addcontentsline{toc}{section}{Part I : LO-distillable entanglement}
\section*{Part I : LO-distillable entanglement}

In the next four sections, we will discuss LO-distillable entanglement in holography and its physical implications. Our central proposal is 
\begin{align}
E_{D}^{[\text{LO}]}(A:C) \approx 0 \qquad \text{if minimal surfaces $\gamma_A$ and $\gamma_C$ are separated}.
\end{align}
In section~\ref{sec:LO_Haar}, we show that Haar random states, when viewed as toy models of holography, indeed satisfy this proposal. 

In section~\ref{sec:LO_Petz}, we derive an alternative bound on $E_{D}^{[\text{LO}]}$ using the Petz map:
\begin{align}
E_{D}^{[\text{LO}]}(A:C) \lesssim \frac{1}{2}I(A:A') \equiv \frac{1}{2}(S_{A} + S_{A'} -S_{AA'})
\end{align}
where $I(A:A')$ is evaluated for the canonical purification of $\rho_{AB}$.
In section~\ref{sec:LO_holography}, we apply this bound to holographic states and obtain
\begin{align}
E_{D}^{[\text{LO}]}(A:C) \lesssim \min\big( J^W(A|C), J^W(C|A) \big)
\end{align}
where $J^W(A|C)= S_{A} - E^W(A:B)$. 
This proves the existence of regimes where $E_{D}^{[\text{LO}]}(A:C)\approx 0$ but with connected entanglement wedge. 

In section~\ref{sec:LO_logical}, we discuss the implications of our result from the perspective of entanglement wedge reconstruction, suggesting the converse of entanglement wedge reconstruction and predicting the shadow of entanglement wedge, an extensive bulk region that cannot be reconstructed on $A$ or $B = A^c$ in bipartition. 

\section{LO-distillable entanglement in Haar random state}\label{sec:LO_Haar}

In this section, we discuss LO-distillable entanglement $E_{D}^{[\text{LO}]}$ in  Haar random states.

\subsection{Toy model of holography}

Consider an $n$-qubit Haar random state $|\psi_{AB}\rangle$ in a bipartition into $A$ and $B$ with $n_A$ and $n_B = n-n_A$ qubits respectively. 
We have 
\begin{align}
S_{A} \approx \min (n_A, n-n_A )  
\end{align}
at leading order in $n$, due to the so-called Page's theorem~\cite{Lubkin:1978nch, Lloyd:1988cn, Page:1993df}.\footnote{  
For $n_A < n_B$, we have $\mathbb{E} \Vert \rho_A - \frac{1}{2^{n_A}}I_A \Vert_1 \lesssim 2^{(n_A-n_B)/2}$ where $\mathbb{E}$ represents Haar average and $\Vert O\Vert_1=\Tr\abs{O}$.
}
This can be interpreted as the RT-like formula with the area ($=$ the number of qubits across the cut) minimization:
\begin{align}
S_{A} \approx \min \big(\figbox{2.0}{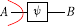} , \figbox{2.0}{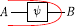} \big).  
\end{align}
For $n_A < n_B$, $A$ is nearly maximally entangled with a $2^{n_A}$-dimensional subspace in $B$, suggesting that $E_{D}^{[\text{LO}]}(A:B)\approx n_A$.
Specifically, since  the reduced state $\rho_A$ has an almost flat spectrum, the Schmidt decomposition takes the approximate form $\ket{\psi}_{AB}\approx (I_A \otimes V_{A^\prime\rightarrow B})\ket{\text{EPR}}^{\otimes n_A}_{AA^\prime} $, where $V$ is an isometry. 
Applying $V^\dag$ then distills $n_A$ EPR pairs.
In fact, one can distill $\approx n_A$ approximate EPR pairs with a vanishing error $\epsilon \rightarrow 0$ (as $n\rightarrow \infty$) by applying the Petz recovery map on $B$. 
This LO-distillability can be understood by overlapping minimal surfaces, namely $\gamma_A = \gamma_B$. 

\begin{figure}
\centering
\raisebox{\height}{a)\hspace{10pt}}\raisebox{-0.8\height}{\includegraphics[width=0.15\textwidth]{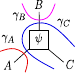}}
\hspace{10pt}
\raisebox{\height}{b)\hspace{10pt}}\raisebox{-0.8\height}{\includegraphics[width=0.15\textwidth]{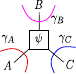}}
\caption{LO-distillable entanglement $E_{D}^{[\text{LO}]}(A:C)$ and minimal surfaces. a) Overlapping minimal surfaces. b) Separated minimal surfaces.
}
\label{fig_Haar_comparison}
\end{figure}

Next, consider a tripartite $n$-qubit Haar random state $|\psi_{ABC}\rangle$ on $A$, $B$, and $C$.
Let us first assume $n_{C} > \frac{n}{2}$. 
We then have 
\begin{align}
S_{C} = S_{AB} \approx \figbox{2.0}{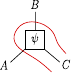} = n_A + n_B 
\end{align}
which suggests that $A$ and $B$ are nearly decoupled from each other, and thus $A$ is nearly maximally entangled with $C$.\footnote{ 
A previous work~\cite{Haar_separable} showed that, for $n_A = (\frac{1}{5}+\epsilon)n$, $n_B = (\frac{1}{5}+ \epsilon)$, and $n_C = (\frac{3}{5}-2\epsilon)n$ ($\epsilon >0$), $\rho_{AB}$ is (almost surely) not separable. 
That $A$ and $B$ are nearly decoupled from each other does not contradict this result. 
Namely, the Page's theorem states that $\rho_{AB}\approx \frac{1}{2^{n_A + n_B}} I_{A} \otimes I_{B}$ with an exponentially small error, leaving a possibility of $\rho_{AB}$ being non-separable.}
In this case, there exists a unitary $U_{C}$ that LO-distills EPR pairs between $A$ and $C$ with $E_{D}^{[\text{LO}]}(A:C)\approx n_A$.
Again, this can be understood as a consequence of overlapping minimal surfaces, as shown in Fig.~\ref{fig_Haar_comparison}(a). 
The same argument applies for $n_A>\frac{n}{2}$ by exchanging $A$ and $C$.
When $n_B>\frac{n}{2}$, $A$ and $C$ are nearly fully decoupled from each other as $I(A:C)\approx 0$ and we have $E_{D}^{[\text{LO}]}(A:C)\approx 0$.

Finally, when $n_A,n_B,n_C < \frac{n}{2}$, we have
\begin{align}
S_{R} \approx  n_R, \qquad R=A,B,C
\end{align}
where the minimal surface $\gamma_R$ of subsystem $R=A,B,C$ does not contain the tensor at the center (Fig.~\ref{fig_Haar_comparison}(b)). 
This mimics the situation with connected wedge as in Fig.~\ref{fig_connected}. Namely, by splitting $B$ into two subsystems, we can schematically draw the minimal surface of $AC$ as follows
\begin{align}
S_{AC} \approx \ \figbox{2.0}{fig_Haar_wedge} = n_B \qquad I(A:C) \approx n_{A} + n_{C} - n_{B} \sim O(n).
\end{align}

\subsection{Atypicality of bipartite entanglement}

The central question is whether one can LO-distill EPR pairs from $\rho_{AC}$ when $n_A,n_B,n_C < \frac{n}{2}$.
Observing that minimal surfaces $\gamma_A, \gamma_C$ are separated by the tensor at the center (Fig.~\ref{fig_Haar_comparison}(b)), our proposal predicts $E_{D}^{[\text{LO}]}(A:C)\approx 0$.
In an accompanying work~\cite{Mori:2025mcd}, we prove that this is indeed the case:

\begin{theorem}\label{theorem:LO-Haar}
(Informal)
Given a tripartite Haar random state $|\psi_{ABC}\rangle$, assume $n_A, n_B, n_C < \frac{n}{2}$. We then have
\begin{align}
E_{D}^{[\mathrm{LO}]}(A:C) \approx 0 \label{eq:Haar_LO}
\end{align}
at leading order in $n$. 
\end{theorem}

A rigorous formulation of this result, along with a formal mathematical proof, is presented in~\cite{Mori:2025mcd}.\footnote{We need to define $E_{D}^{[\text{LO}]}(A:C)$ rigorously with the EPR fidelity tolerance and bound the probability of sampling a state with approximate EPR pairs.}
In this paper, we present a heuristic counting argument by focusing on local unitary (LU) operations, instead of generic LOs. 
The essential difference between LU and LO is whether one allows the use of ancilla qubits or not. 
The proof for LOs requires a strengthening of the argument with the measure concentration as shown in~\cite{Mori:2025mcd}.

We begin by presenting a useful insight concerning Haar random states by following~\cite{Roberts:2016hpo}.
In an $n$-qubit system, there are $2^n$ mutually orthogonal states which may be labeled by $|j\rangle$ with $j=1,\cdots, 2^n$. 
But if we relax the orthogonality condition, one can show that there are doubly-exponentially ($e^{e^{O(n)}}$) many states that are nearly orthogonal to each other.
This can be readily understood by considering the following family of $n$-qubit quantum states
\begin{align}
|\psi(c)\rangle = \frac{1}{2^{n/2}}( c_{1} |1 \rangle + c_2 |2 \rangle + \cdots + c_{2^n} |2^n \rangle), \quad c_j = \pm 1,
\end{align}
where $c = (c_1, \cdots, c_{2^n})$. 
By choosing $c$ and $c'$ randomly, we find 
\begin{align}
|\langle \psi(c) | \psi(c') \rangle| =  \Big|\frac{1}{2^{n}}\sum_{j=1}^{2^n} c_j c_j'\Big| \approx \frac{1}{2^{n/2}} \underset{n\rightarrow \infty}{\longrightarrow} 0.
\end{align}
This suggests that there are at least doubly-exponential, \emph{nearly} orthogonal quantum states in the Hilbert space.
The upshot is that choosing a Haar random state essentially means picking a state from a set of doubly-exponentially many nearly orthogonal quantum states.
This heuristics can be made rigorous by introducing a tolerance $\epsilon$ on fidelity overlaps, i.e. an $\epsilon$-net~\cite{Hayden_2004}. 

\begin{figure}
\centering
\includegraphics[width=0.23\textwidth]{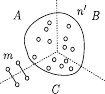}
\caption{Tripartite state $|\psi_{ABC}\rangle$ with LU-distillable EPR pairs. 
}
\label{fig_counting}
\end{figure}

Given a quantum state $|\psi_{ABC}\rangle$ with $n_{A},n_B,n_C < \frac{n}{2}$, suppose that it is possible to distill $m$ EPR pairs applying local unitaries $U_A \otimes U_{C}$. 
This leaves $n'= n - 2m$ qubits decoupled from EPR pairs as shown in Fig.~\ref{fig_counting}. The decoupled $n'$-qubit state can be arbitrary. 
We now argue that such quantum states with LU-distillable EPR pairs are extremely rare.
Denote the total number of nearly orthogonal states with $\sim \epsilon$ mutual overlaps by $\Phi_{\text{state}}(n)$. 
Precise form of $\Phi_{\text{state}}(n)$ is not essential, and we only need that $\Phi_{\text{state}}(n)$ scales doubly-exponentially. 
Similarly, let us denote the total number of $n$-qubit unitary operators by $\Phi_{\text{unitary}}(n)$.
Recalling that an $n$-qubit unitary can be viewed as a $2n$-qubit state via the Choi isomorphism, we have 
\begin{align}
\Phi_{\text{unitary}}(n) < \Phi_{\text{state}}(2n).
\end{align}

Let us estimate the total number of states with LU-distillable EPR pairs. First, one can unitarily rotate $A,C$ by $U_A \otimes U_{C}$. The number of such unitary operators is given by 
\begin{align}
\Phi_{\text{unitary}}(n_A) \Phi_{\text{unitary}}(n_C) <  
\Phi_{\text{unitary}}(n_R + c), \qquad n_{R} = \max\big( n_A, n_C \big)
\end{align}
where $c > 0$ is an $O(1)$ constant. The upper bound comes from the fact $\Phi_{\text{unitary}}$ being doubly exponential.
Since the number of decoupled $n'$-qubit states is given by $\Phi_{\text{state}}(n')$, the total number of states with LU-distillable EPR pairs is upper bounded by 
\begin{align}
\Phi_{\text{state}}(2n_R + 2c)  \Phi_{\text{state}}(n') 
\end{align}
which is much less than $\Phi_{\text{state}}(n)$ as long as $n_R < \frac{n}{2}$. 
Hence, by randomly choosing a state from a set of $\Phi_{\text{state}}(n)$ nearly orthogonal states, it is extremely unlikely to obtain a state with LU-distillable entanglement, suggesting $E_{D}^{[\text{LU}]}(A:C) \approx 0$ at leading order in $n$.
Although we considered the LU-distillability of perfect EPR pairs in the above analysis, relaxing this condition to admit approximate EPR pairs does not significantly change the analysis. 
It is worth noting that this argument breaks down when $n' \approx n$ (i.e. no EPR pairs) or $2n_R \approx n$ (i.e. one subsystem contains more than half of the system). 
In the latter case, we can indeed perform LU-distillation of EPR pairs. 

\section{A pretty good bound for LO-distillable entanglement}\label{sec:LO_Petz}

Next, we provide another bound that extends a powerful result due to Barnum and Knill, concerning entanglement fidelity in quantum error corrections~\cite{Barnum:2002bfd}. 
In a nutshell, our argument relies on the fact that the Petz recovery map is a \emph{pretty good} decoder, and thus it suffices to study the decoding performance of the Petz recovery map in discussing LO-distillable entanglement.  

One merit of this approach is that it uses entropic quantities without relying on the counting argument, and thus is readily applicable to holographic states.
Another benefit is that it provides a concrete operational meaning to the reflected entropy. 
The downside is that it provides a weaker upper bound, namely $E_{D}^{[\text{LO}]}(A:C)\lesssim \min(\max(0, S_{A}-S_{AC}), \max(0,S_{C}-S_{AC}))$...!
It is still worth emphasizing that this bound is stronger than the naive bounds from the mutual information $I(A:C)$ or the logarithmic negativity $E_N(A:C)$. 

\subsection{Petz map is pretty good}

We begin by interpreting LO-distillable entanglement as the decodability in a quantum error correcting code. 
Recall that, given a Haar random state $|\psi_{ABC}\rangle$, one can view it as a quantum error-correcting code with an approximate encoding isometry $\Lambda: A \rightarrow BC$ via the state-channel duality as $\rho_A$ can be approximated by the maximally mixed state in trace distance.
Namely, we can write 
\begin{align}
|\psi_{ABC}\rangle \approx \Lambda \otimes I_{A'} |\text{EPR}\rangle_{A'A} 
= \figbox{2.0}{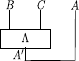}
\end{align}
where $\Lambda$ can be treated as a Haar random isometry and the initial state on $A'A$ is chosen to be a canonical purification of $\rho_A\approx\frac{I_A}{d_A}$, namely $|\text{EPR}\rangle_{A'A}$. Henceforth, we denote the dimension of a subsystem $X$ containing $n_X$ qubits by $d_X=2^{n_X}$.
Below, we will mostly focus on LU-distillable entanglement for simplicity of discussion.\footnote{Extension to LO-distillable entanglement requires a slight modification of the original Barnum-Knill argument.}

\begin{figure}
\centering
\includegraphics[width=0.4\textwidth]{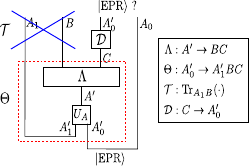}
\caption{
Entanglement distillation as decoding problem in a quantum error correcting code. The unitary $U_A$ is a basis transformation corresponding to a decomposition $A^\prime\rightarrow A_0^\prime A_1^\prime$ and $\mathcal{D}$ is a decoder acting on $C$.
}
\label{fig_BK_theorem}
\end{figure}

Suppose that $n_{A_0}$ EPR pairs can be LU-distilled from $\rho_{AC}$ by applying some local unitary operator $U_A \otimes U_C$. 
Using the state-channel duality, this can be viewed as a decoding problem in a quantum error code as shown in Fig.~\ref{fig_BK_theorem} where EPR pairs are to be prepared on $A_0$ and $A_0'$. 
Specifically, we have the following processes:
\begin{enumerate}[i)]
\item \emph{Encoding}: An isometry $\Theta:A_0' \rightarrow A_1 BC$ encodes  an $n_{A_0}$-qubit input state into an $(n_{A_1}+n_B + n_C)$-qubit output state.
\item \emph{Noise}: The system undergoes an erasure noise channel $\mathcal{T}(\cdot) = \Tr_{A_1 B}(\cdot)$.
\item \emph{Decoding}: A decoding channel $\mathcal{D}: C\rightarrow A_0'$ is applied on $C$ to generate EPR pairs on $A_0' A_0$. 
\end{enumerate}

In this interpretation, the \emph{average decoding success} can be quantified by the entanglement fidelity 
\begin{align}
F_{\mathcal{D}}\equiv 
\langle \text{EPR}_{A_0 A_0'} | \mathcal{I}_{A_0} \otimes \mathcal{D}_{C\rightarrow A_0'}(\rho_{A_0 C}) |  \text{EPR}_{A_0 A'_0} \rangle.
\end{align}
If (approximate) EPR pairs can be prepared on $A_0A_0'$, we would have $F_{\mathcal{D}} \geq 1 - \epsilon$ with some $\epsilon \rightarrow 0$ as $n \rightarrow \infty$. 
(Recall Eq.~\eqref{eq:ED_def2}.)
Below, we will prove that such $\epsilon$ cannot exist for $n_{A_0} \gtrsim \max(0, n_A -n_B )$, suggesting $E_{D}^{[\text{LU}]}(A:C)\lesssim \max(0, n_A -n_B )$.

In~\cite{Barnum:2002bfd}, Barnum and Knill proved the following inequality.

\begin{theorem}\label{thm:barnum-knill}
For any decoder $\mathcal{D}$, we have
\begin{align}
F_{\mathcal{D}_{\mathrm{Petz}}} \geq F_{\mathcal{D}}^2 \label{eq:thm-2}
\end{align}
where $\mathcal{D}_{\mathrm{Petz}}$ denotes the Petz recovery map
\begin{align}
\mathcal{D}_{\mathrm{Petz}}(\cdot) \equiv
\rho^{1/2}\mathcal{N}^{\dagger}
\big[ \mathcal{N}(\rho)^{-1/2} (\cdot) \mathcal{N}(\rho)^{-1/2} \big]   
\rho^{1/2}.
\end{align}
Here $\mathcal{N}$ is a quantum channel representing both encoding and noise ($\mathcal{N}=\mathcal{T}\circ \Theta$ in our setup) and $\rho$ denotes the reference state ($\rho_{A_0} \approx \frac{I_{A_0}}{d_{A_0}}$ in our setup).
\end{theorem}

The upshot of this result is that, if there exists a good decoder $\mathcal{D}$ which distills EPR pairs with high fidelity, then the Petz map will also distill EPR pairs with reasonably high fidelity. 
Namely, if $F_{\mathcal{D}}= 1-\epsilon$ with small $\epsilon$, $F_{\mathcal{D}_{\text{Petz}}}\geq 1 - 2\epsilon +O(\epsilon^2)$.
In other words, the Petz map is a \emph{pretty good} (if not the best) decoder. 
Hence, as long as one knows that entanglement distillation is possible by some protocol, one can also use the Petz recovery map to distill entanglement.
Considering a contraposition of this statement, one can then prove that EPR pairs cannot be LU-distilled by verifying that the Petz recovery map fails to distill EPR pairs.  

\subsection{Double-copy state and reflected entropy}

Let us apply the Barnum-Knill bound to Haar random states.
For a Haar random state $|\psi_{ABC}\rangle$, the action of the Petz map significantly simplifies as marginal density matrices are nearly maximally mixed. 
We then find that the Petz map generates the following state
\begin{align}
\sigma_{A_0 A_0'} = \Tr_{A_1 A^\prime_{1}} (\sigma_{AA'})
\end{align}
where $\sigma_{AA'}$ is a reduced density matrix of the \emph{double-copy state}:
\begin{align}
\sigma_{AA'}=\Tr_{BB'}\big(\big|\Phi^{(\text{double})}_{ABA'B'}\big\rangle\big\langle \Phi^{(\text{double})}_{ABA'B'}\big| \big), \qquad \big|\Phi^{(\text{double})}_{ABA'B'}\big\rangle \approx \sqrt{d_C}\cdot \figbox{2.0}{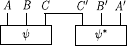}
\end{align}
where two copies $|\psi_{ABC}\rangle\otimes |\psi_{A'B'C'}^*\rangle$ are prepared and $CC'$ are projected onto $|\text{EPR}\rangle_{CC'}$. 

Note that the double-copy state $\big|\Phi^{(\text{double})}_{ABA'B'}\big\rangle$ is identical to the so-called \emph{canonical purification} of $\rho_{AB}$, namely $|\sqrt{\rho_{AB}}\rangle$. 
This is due to that the spectrum of $\rho_{AB}$ (and $\rho_C$) is flat. 
It is worth noting that the Grover recovery algorithm from~\cite{Yoshida:2017non} unitarily prepares an approximation of the double-copy state. See~\cite{Nakayama:2023kgr,Li:2024tcm} for relevant observations, and see also~\cite{Gilyen:2020gmg, Utsumi:2024bjx} for generalization of such algorithms using quantum singular value transformation.

We now study if $\rho_{AA'}$ contains LO-distillable entanglement or not by evaluating the mutual information $I(A:A')$ which can be written explicitly as
\begin{align}
I(A:A') = S_{A} + S_{A'} - S_{AA'} \qquad \text{(evaluated for $\big|\Phi^{(\text{double})}_{ABA'B'}\big\rangle$)},
\end{align}
where $S_{A}=S_{A'} \approx n_{A}$. The third term $S_{AA'}$ is called the \emph{reflected entropy}~\cite{Dutta:2019gen} of $\rho_{AB}$. 

Previous works~\cite{Akers:2021pvd, Akers:2022zxr, Czech:2023rbh} performed careful and detailed studies of entanglement properties of the double-copy state $\big|\Phi^{(\text{double})}_{ABA'B'}\big\rangle$ constructed from a Haar random state. 
Their main finding is 
\begin{align}
S_{AA'} \approx 2 \min(n_A,n_B) = \min\qty(\figbox{0.2}{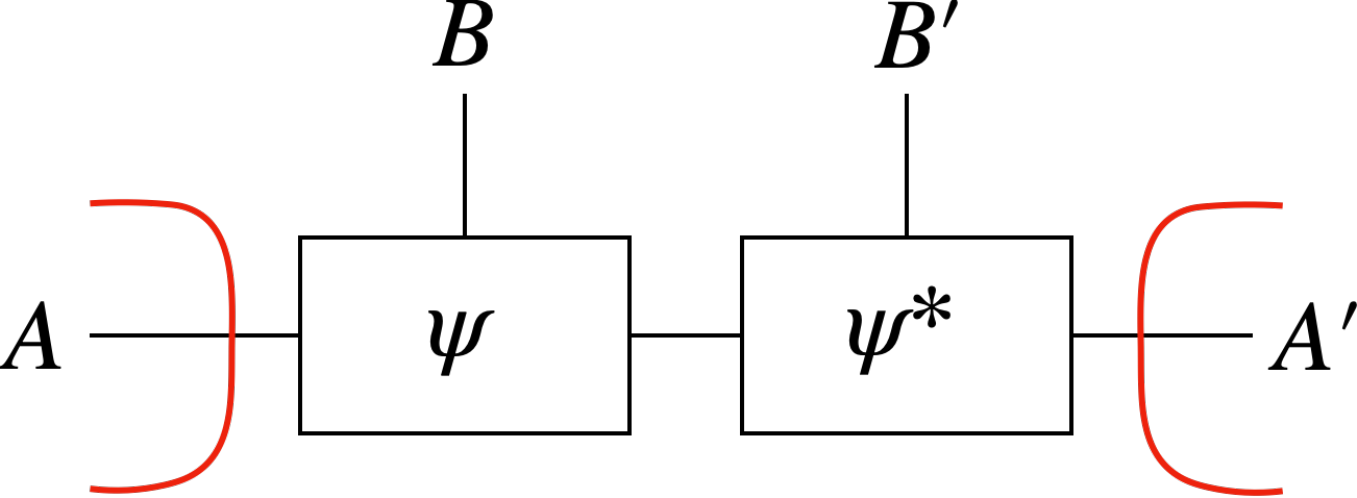}\, , \figbox{0.2}{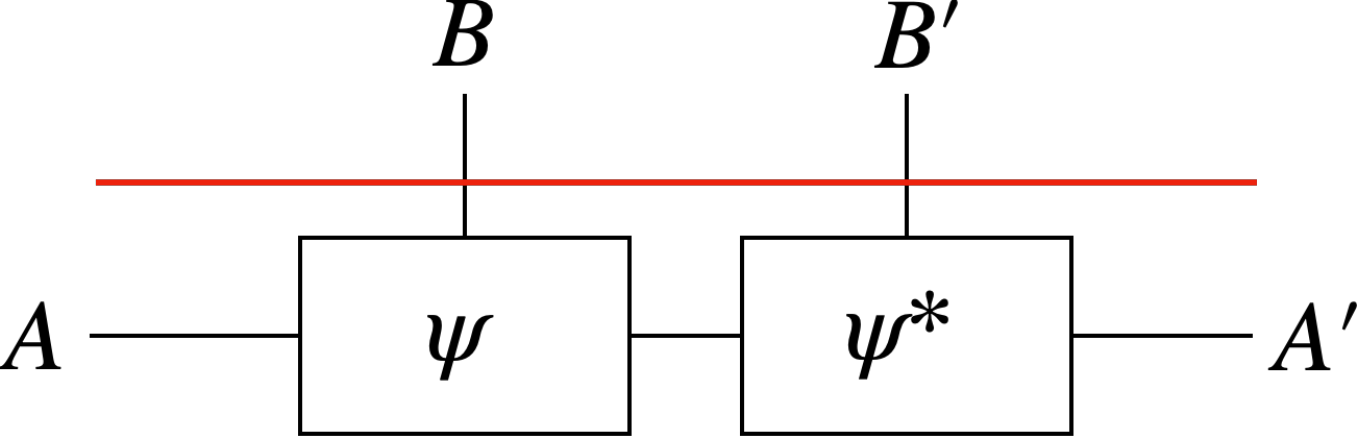})
\end{align}
where entanglement entropy is given by the RT-like formula in the double-copy geometry. 
That $S_{AA'}$ obeys the RT-like formula is a non-trivial statement as $\big|\Phi^{(\text{double})}_{ABA'B'}\big\rangle$ involves two identical random states. 
In Appendix~\ref{app:Haar}, we will sketch the derivation of this result. 
Relying on this result, we find
\begin{align}
\frac{1}{2}I(A:A') \approx \max(0, n_A - n_B ).\label{eq:ABC1}
\end{align}

As we will discuss further below, this enables us to conclude $E_{D}^{[\text{LO}]}(A:C) \lesssim  \max(0, n_A - n_B )$ when the EPR fidelity is close to unity. Repeating the same analysis by exchanging $A$ and $C$, we arrive at the following bound
\begin{align}
\boxed{ \ 
E_{D}^{[\text{LO}]}(A:C) \lesssim \min(\max(0, n_A - n_{B} ), \max(0, n_C - n_{B} )). \label{eq:LO-bound}
\ }
\end{align}
While this bound is weaker than theorem~\ref{theorem:LO-Haar}, it has an advantage of using entropic quantities and not heavily relying on Haar randomness. 
Crucially, this approach can potentially avoid the use of counting arguments. 
This bound remains valid in the \emph{asymptotic} setting as well. 

\subsection{Pretty good bound}

When deriving Eq.~\eqref{eq:LO-bound}, we have implicitly assumed that $E_{D}^{[\text{LO}]}(A:C) \lesssim \frac{1}{2}I(A:A')$. 
Strictly speaking, in one-shot settings, such an inequality needs to be established along with the EPR fidelity tolerance $F$. 

\begin{theorem}\label{thm:pretty_good}
Given a tripartite state $|\psi_{ABC}\rangle$ with $\Vert\rho_{A}-\frac{I_A}{d_A}\Vert_1=o(\frac{1}{n_A})$, let $0< F \leq 1$ be the EPR fidelity tolerance for $E_{D}^{[\mathrm{LO}]}(A:C)$. We then have 
\begin{align}
F^2 E_{D}^{[\mathrm{LO}]}(A:C) \le \frac{1}{2}I(A:A') +O(1).
\end{align}
Here, $\frac{1}{2}I(A:A')$ is evaluated for $\big|\Phi^{(\mathrm{double})}_{ABA'B'}\big\rangle$, an output of the Petz recovery map
\begin{align}
\big|\Phi^{(\mathrm{double})}_{ABA'B'}\big\rangle \equiv 
(I_{AB} \otimes \mathcal{D}_{\mathrm{Petz}(C\rightarrow A'B')})\big( |\psi_{ABC}\rangle \langle \psi_{ABC}| \big).
\end{align}
\end{theorem}

Here, $F$ controls the quality of distilled EPR pairs where $F=1$ corresponds to perfect EPR pairs. 
We are primarily interested in $F\approx 1$ cases where the above bound leads to $E_{D}^{[\mathrm{LO}]}(A:C) \lesssim \frac{1}{2}I(A:A')$, leading to Eq.~\eqref{eq:LO-bound}.\footnote{
This bound becomes meaningful for $F\sim 1$. In contrast, the bound from theorem~\ref{theorem:LO-Haar} works well for $F \sim \frac{1}{2^m}$ with $m$ being the number of distilled EPR pairs. See~\cite{Mori:2025mcd} for details. 
}

\begin{proof}
Let us focus on LU-distillable entanglement.
From theorem~\ref{thm:barnum-knill}, there exists a decoder $\mathcal{D}$ which distills $n_{A_0} = E_{D}^{[\mathrm{LO}]}(A:C)$ copies of EPR pairs with the EPR fidelity $F$, satisfying
\begin{align}
F_{\mathrm{Petz}}\equiv \langle \text{EPR}_{A_0 A_0'} | \rho_{A_0 A_0'} | \text{EPR}_{A_0 A_0'} \rangle \geq F^2. 
\end{align}
From this, $I(A_0 : A_0')$ can be lower bounded. 
Let us consider the following two-fold Haar twirl
\begin{align}
\Phi_{\text{Haar}}^{(2)}(\cdot ) \equiv
\int dU (U_{A_0}\otimes U_{A_0'}^*) (\cdot)  (U_{A_0}\otimes U_{A_0'}^*)^{\dagger}. \label{eq:twirl}
\end{align}
Note that this quantum channel acts on $A_0A_0'$ and fully depolarizes any states orthogonal to $| \text{EPR}_{A_0 A_0'} \rangle$.
Hence, the entropy of $\Phi_{\text{Haar}}^{(2)}(\rho_{A_0 A_0'}) $ is given by
\begin{align}
S_{A_0A_0'}\big( \Phi_{\text{Haar}}^{(2)}(\rho_{A_0 A_0'})  \big)= 2 (1-F_{\mathrm{Petz}}) n_A + O\Big(\frac{n_{A_{0}}}{d_{A_0}}\Big).
\end{align}
From the concavity of von Neumann entropy, we have
\begin{equation}
S_{A_0 A_0'}(\rho_{A_0 A_0'})\leq S_{A_0A_0'}\big( \Phi_{\text{Haar}}^{(2)}(\rho_{A_0 A_0'})  \big)  \leq 2 (1 - F^2)n_{A_{0}} + O\Big(\frac{n_{A_{0}}}{d_{A_0}}\Big).
\end{equation}
Also, from the continuity of entropies and $\Vert\rho_{A}-\frac{I_A}{d_A}\Vert_1=o(\frac{1}{n_A})$, we have $n_{A_0} - S(\rho_{A_0})\le O(1)$. 
We thus find 
\begin{align}
 F^2 n_{A_0} \leq \frac{1}{2}I(A_0 : A_0') +O(1) \le \frac{1}{2}I(A:A^\prime) + O(1),
 \label{eq:contra-thm3}
\end{align}
where the rightmost inequality comes from the monotonicity of mutual information under partial trace $\Tr_{A_1}$. 
\end{proof}

\subsection{Isotropic state from Petz map}

Further intuition about LO-distillable entanglement in Haar random state can be obtained by explicitly finding the expression of $\rho_{AA'}$. 

Let us first focus on the regime with $n_A < n_B$ (and $n_A,n_B,n_C<\frac{n}{2}$). 
In Appendix~\ref{app:Haar}, we will show that $\rho_{AA'}$ takes the following form:
\begin{align}
\rho_{AA'} \approx 2^{-\Delta}|\text{EPR}\rangle \langle \text{EPR}|_{AA'}  + (1-2^{-\Delta})\mu_{\text{max}}, \qquad \Delta = n_A + n_B - n_C  > 0, \label{eq:double_sec2}
\end{align}
where $\mu_{\text{max}}$ is the maximally mixed state on $AA'$. 
The quantum state on the RHS of Eq.~\eqref{eq:double_sec2} is called an \emph{isotropic state} since it is invariant under $U_A \otimes U_{A'}^*$ for arbitrary $U$.\footnote{
Entanglement properties of isotropic states have been studied in the literature, see~\cite{terhal2000entanglement} for instance. 
}

\begin{figure}
\centering
\includegraphics[width=0.43\textwidth]{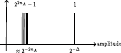}
\caption{The spectrum of $\rho_{AA'}$ with a single peak of $|\text{EPR}\rangle$ and background flat spectrum.
}
\label{fig_spectrum}
\end{figure}

Here it is useful to expand $\rho_{AA'}$ explicitly as
\begin{align}
\rho_{AA'} \approx 2^{-\Delta}|\text{EPR}\rangle \langle \text{EPR}|_{AA'} + 2^{-2n_{A}} 
\sum_{j=1}^{2^{2 n_A}-1}|\psi_j\rangle\langle \psi_j|
\label{eq:hol-iso}
\end{align}
where $|\psi_j\rangle$'s are states orthogonal to $|\text{EPR}\rangle$. 
Since $ 2^{-\Delta} \gg 2^{-2n_{A}}$, the spectrum of $\rho_{AA'}$ consists of a single peak of $|\text{EPR}\rangle$ and a flat background with much smaller amplitudes as depicted in Fig.~\ref{fig_spectrum}. 
See~\cite{Akers:2021pvd, Akers:2022zxr,  Czech:2023rbh} for previous works on this spectral property. 
While $|\text{EPR}\rangle$ might appear as the most probable state in $\rho_{AA'}$, its probability amplitude is suppressed by $2^{-\Delta}$, suggesting that the Petz map fails to distill EPR pairs. 
While we have focused on the cases where $n_A, n_B, n_C < \frac{n}{2}$, it is useful to study $\rho_{AA'}$ when $n_C$ approaches $n_C \approx \frac{n}{2}$.
In this limit, we have $\Delta \approx 0$, and thus $\rho_{AA'}$ will be dominated by $|\text{EPR}\rangle\langle \text{EPR}|_{AA'}$. 
This is consistent with the fact that, for $n_C > \frac{n}{2}$, the Petz map distills EPR pairs between $A$ and $A'$ as $A$ is (nearly) maximally entangled with $C$. 

Finally, when $n_A > n_B$, we find that the form of Eq.~\eqref{eq:double_sec2} remains valid with $\mu_{\text{max}}$ replaced by a maximally mixed state in some random $2^{2n_B}$-dimensional subspace. 
Namely, the spectral statistics of $\rho_{AA'}$ remains the same except that the background spectrum now has only $2^{2n_B}-1$ states.
In this case, the evaluation of $I(A:A')$ is not sufficient to establish the desired result $E_{D}^{[\text{LO}]}(A:C)\approx 0$ due to that the background state $\mu_{\text{max}}$ may be entangled. 

\begin{figure}
    \centering
    \raisebox{\height}{a)}\raisebox{-0.75\height}{\includegraphics[width=0.3\linewidth]{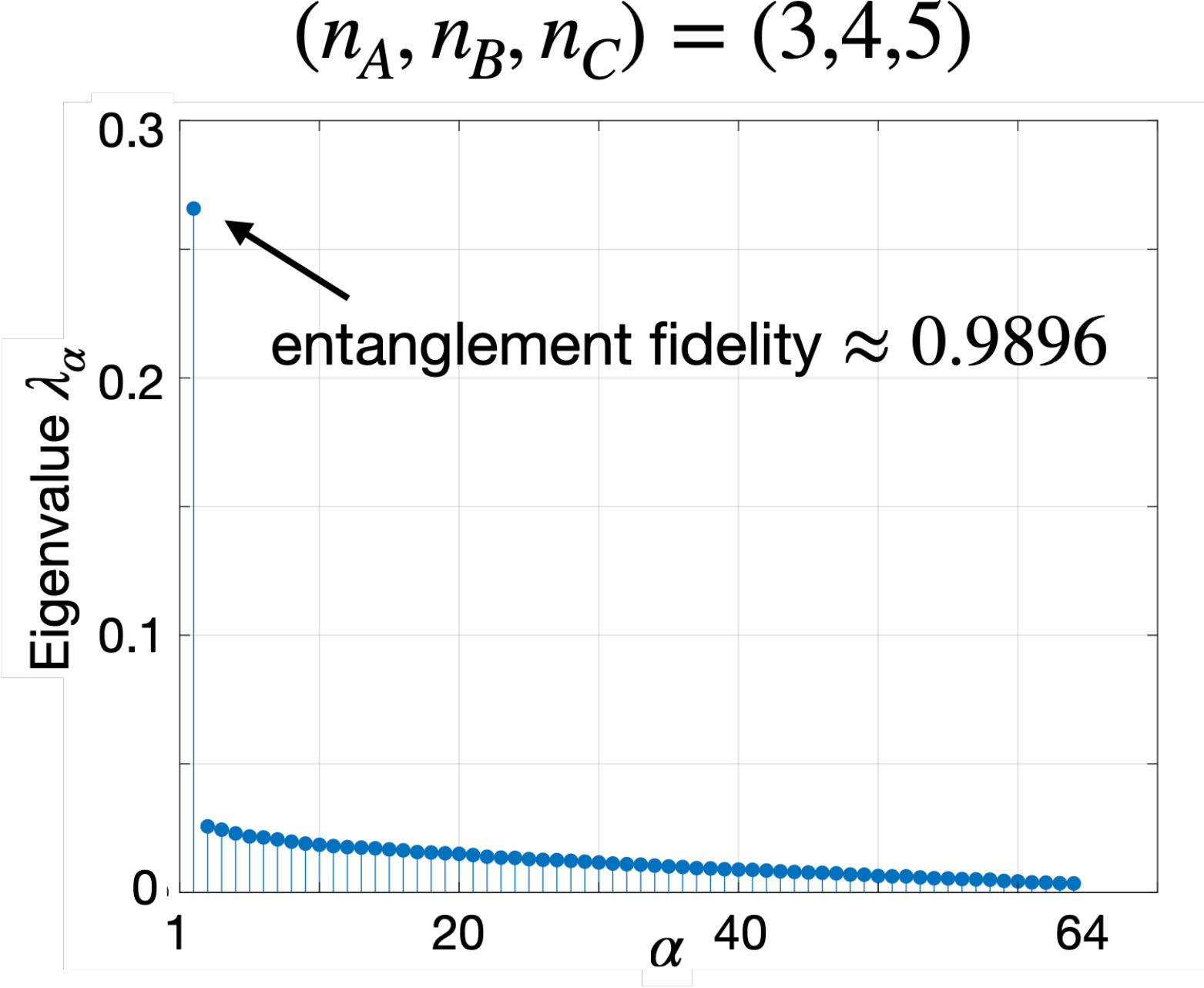}}
    \hspace{2pt}
    \raisebox{\height}{b)}\raisebox{-0.75\height}{\includegraphics[width=0.3\linewidth]{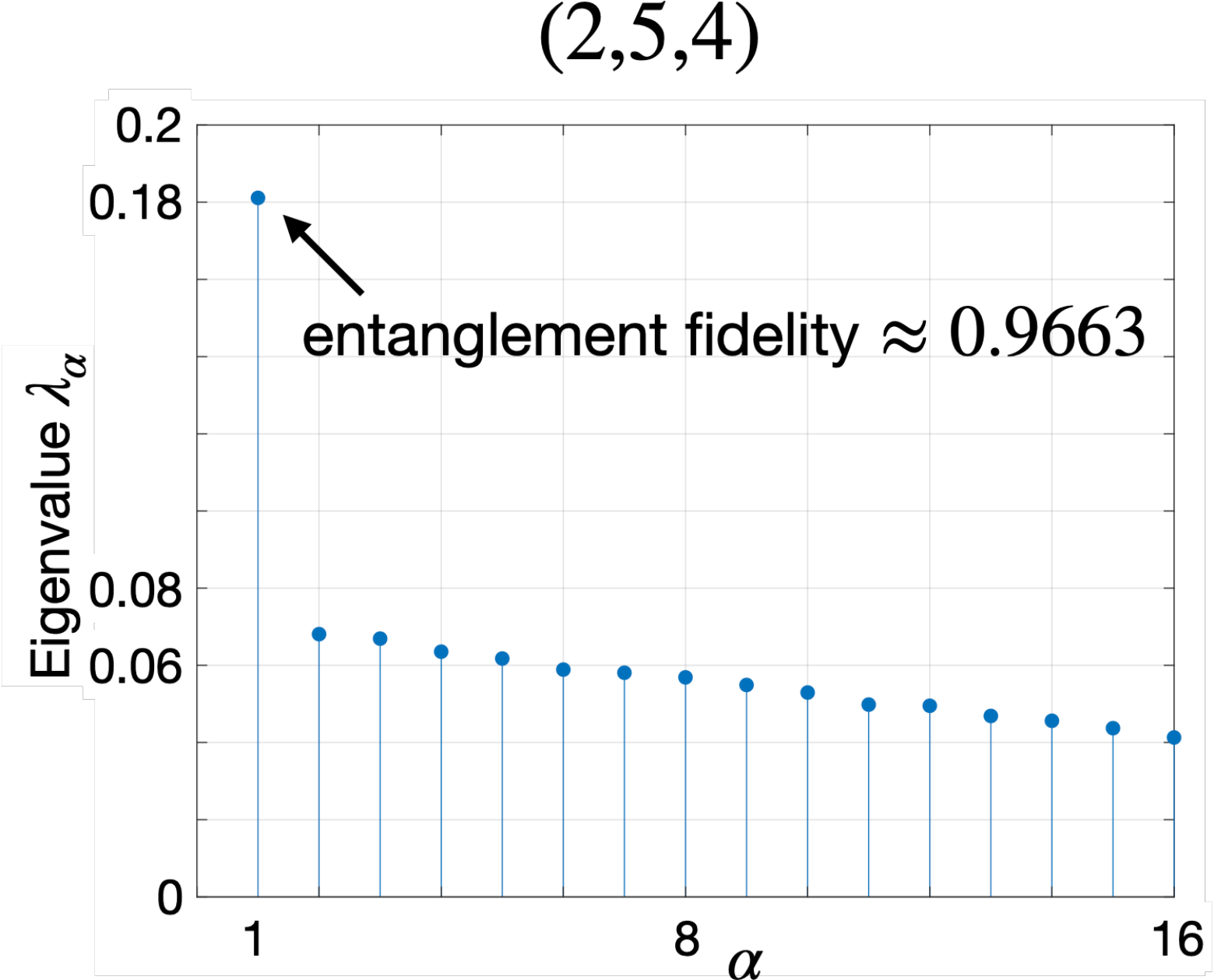}}
    \hspace{2pt}
    \raisebox{\height}{c)}\raisebox{-0.75\height}{\includegraphics[width=0.3\linewidth]{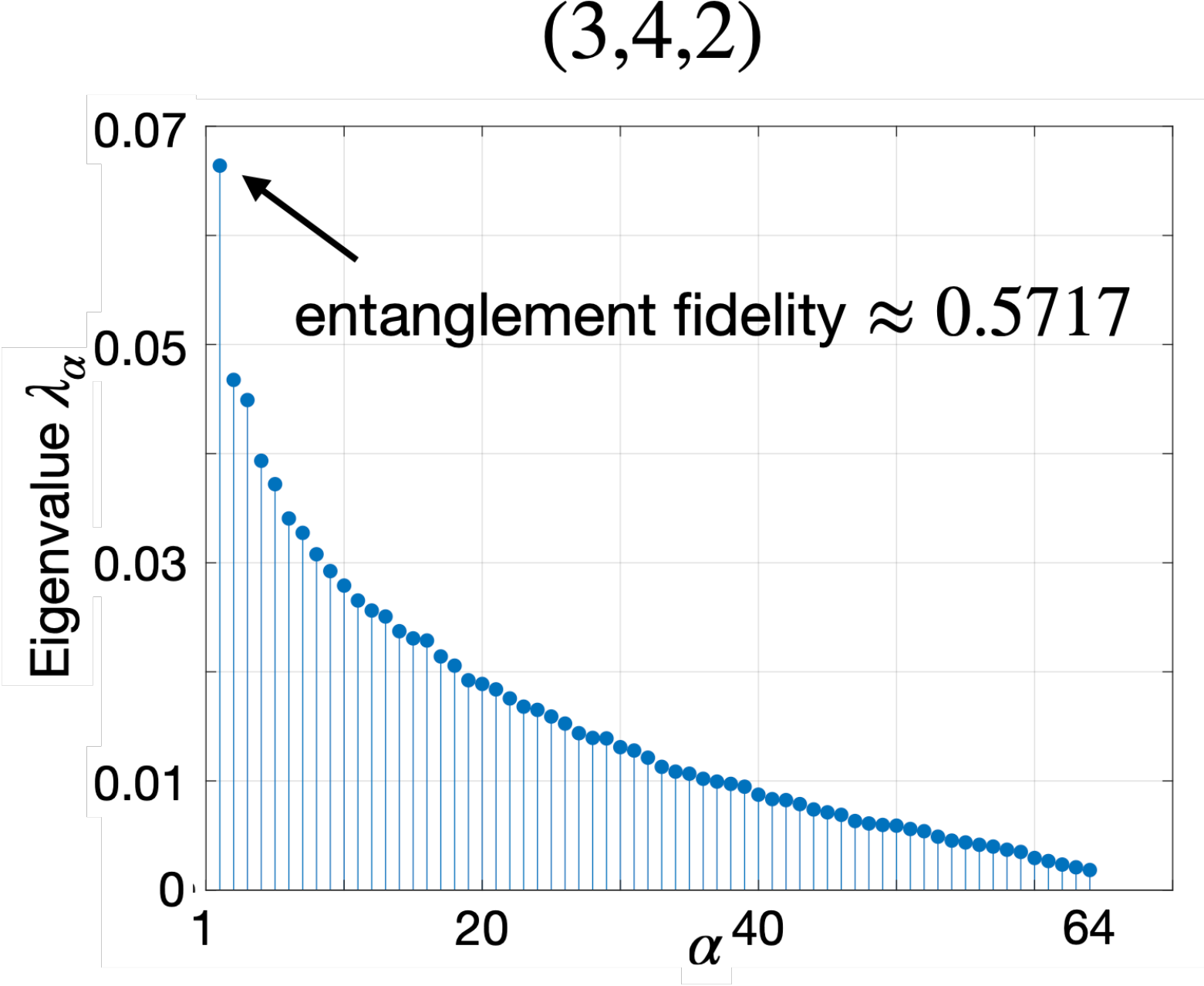}}
    \caption{
    Spectra of the reduced density matrix $\rho_{AA^\prime}$ of the double-copy state and entanglement fidelities of the eigenstate with the highest eigenvalue.
   }
    \label{fig:numeric}
\end{figure}

We have numerically verified our claim, concerning the peak state being $|\text{EPR}\rangle$, as shown in Fig.~\ref{fig:numeric}.
Specifically, we sampled a random tripartite state $V_{C^\prime \rightarrow AB}\ket{\text{EPR}}_{C^\prime C}$ where $V_{C^\prime \rightarrow AB}$ is a random isometry which maps the $d_C$-dimensional Hilbert space to the larger Hilbert space of dimension $d_{A}d_B$.
To reduce the computational cost, we used a random isometry $V$ instead of a Haar random state $|\psi_{ABC}\rangle$.
The plots show the eigenvalue spectra of $\rho_{AA^\prime}$ and the entanglement fidelity of the eigenvector with the highest eigenvalue. 

Fig.~\ref{fig:numeric}(a) shows the spectrum for a $12$-qubit random state where $(n_A,n_B,n_C)=(3,4,5)$. 
In this regime with $n_A < n_B$, we expect that the resulting state $\rho_{AA'}$ is close to an isotropic state. 
We indeed found that the spectrum has a single peak and otherwise the spectrum is almost flat. 
Furthermore, the peak amplitude ($\approx 0.27$) is significantly larger than the background amplitudes ($\lesssim 0.03$).
Finally, the entanglement fidelity of the peak is indeed close to unity ($\approx 0.9896$), confirming our claim. 

Fig.~\ref{fig:numeric}(b) shows the spectrum for a $11$-qubit random state where $(n_A,n_B,n_C)=(2,5,4)$. 
In this regime, we expect that the resulting state $\rho_{AA'}$ is still close to an isotropic state as, but with a smaller peak amplitude. 
But it will be separable with $n_C<n_B$ due to a smaller peak amplitude.
We indeed found that the spectrum has a peak with the entanglement fidelity $0.9663$ and otherwise the spectrum is close to flat. 
However, as expected, we have a smaller EPR peak ($\approx 0.18$). 

Finally, Fig.~\ref{fig:numeric}(c) shows the spectrum for a $9$ qubit random state where  $(n_A,n_B,n_C)=(3,4,2)$.
In this regime, we still expect to obtain an isotropic state with an even smaller peak. 
There are however a few features that appear to deviate from an isotropic state.
First, we found that the rest of the spectrum is decaying slowly, rather than being flat. 
Additionally, the entanglement fidelity of the peak state is $\approx 0.5717$.
These however may be due to finite size effects. 
Leading order estimate of the peak amplitude is $2^{-\Delta}= 2^{-5}$ with $\Delta = n_A+n_B-n_C = 5$, which is comparable to the estimate of the background amplitude $2^{-2n_A} = 2^{-6}$. 

\section{LO-distillable entanglement in holography}\label{sec:LO_holography}

In this section, we present an upper bound on $E_{D}^{[\text{LO}]}(A:C)$ for holographic states by using the Petz map.
We focus on the so-called fixed-area states, where subleading fluctuations of area operators are neglected. This approach is essentially equivalent to random tensor network states, in which the spectrum of reduced states is nearly flat. 

Applying the Petz map generates the double-copy state of the following form:
\begin{align}
\figbox{1.7}{fig_TN_glue}
\end{align}
where two copies $|\psi_{ABC}\rangle$ and $|\psi_{ABC}^*\rangle$ are glued at the minimal surface $\gamma_C$. 

Let us evaluate the mutual information $I(A:A')$. We find 
\begin{align}
S_{A} \approx \figbox{1.7}{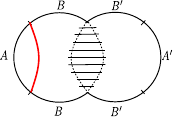}, \qquad 
S_{A'} \approx \figbox{1.7}{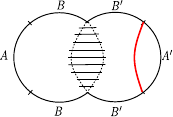}
\end{align}
whereas 
\begin{align}
S_{AA'} \approx \min\qty[ \figbox{1.7}{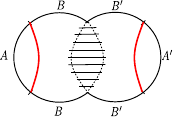}, \ \figbox{1.7}{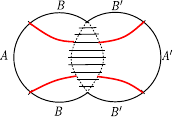}  ] \ . 
\end{align}
In other words, $S_{AA'}$ is given by the minimal area surface in the doubled geometry. 
We will confirm this for random tensor network states in Appendix~\ref{app:TN}.\footnote{
The sub-AdS scale is a subtle issue in tensor networks. 
In this paper, we simply consider tiling Haar random tensors down to the sub-AdS scale. 
}
See~\cite{Dutta:2019gen, Chandrasekaran:2020qtn, Akers:2021pvd, Akers:2022zxr} for relevant computations for reflected entropies.  
Recalling the definition of entanglement wedge cross section $E^W(A:B)$, we have 
\begin{align}
S_{AA'} \approx 2 E^W(A:B) 
\end{align}
and thus
\begin{align}
\frac{1}{2}I(A:A') \approx J^{W}(A|C) \equiv S_A - E^W(A:B). 
\end{align}
Hence, $E_{D}^{[\text{LO}]}(A:C)\lesssim J^{W}(A|C)$. Repeating the same argument by exchanging $A$ and $C$, we arrive at the following result.

\begin{claim}
Given a holographic state $|\psi\rangle_{ABC}$, we have
\begin{align}
E_{D}^{[\mathrm{LO}]}(A:C) \lesssim \min( J^{W}(A|C), J^{W}(C|A) ) 
\label{eq:petz-lo-hol}
\end{align}
where $J^{W}(A|C) \equiv S_A - E^W(A:B)$.
\end{claim}

While this bound is weaker than our proposal (namely $E_{D}^{[\text{LO}]}(A:C)\approx 0$ for non-overlapping $\gamma_A$ and $\gamma_C$), this rigorously establishes the existence of holographic $\rho_{AC}$ satisfying
\begin{align}
E_{D}^{[\text{LO}]}(A:C)\approx 0 ,\qquad I(A:C) \sim O(1/G_{N}).
\end{align}
Also, we will later show
\begin{align}
E_{D}^{[\text{1WAY LOCC}]}(A:C) \approx \max( J^{W}(A|C), J^{W}(C|A) ).
\end{align}
Hence, there exists a regime in holography where 
\begin{align}
E_{D}^{[\text{LO}]}(A:C) < E_{D}^{[\text{1WAY LOCC}]}(A:C) \leq E_{D}^{[\text{LOCC}]}(A:C) 
\end{align}
with an $O(1/G_N)$ gap between LO and LOCC distillable entanglement. 

\section{Shadow of entanglement wedge}\label{sec:LO_logical}

In this section, we explore the implications of our result on $E_D^{[\text{LO}]}(A:C)$ from the perspective of entanglement wedge reconstruction. Specifically, it predicts the possible existence of extensive bulk regions whose DOFs cannot be reconstructed on $A$ or $B = A^c$ in a boundary bipartition. 
This section partially overlaps with section 5.2 of~\cite{Mori:2025mcd}

\subsection{Entanglement wedge reconstruction (and its converse)}

In the AdS/CFT correspondence, the physics of bulk quantum gravity is holographically encoded into boundary quantum systems, akin to a quantum error-correcting code. The central concept supporting this view is \emph{entanglement wedge reconstruction}~\cite{Almheiri:2014lwa}, which states that, \emph{if} a bulk operator $\phi$ lies inside the entanglement wedge $\mathcal{E}_A$ of a boundary subsystem $A$, then it can be expressed as a boundary operator $O_A$ supported entirely on $A$: \begin{align}\label{eq:ifstatement} 
\text{$\phi$ can be reconstructed on $A$} \ \Leftarrow \ \text{$\phi$ is inside $\mathcal{E}_A$}. 
\end{align}
While the microscopic mechanisms underlying bulk reconstruction remain somewhat mysterious, random tensor network toy models offer valuable insight into how bulk operators might be reconstructed on boundary subsystems. Consider, for example, a Haar random state $|\Psi\rangle$ as a minimal toy model. Suppose the bulk consists of a single qubit encoded into $n-1$ boundary qubits, by interpreting an $n$-qubit Haar random state $|\Psi\rangle$ as an encoding isometry $1 \to n-1$, as illustrated schematically: \begin{align} \figbox{2.0}{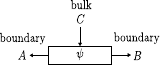}
\end{align} Here, the bulk qubit is denoted by $C$, and the boundary qubits are partitioned into $A$ and $B$.

In this toy model, the question of whether the bulk unitary operator $U_C$ can be reconstructed on a subsystem $A$ translates to whether a logical unitary operator $\overline{U_C}$ can be supported on $A$ in the $C \to AB$ quantum error-correcting code. 
It is well known that $A$ supports a non-trivial logical operator $\overline{U_C}$ when $n_A > \frac{n}{2}$ (and does not when $n_A < \frac{n}{2}$).
This standard result in Haar encoding can be understood via entanglement wedge reconstruction. 
For static cases in the AdS/CFT, the entanglement wedge is computed by minimizing the generalized entropy: 
\begin{align} 
S_A = \min_{\gamma_A} \qty(\frac{ \text{Area}(\gamma_A) }{4 G_N} + S_{\text{bulk}}), 
\end{align} 
where $S_{\text{bulk}}$ is the bulk entropy in a region bounded by $\gamma_A$. 
When $A$ occupies more than half of the total system, we find: 
\begin{align} 
S_A = \figbox{2.0}{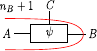} 
\end{align} 
with the bulk qubit $C$ lying within $\mathcal{E}_A$, indicating its recoverability on $A$. 
Here, $+1$ arises from $S_{\text{bulk}} = S_C = 1$.
Conversely, when $A$ occupies less than half of the system, we find
\begin{align} 
S_A = \figbox{2.0}{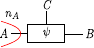}
\end{align} 
with the bulk qubit now outside $\mathcal{E}_A$, but within $\mathcal{E}_B$, suggesting recoverability from $B$. Hence, the bulk information in $C$ can be reconstructed from either $A$ or $B$—unless $|A| = |B|$ exactly.

\begin{figure}[h] \centering a)\raisebox{5.5mm}{\includegraphics[width=0.25\textwidth]{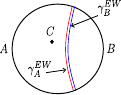}} b)\includegraphics[width=0.28\textwidth]{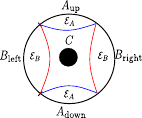} \caption{ Entanglement wedge reconstruction. a) When $C$ carries subleading entropy, $\gamma_A^{EW}$ and $\gamma_B^{EW}$ coincide, and $C$ can be reconstructed on only one of $A$ or $B$. b) Here, $A = A_{\text{up}} \cup A_{\text{down}}$ and $B = B_{\text{left}} \cup B_{\text{right}}$. If $C$ carries leading-order entropy, $\gamma_A^{EW}$ and $\gamma_B^{EW}$ may differ, and $C$ may lie outside both wedges. Can $A$ or $B$ reconstruct $C$? } \label{fig_reconstruction} \end{figure}

A similar setup applies in the AdS$_{3}$/CFT$_{2}$, as shown in Fig.~\ref{fig_reconstruction}(a). 
When the bulk DOF $C$ carries subleading entropy, the minimal surface $\gamma_A^{EW}$ defining $\mathcal{E}_A$ coincides with the RT surface $\gamma_A^{RT}$: 
\begin{align} 
\gamma_A^{EW} \approx \gamma_A^{RT} \qquad \text{(at leading order in $1/G_N$)}. 
\end{align} 
Thus, if $C$ is enclosed by $\gamma_A^{RT}$, an operator on $C$ can be reconstructed on $A$. 
If not, it is enclosed by $\gamma_B^{RT}$ (unless the partition is fine-tuned), and reconstruction is possible on $B$ instead.
By the no-cloning theorem, reconstruction on both $A$ and $B$ is forbidden. 
Hence, unless $C$ lies exactly on the minimal surface $\gamma_A^{EW}$, it can be reconstructed on \emph{one and only one} subsystem—either $A$ or $B$.

So far, we observed that when the bulk $C$ carries subleading entropy, entanglement wedge reconstruction becomes an \emph{if and only if} statement at leading order in $1/G_N$ (or $n$): 
\begin{align} 
\text{$\phi$ can be reconstructed on $A$} \ \Leftrightarrow \ \text{$\phi$ is inside $\mathcal{E}_{A}$} \qquad (\text{for subleading $S_{\text{bulk}}$}). 
\end{align} 
We then naturally wonder: does this equivalence hold when $C$ carries leading-order entropy?
It is worth emphasizing that the original formulation of entanglement wedge reconstruction is an \emph{if} statement~\cite{Dong:2016fnf}. 
Some of previous works shows that the converse statement also holds if the complementary recovery is possible~\cite{Akers:2021fut}.
However, the complementary recovery generally fails when the bulk region $C$ carries leading order entropies.
In such cases, physical or analytical evidence supporting the converse statement has thus far been lacking.
That is, it remains unclear whether a bulk operator outside $\mathcal{E}_A$ can be reconstructed on $A$ when $S_C = O(1/G_N)$: 
\begin{align} 
\text{$\phi$ can be reconstructed on $A$} \ \overset{?}{\Rightarrow} \ \text{$\phi$ is inside $\mathcal{E}_A$}. 
\end{align} 
When $C$ is subleading, the converse holds because $\gamma_A^{EW} = \gamma_B^{EW}$ at leading order. But this can fail when $C$ is not subleading.

A specific holographic example highlighting this subtlety, concerning the converse of entanglement wedge reconstruction, was studied in~\cite{Akers:2020pmf} (Fig.\ref{fig_reconstruction}(b)). 
The boundary is divided into four segments of roughly equal size and grouped into $A$ and $B$. 
In the absence of bulk DOFs, the minimal surfaces for $A$ and $B$ coincide (i.e., the shorter of the red and blue geodesics). 
However, when $S_C = O(1/G_N)$, the location of the minimal surface can change at leading order, since $S_{\text{bulk}} = S_C$ contributes to the generalized entropy. 
In particular, $\gamma_A^{EW} \neq \gamma_B^{EW}$ when the area difference between red and blue geodesics (in units of $1/4G_N$) is smaller than $S_C$. In this case, $C$ lies outside both $\mathcal{E}_A$ and $\mathcal{E}_B$.\footnote{The four-segment partition leaves a sufficiently large AdS-scale bulk region where a DOF with $1/G_N$ entropy can be placed without excessive backreaction. One could replace $C$ with a small black hole or conical defect to account for backreaction explicitly. This setup requires no fine-tuning of $A$ and $B$'s sizes, as long as the geodesic-length condition is satisfied.}

\subsection{No logical operators in bipartition}

The key question is whether bulk DOFs $C$ can still be reconstructed on $A$ or $B$. The converse of entanglement wedge reconstruction would suggest the answer is no, since $C$ lies outside both $\mathcal{E}_A$ and $\mathcal{E}_B$. However, one finds that $I(C:A), I(C:B) = O(1/G_N)$ (in the state-channel duality), suggesting leading-order correlations between $C$ and both $A$ and $B$.

Our result on LO-distillable entanglement in Haar random states provides an evidence supporting the converse and as such, the presence of an extensive bulk region that cannot be reconstructed from either $A$ or $B$. 
Recall that the relation between the LO-distillation problem and entanglement wedge reconstruction becomes evident by considering a pair of anti-commuting Pauli logical operators. 
Namely, if Pauli logical operators $\overline{X}, \overline{Z}$ could be supported on $A$, it would imply that an EPR pair could be LU-distilled between $A$ and $C$. 
Hence, one can deduce that logical Pauli operators $\overline{X}, \overline{Z}$ cannot be supported inside $A$. 
This observation enables us to establish the following no-go result on random encodings. 

\begin{theorem}[informal]\label{theorem_logical}
    Consider a random encoding $C\to AB$. If $n_C<n_A+n_B$ and $n_A<n_B+n_C$, then $A$ contains no quantum information about $C$.
    Namely, $A$ does not support any non-trivial logical unitary operator. 
\end{theorem}

A rigorous formulation of this result, along with a formal mathematical proof, is presented in~\cite{Mori:2025mcd}. 
This theorem supports the converse of entanglement wedge reconstruction. Namely, a Haar-random encoding with $n_A, n_B, n_C < \frac{n}{2}$ closely mimics the geometry of Fig.\ref{fig_reconstruction}(b), as schematically depicted below
\begin{align}
\figbox{2.0}{fig_reconstruction_shadow_A}\ .
\end{align} 
where $\mathcal{E}_A$ and $\mathcal{E}_B$ do not contain $C$. Theorem~\ref{theorem_logical} then implies that no non-trivial logical unitary operator $U_C$ can be reconstructed on either $A$ or $B$. 

\begin{proposal}\label{claim_converse}
The converse of entanglement wedge reconstruction holds, namely
\begin{align} 
\text{$\phi$ can be reconstructed on $A$} \ \Leftrightarrow \ \text{$\phi$ is inside $\mathcal{E}_{A}$}  
\end{align} 
even when the bulk carries $O(1/G_{N})$ entropies. Furthermore, given a bipartition of the boundary into $A$ and $B$, there can exist an extensive bulk region $C$ whose unitary operators cannot be reconstructed on either $A$ or $B$, as in Fig.\ref{fig_reconstruction}(b). 
\end{proposal}

In the above discussion, it is crucial that we consider unitary logical operators $\overline{U_C}$. 
In fact, some \emph{non-unitary} logical operators can be constructed on $A$ or $B$. 
For instance, let us further decompose $C$ into two subsystems $C=C_{0}C_1$ where $C_0$ consists of a single qubit while $C_1$ consists of the remaining $n_C-1$ qubits. 
Consider the following operator
\begin{align}\label{eq:nonUlogical}
V_{C} = X_{C_0} \otimes |0\rangle\langle 0 |_{C_{1}}.
\end{align}
where $|0\rangle\langle 0 |_{C_{1}}$ is a projection acting on $C_1$. 
Observe that $|0\rangle\langle 0 |_{C_{1}}$ acting on a Haar random state $|\Psi_{ABC}\rangle$ effectively produces a new random state $|\Phi_{ABC_0}\rangle \propto  |0\rangle\langle 0 |_{C_{1}} |\Psi_{ABC}\rangle$ after appropriate normalization. 
Then, constructing a logical operator $\overline{V_{C}}$ for  $|\Psi_{ABC}\rangle$ reduces to constructing $\overline{X_{C_0}}$ for  $|\Phi_{ABC_0}\rangle$, i.e. to EPR distillation in the projected state.  
If $A$ contains more than half of $ABC_{0}$, i.e. $n_{A} > n_{B} +1$, then $\overline{X_{C_0}}$ can be supported on $A$ even if no logical \emph{unitary} operator can. 
In holographic context, this corresponds to gravitational backreaction induced by holographic measurements, which changes the entanglement wedge.
This point will be further discussed in section~\ref{sec:LOCC_Haar},~\ref{sec:GLOCC_holography}.
In particular, the reconstruction of a non-unitary logical operator $\overline{V_C}$ on $A$ can be interpreted as a 1WAY LOCC entanglement distillation where i) one performs projective measurements $\{|i\rangle\langle i| \}_{C_1}$ on $C_1$, ii) sends the measurement outcome $i$ to $A$, and iii) applies an appropriate LU on $A$ to prepare an EPR pair between $A$ and $C_{0}$. 

We expect that our proposal can provide useful insights into DOFs behind the horizons, including the black hole interior~\cite{Yoshida:2019qqw, Yoshida:2019kyp} and the spacetime outside the cosmological horizon~\cite{Franken:2024wmh,Shaghoulian:2022fop}, which we hope to explore in the future work.

\subsection{On Cleaning lemma}

We have established that, when $|\Psi_{ABC}\rangle$ is a Haar random state (or when $V : C \to AB$ is a random isometry), encoded logical qubits cannot be recovered from subsystems $A$ or $B$ assuming $n_A, n_B, n_C<\frac{1}{2}(n_A+n_B+n_C)$. 
One interesting corollary of this result is that the so-called cleaning lemma does not necessarily extend to non-stabilizer codes. To recap, the cleaning lemma for a stabilizer code asserts that, if a subsystem $A$ supports no non-trivial logical operators, then the complementary subsystem $B=A^c$ supports all the logical operators of the code~\cite{Bravyi:2009zzh}. This fundamental result is central in establishing the fault-tolerance of topological stabilizer codes (those with geometrically local generators), as it ensures that logical operators can be supported on regions that avoid damaged qubits. While the original formulation is restricted to stabilizer codes, analogous properties, such as the deformability of string-like logical operators, are known to hold in various models of topological phases beyond the stabilizer formalism. Despite these examples, our result indicates that the cleaning lemma, in its original formulation given by~\cite{Bravyi:2009zzh}, does not extend to general non-stabilizer quantum error-correcting codes.

Furthermore, when $|\Psi_{ABC}\rangle$ is a random \emph{stabilizer} state, part of the encoded logical qubits can be recovered from subsystems $A$ and $B$.
Namely, let $g_R$ be the number of independent non-trivial logical operators supported on a subsystem $R$; the following relations are well known~\cite{PhysRevA.81.052302}: 
\begin{align}
g_A = I(A:C)\approx n_A+n_C-n_B, \qquad g_B = I(B:C) \approx n_B+n_C-n_A, \qquad g_A + g_B = 2k
\end{align}
where $k=n_C$ in our setting and the approximations hold when $n_R<\frac{n}{2}$ for $R=A,B,C$. 
When the Clifford encoding isometry $V: C\to AB$ is chosen randomly, it is likely that a given logical operator $\ell_A$ on $A$ can find some other logical operator $r_A$ on $A$ that anti-commutes with $\ell_A$. 
As such, one can choose $\frac{g_A}{2} - o(1)$ pairs of mutually anti-commuting basis logical operators on $A$, suggesting that $\frac{g_A}{2} - o(1)$ logical qubits can be recovered from $A$. 
This behavior is closely related to the observation that random stabilizer states $|\psi_{ABC}\rangle$ predominantly exhibit bipartite entanglement. 


\addcontentsline{toc}{section}{Part II : LOCC-distillable entanglement}

\section*{Part II : LOCC-distillable entanglement}

In the next four sections, we discuss LOCC-distillable entanglement. Our main proposal is
\begin{equation}
\begin{split}
&E_{D}^{[\text{1WAY LOCC}]}(A\leftarrow C) \approx J^W(A|C) \equiv S_{A} - E^{W}(A:B) \\
&E_{D}^{[\text{1WAY LOCC}]}(A:C) \approx J^W(A:C) = \max \big(J^W(A|C),J^W(C|A) \big).
\end{split}
\end{equation}
We also claim the holographic relation for $E_F$:
\begin{align}
E_{F}(A:C) \approx E^W(A:C).
\end{align}
In section~\ref{sec:LOCC_Haar}, we prove these proposals for Haar random states.
In section~\ref{sec:GLOCC_holography}, we show that these proposals hold under holographic measurements that place EoW brane-like objects in the bulk. 
Furthermore, we prove that the proposals also hold under LOCCs, assuming the holographic relation $E_{F}\approx E^W$ (or equivalently $J(A|C)\approx J^W(A|C)$). In section~\ref{sec:EF}, \ref{sec:JW}, we provide supporting arguments for the proposals $E_{F}\approx E^W$ and $J^W\approx J^W$.  
In section~\ref{sec:bound_entanglement}, we discuss the implications of our results from the perspective of bound entanglement. 

\section{LOCC-distillable entanglement in Haar random state}\label{sec:LOCC_Haar}

In this section, we discuss the 1WAY LOCC-distillable entanglement $E_{D}^{[\text{1WAY LOCC}]}(A:C)$ for Haar random states. 
We also demonstrate that the proposal $E_{F}\approx E^W$ holds for a Haar random state by identifying the counterpart of $E^W$. 

\subsection{LOCC protocol}

\begin{figure}
\centering
\raisebox{\height}{a)\hspace{10pt}}\raisebox{-0.75\height}{\includegraphics[width=0.22\textwidth]{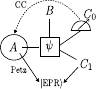}}
\hspace{10pt}
\raisebox{\height}{b)\hspace{10pt}}\raisebox{-0.75\height}{\includegraphics[width=0.18\textwidth]{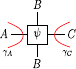}}
\hspace{10pt}
\raisebox{\height}{c)\hspace{10pt}}\raisebox{-0.75\height}{\includegraphics[width=0.18\textwidth]{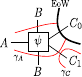}}
\caption{a) A 1WAY LOCC distillation protocol for a Haar random state.  b) Minimal surfaces before a projective measurement. c) Minimal surfaces after a projective measurement, represented by an EoW brane-like object, in the case where $n_A>n_B$.
}
\label{fig_LOCC}
\end{figure}

We present a 1WAY LOCC ($A\leftarrow C$) protocol that distills $\approx n_A-n_B$ EPR pairs (assuming $n_A > n_B$). 
The protocol performs random projective measurements on $C$ and sends the measurement outcomes to $A$ (Fig.~\ref{fig_LOCC}(a)) .
By exchanging $A$ and $C$, the protocol distills $\approx n_C-n_B$ EPR pairs when $n_C > n_B$.
Recalling $\text{hash}(A:C)\equiv \max(S_{A} - S_{AC}, S_{C} - S_{AC}, 0)$, this protocol distills $\approx \text{hash}(A:C)$ EPR pairs.

\begin{enumerate}[1)]
\item Perform random projective measurements on $|C_{0}| \approx n_B + n_C - n_A$ qubits in a subset $C_{0}\subset C$, leaving the remaining $|C_{1}| \approx n_{A}-n_B$ qubits on $C_{1}$ untouched. 
\item Send the measurement outcome from $C$ to $A$. 
\item Apply the Petz recovery map on $A$ to distill $|C_{1}| \approx n_{A}-n_B$  EPR pairs.
\end{enumerate}
This protocol effectively reduces the Hilbert space size of $C$ by projective measurements so that $|C_1| + |B| \approx |A|$ in the post measurement state, suggesting that $C_{1}$ is nearly maximally entangled with $A$. 
Classical communication is crucial - without receiving the measurement outcome, the other party $A$ cannot identify the subspace of $C$ that is entangled with $A$.

A holographic interpretation of this protocol arises by viewing projective measurements as placing an EoW brane-like object. 
Measuring $C_0$ effectively removes the qubits in $C_0$ from $|\psi\rangle$, analogous to how an End-of-World (EoW) brane terminates spacetime. 
Recall that minimal surfaces $\gamma_A, \gamma_C$ are separated by the tensor at the center as in Fig.~\ref{fig_LOCC}(b). 
By placing an EoW brane-like object on $C_{0}$, $\gamma_A$ changes its profile and contains the tensor as in Fig.~\ref{fig_LOCC}(c). 
As a result, $\gamma_A$ and $\gamma_C$ now overlap, allowing EPR pairs to be distilled between $C_{1}$ and $A$ using the Petz map on $A$. 

Using this interpretation, entanglement cross section $E^W(A:B)$ can be identified as
\begin{align}
E^W(A:B) = \min \left( \figbox{2.0}{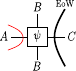} \ \figbox{2.0}{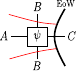} \right) \ = \min (n_{A}, n_{B} ).
\end{align}
Furthermore, we obtain
\begin{equation}
\begin{split}
&J^W(A|C) \equiv S_{A} - E^W(A:B) \approx \max (0, n_A -n_{B} )\\ 
&J^W(A:C) \equiv \max \big( J^W(A|C), J^W(C|A) \big)  \approx \max (0, n_A -n_{B}, n_C-n_B ) \approx \text{hash}(A:C). 
\end{split}
\end{equation}
Hence, we find 
\begin{equation}
\begin{split}
&E_{D}^{[\text{1WAY LOCC}]}(A\leftarrow C) \gtrsim J^W(A|C) \\
&E_{D}^{[\text{1WAY LOCC}]}(A:C) \gtrsim J^W(A:C) \approx \text{hash}(A:C). 
\end{split}
\end{equation}
Below, we show that these lower bounds are in fact tight.

\subsection{Entanglement of formation}

Next, we discuss entanglement of formation $E_{F}(A:B)$ and locally accessible information $J(A|C)$. Specifically, we point out that holographic proposals $E_{F}(A:B)\approx E^W(A:B)$ and $J(A|C)\approx J^W(A|C)$ hold for Haar random states. 
This is immediate from previous work~\cite{Hayden_2006}, which obtained the following expression for $E_{F}(A:B)$ (and $J(A|C)$).

\begin{theorem}[informal]
Given a Haar random state $|\psi_{ABC}\rangle$, we almost surely have
\begin{equation}
\begin{split}
E_{F}(A:B) &\approx \min(n_A,n_B)  \qquad n_{C} < n_{A} + n_{B}  \\ 
&\approx 0  \qquad \qquad \qquad  \ \ n_{C} > n_{A} + n_{B} .
\end{split}
\end{equation}
Also, by the Koashi-Winter relation $J(A|C)= S_{A} - E_{F}(A:B)$, we have 
\begin{equation}
\begin{split}
J(A|C) &\approx \max(0, n_A - n_B )  \qquad n_{C} < n_{A} + n_{B}  \\ 
&\approx n_A  \qquad \qquad \qquad  \qquad  n_{C} > n_{A} + n_{B} .
\end{split}
\end{equation}
\end{theorem}

It is worth presenting a heuristic argument behind this result. 
Let us focus on the cases where $n_{C} < n_{A} + n_{B}$.
Considering all the possible convex decompositions $\rho_{AB}=\sum_{j}p_j|\psi_j\rangle \langle \psi_j | $, $E_{F}(A:B)$ corresponds to the minimal entanglement of $A$, namely $\min \sum_j p_j S(\rho_{A}^j)$.
Observe that $\rho_{AB}$ is supported on a $2^{n_C}$-dimensional random subspace of $AB$. This subspace contains doubly exponentially many states with respect to $n_C$ (see Section~\ref{sec:LO_Haar}). 

It turns out that all pure states in this subspace are nearly maximally entangled across $AB$.
Given a Haar random state $|\phi_{AB}\rangle$ on $AB$ (with $n_A < n_B$), the probability of having $S_{A} < n_{A} - \epsilon$ is exponentially suppressed as $\sim \exp( - \frac{c d_A d_B \epsilon^2}{n_A^2} )$ with some positive constant $c>0$~\cite{Hayden_2006} due to the measure concentration.
The total probability of a random subspace containing a state with $S_{A} < n_{A} - O(\epsilon)$ is then suppressed by $\sim \Phi_{\text{state}}(n_C) \exp( - \frac{c d_A d_B \epsilon^2}{n_A^2} ) \rightarrow 0$ for $n_C < n_A + n_B$. 
Hence, the subspace supporting $\rho_{AB}$ contains nearly maximally entangled states only, establishing that $E_{F}(A:B) \approx n_A$ when $n_A < n_{B}$. 

An important corollary of this result is the holographic proposals of $E_{F}(A:B)$ and $J(A|C) $ for Haar random states.

\begin{corollary}
For a Haar random state $|\psi_{ABC}\rangle$, we have 
\begin{equation}
\begin{split}
&E_{F}(A:B) \approx E^W(A:B) \\
&J(A|C) \approx J^W(A|C) \equiv S_{A} - E^W(A:B).
\end{split}
\end{equation}
In particular, $E^W(A:B)=\min (n_{A}, n_{B} ),\, J^W(A|C) \approx \max ( 0, n_{A} - n_B )$ when $n_C < n_A + n_B$, and $E^W(A:B)=0,\, J^W(A|C)\approx n_A$ when $n_C>n_A+n_B$.
\end{corollary}

\subsection{LOCC-distillable entanglement}

Finally, we show that $E_{D}^{[\text{1WAY LOCC}]}(A \leftarrow C) \approx J^W(A|C)$, and thus $E_{D}^{[\text{1WAY LOCC}]}(A:C) \approx \text{hash}(A:C)$. 

\begin{claim}\label{claim_JCA}
We have
\begin{align}
E_{D}^{[\mathrm{1WAY\ LOCC}]}(A \leftarrow C)   \leq J(A|C) \label{eq:ED_JAC} 
\end{align} 
in the one-shot setting. 
\end{claim}

We prove this claim for the case where the fidelity of distilled EPR pairs is close to unity. 
By definition, there must exist some 1WAY LOCC protocol which performs a POVM acting on $C$, followed by LOs on $A$ and $C$, and distill $E_{D}^{[\text{1WAY LOCC}]}(A \leftarrow C)$ copies of EPR pairs. 
Recall that POVMs acting on $C$ does not increase $S_{A}$ on average since
\begin{equation}
\begin{split}
S_{A}^{\text{before}}  - \mathbb{E}(S_{A}^{\text{after}}) &=S_{A}^{\text{before}} -  \sum_{j} p_j S_{A}(\rho_{A}^j), \qquad p_j = \Tr(\Pi_{C}^j \rho_{C}),\quad \rho_{A}^j = \frac{\Tr_C(\Pi_C^j \rho_{AC})}{p_j} \\
&=  \sum_{j}p_j S(\rho_A^j||\rho_A) \geq 0
\end{split}
\end{equation}
where we used the non-negativity of relative entropy. 
Also recall that LOs can be implemented by LUs with additions of ancilla qubits and tracing out subsystems. 
Therefore, one can convert $\rho_{AC}$ to $\sigma \otimes \dyad{\text{EPR}}^{\otimes E_D}$ with some $\sigma$ via a POVM on $C$, followed by some LUs on $A$ and $C$ including ancilla qubits.
Here, omitting the trace operations does not affect the distillability since EPR pairs are produced on qubits that are decoupled from the rest. 
Hence, this 1WAY LOCC protocol prepares $E_{D}^{[\text{1WAY LOCC}]}(A \leftarrow C)$ copies of approximate EPR pairs without increasing  $S_A$. 
Measuring distilled EPR pairs on $C$ can decrease $S_{A}$ at least by $E_{D}^{[\text{1WAY LOCC}]}(A \leftarrow C)$, suggesting the desired inequality Eq.~\eqref{eq:ED_JAC}.

We have already seen that $E_{D}^{[\text{1WAY LOCC}]}(A \leftarrow C)   \leq J^W(A|C)$ and $J(A|C)\approx J^W(A|C)$. 
Hence, we confirm our proposal for $E_{D}^{[\text{1WAY LOCC}]}$ in Haar random states. 

\begin{claim}[informal]
Given a Haar random state $|\psi_{ABC}\rangle$ with $n_{C} > n_{A} + n_{B}$, we have 
\begin{equation}
\begin{split}
&E_{D}^{[\mathrm{1WAY \ LOCC}](A\leftarrow C )}(A:C) \approx J^W(A|C) = \max(0, n_A- n_B)\\
&E_{D}^{[\mathrm{1WAY\ LOCC}]}(A:C) \approx J^W(A:C) =  \max(0, n_A- n_B, n_C -n_B)  
\end{split}
\end{equation}
where $J^W(A:C)\approx \mathrm{hash}(A:C)$. 
\end{claim}

\section{LOCC-distillable entanglement in holography}\label{sec:GLOCC_holography}

In this section, we discuss $E_{D}^{[\text{1WAY LOCC}]}(A:C)$ in holography. 

\subsection{Holographic measurement}\label{sec:hol-meas}

We begin by clarifying the class of projective measurements we will employ in the LOCC protocol. 
In the AdS/CFT correspondence, degrees of freedom (DOFs) within the entanglement wedge $\mathcal{E}_A$ of a boundary region $A$ can be reconstructed from operators acting on $A$, a fact known as the entanglement wedge reconstruction.
This implies that one can, in principle, simulate the insertion of bulk objects, such as End-of-the-World (EoW) branes, by acting locally on $A$.
An EoW brane-like object effectively terminates the bulk geometry and does not contribute further to the entanglement entropy where the minimal surface may end on such objects~\cite{Takayanagi:2011zk, Fujita:2011fp}.

Motivated by this, we introduce the following class of projective measurements, which we term \emph{holographic measurements}:
\begin{itemize}
\item[] Let $A$ be a boundary subsystem with entanglement wedge $\mathcal{E}_A$, and let $\Sigma$ be a convex bulk surface homologous to $A$ and lying within $\mathcal{E}_A$.\footnote{
One might question whether an EoW brane could extend beyond $\mathcal{E}_A$ by allowing negative tension.
However, as argued in~\cite{Mori:2023swn}, quantum information constraints, as well as energy constraints, prevent EoW branes from being placed outside the entanglement wedge.
}
Then, there exists a disentangled projective measurement basis on $A$ such that the post-measurement states almost surely admit semiclassical duals with an EoW brane-like object placed along some portion of $\Sigma$.
\end{itemize}

We should note that the existence of a disentangled basis along a bulk surface is not immediately manifest from the perspective of conformal field theories. 
Namely, accessing such DOFs on a bulk surface requires a coarse-graining of the boundary Hilbert space, but explicit procedures are only known in certain limited cases~\cite{McGough:2016lol,Miyaji:2015fia,Bao:2024ixc}. 
A further challenge lies in identifying complete set of disentangled basis states.
A standard example of disentangled states in conformal field theories (CFT) is the Cardy boundary state. However, each Cardy state is a linear combination of Ishibashi states, which are maximally entangled holomorphic and anti-holomorphic sectors, and therefore span only half of the full Hilbert space~\cite{Miyaji:2021ktr}.
While it may be possible to construct a disentangled basis via triangulation techniques~\cite{Hung:2024gma,Geng:2025efs}, we leave such an explicit construction to future work.

\begin{figure}
\centering
\includegraphics[width=0.3\textwidth]{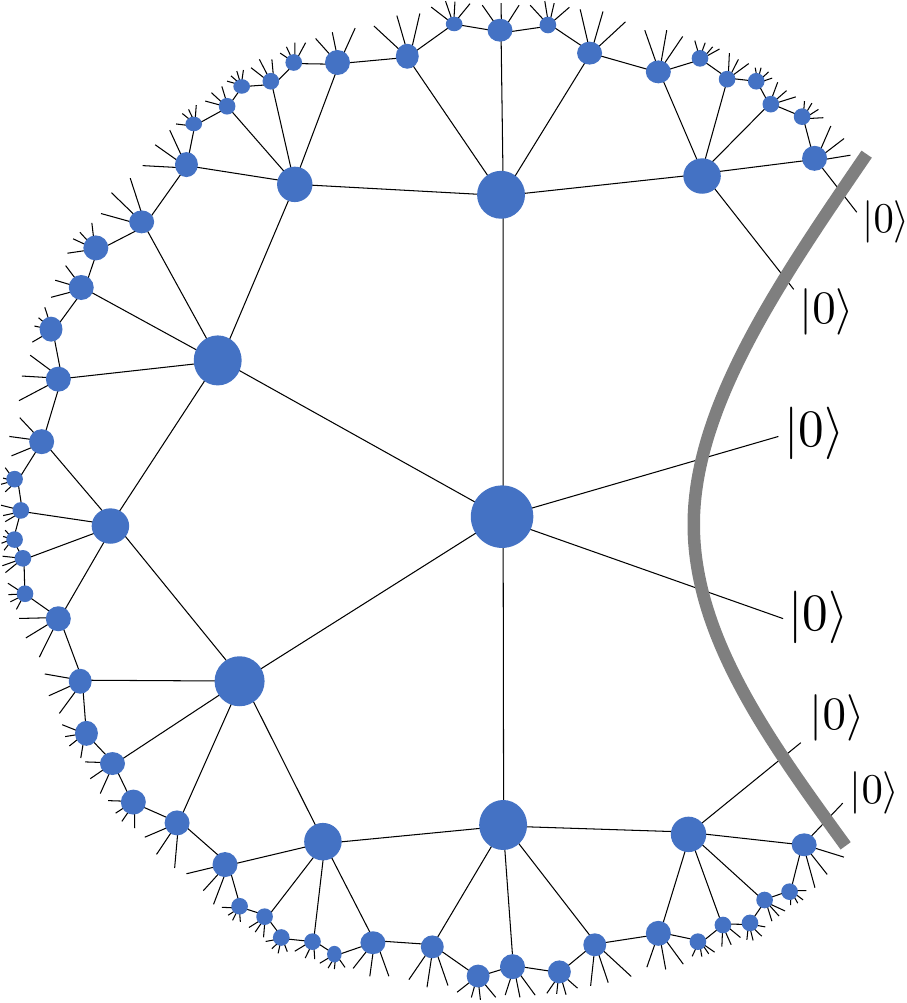}
\caption{Tensor network picture for placing EoW brane-like object by performing projective measurement in disentangled basis states. In this paper, we consider a random tensor network, where each tensor denoted by a dot is a Haar random state and the tensor network is obtained by contracting the neighboring tensor legs.} 
\label{fig_TN_EoW}
\end{figure}

Nevertheless, a useful picture arises in tensor network models~\cite{Pastawski:2015qua, Hayden:2016cfa, Almheiri:2018xdw, Antonini:2022sfm, Mori:2023swn}, as shown in Fig.~\ref{fig_TN_EoW}.
By coarse-graining in the radial direction, one can perform projective measurements on DOFs associated with a convex surface $\Sigma \subset \mathcal{E}_A$.
Projective measurements in a product basis along $\Sigma$ lead to post-measurement states whose geometry includes an EoW brane-like termination along $\Sigma$.\footnote{We refer to such surfaces as EoW brane-like objects, rather than EoW branes, because they do not necessarily obey Neumann boundary conditions, nor do the measurement bases correspond to conformal boundary states.
Furthermore, unlike traditional EoW branes, their effective tension may vary.
Note also that these post-measurement states do not necessarily correspond to static geometries.
Nevertheless, as we will show, the optimal configurations for evaluating some of entanglement measures involve EoW brane-like objects located precisely at the boundary of the entanglement wedge.
Since these boundaries are extremal surfaces, they have vanishing extrinsic curvature and hence correspond to branes with zero tension.}
Post-measurement geometries are expected to be independent of the specific measurement outcome, with fluctuations suppressed in $G_N$.

\subsection{LOCC distillation via holographic measurement}

For simplicity of presentation, we focus on the pure AdS$_{3}$ setup illustrated in Fig.~\ref{fig_connected}.
Furthermore, we work in a regime with
\begin{align}
J^{W}(A|C) \equiv S_{A} - E^{W}(A:B) >0.
\end{align}
This condition can be schematically depicted as 
\begin{align}
\figbox{1.7}{fig_SA}\ > \ \figbox{1.7}{fig_EWAB}.
\end{align}

\begin{figure}
\centering
\includegraphics[width=0.25\textwidth]{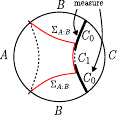}
\caption{An LOCC entanglement distillation protocol in holography. 
The protocol places EoW brane-like objects on a portion of $\gamma_C$.
}
\label{fig_distillation_tensor}
\end{figure}

The protocol performs projective measurements on DOFs associated with a portion of the minimal surface $\gamma_C$. 
Let us split $\gamma_C$ into two parts $\gamma_C = \gamma_{C_{0}}\cup \gamma_{C_1}$ as depicted in Fig.~\ref{fig_distillation_tensor}.\footnote{There are many other measurement configurations that achieve the same amount of distillable entanglement.}
\begin{enumerate}[1)]
\item Perform projective measurements on $\gamma_{C_{0}}$ in a disentangled basis, and leave the remaining part $\gamma_{C_1}$ untouched. 

\item Send the measurement outcome from $C$ to $A$. 

\item Apply the Petz recovery map on $A$ to distill $\frac{1}{4G_{N}}\text{Area}(\gamma_{C_{1}})$  copies of EPR pairs.
\end{enumerate}

Here we take $\gamma_{C_{0}}$ large enough such that the entanglement wedge cross section $\Sigma_{A:B}$ anchors on $\gamma_{C_0}$ as shown in Fig.~\ref{fig_distillation_tensor}. 
The post-measurement state has a semiclassical dual geometry with EoW brane-like objects. 
Namely, regardless of the measurement outcomes, we obtain the same geometry at leading order.
We also choose the area (length) of $\gamma_{C_{1}}$ so that the remaining portion $\gamma_{C_{1}}$ carries $\frac{1}{4G_{N}}\text{Area}(\gamma_{C_{1}})\approx J^W(A|C)$ entropy. 
This is possible since, otherwise, $\gamma_A$ would not be the minimal surface of $A$.

The third step distills EPR pairs because, in the post-measurement state, the minimal surface $\gamma_A$ changes its profile and overlaps with $\gamma_{C_1}$ due to projective measurements. 
Namely, two candidate surfaces for $A$ satisfy 
\begin{align}
 \figbox{1.7}{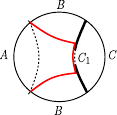} \ \approx \ \figbox{1.7}{fig_SA} 
 \label{eq:EW-trans}
\end{align}
since we choose $\frac{1}{4G_{N}}\text{Area}(\gamma_{C_{1}}) \approx J^W(A|C)$. Note that there is no contribution from a portion of the curve overlapping with the thick black curves, which denote the EoW brane-like objects.

It is worth noting that this protocol beats the hashing bound. 
Namely, we have 
\begin{align}
S_{AC}= \figbox{1.7}{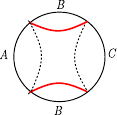} \ \geq E^{W}(A:B) = \figbox{1.7}{fig_EWAB}
\end{align}
by the definition of $E^{W}$. This implies $S_{A} - S_{AC} \leq S_{A} - E^W(B:C)$. Hence, we have 
\begin{align}
\text{hash}(A:C)\leq J^W(A:C).
\label{eq:Hol-hash}
\end{align}

\subsection{Optimality under holographic measurement}

Here, we prove that the aforementioned protocol is optimal under holographic measurements.

\begin{claim}
In holography, the 1WAY LOCC distillable entanglement is given by
\begin{equation}
\begin{split}
&E_{D}^{[\mathrm{1WAY \ LOCC}]}  (A \leftarrow C) \approx J^W(A|C)\\
&E_{D}^{[\mathrm{1WAY \ LOCC}]}  (A:C) \approx J^W(A:C)
\label{eq:ED-1-G-AC}
\end{split}
\end{equation}
if LOCC is restricted to holographic measurements.
\end{claim}

To prove this statement, let us suppose that there exists a 1WAY holographic measurement protocol that distills more than $J^W(A|C)$ EPR pairs by measuring $C_0 \subset C$. 
Let $\sigma_{AC}^{j_C}$ be a post-measurement state with $j_C$ denoting the measurement outcome. 
In the previous section, we derived an upper bound (Eq.~\eqref{eq:petz-lo-hol}) on $E_{D}^{[\text{LO}]}(A:C)$ by using the performance of the Petz map. 
Since $\sigma_{AC}^{j_C}$ is also a holographic state with semiclassical dual, we can apply this bound to $\sigma_{AC}^{j_C}$ and obtain
\begin{align}
E_{D}^{[\text{LO}]}(A:C)(\sigma_{AC}^{j_C}) \lesssim J^W(A|C)(\sigma_{AC}^{j_C})  .
\end{align}
Here, we claim the following inequality, which will be proven shortly:
\begin{align}
J^W(A|C) (\sigma_{AC}^{j_C}) \lesssim J^W(A|C) (\rho_{AC}). \label{eq:1WAY_claim}
\end{align}
This essentially states that holographic measurements on $A$ do not increase $J^W(A|C)$. 
This inequality then suggests $E_{D}^{[\text{LO}]}(A:C)(\sigma_{AC}^{j_C})\lesssim J^W(A|C)(\rho_{AC})$, which leads to a contradiction.

The remaining task is to prove the inequality in Eq.~\eqref{eq:1WAY_claim}.
Recall that 
\begin{equation}
\begin{split}
J^W(A|C)(\rho_{AC}) &\equiv S_{A}(\rho_{A}) - E^W (A:B) (\rho_{AB}) \\
J^W(A|C)(\sigma_{AC}^{j_C}) &\equiv S_{A}(\sigma_{A}^{j_C}) - E^W (A:B) (\sigma_{AB}^{j_C}).
\end{split}
\end{equation}
One can show 
\begin{align}
S_{A}(\sigma_{A}^{j_C}) \leq S_{A}(\rho_{A}) 
\end{align}
by writing the average entropy drop due to measurements on $C_0$ as
\begin{align}
\Delta S_{A} = S_A(\rho_{A}) - \sum_{j_C}p_{j_C} S_A( \sigma_A^{j_C})= \sum_j p_{j_C} S( \sigma_A^{j_C} \Vert \rho_A ) \geq 0, \label{eq:LAI-rel}
\end{align}
using the non-negativity of the relative entropy, and observing that $S_A( \sigma_A^{j_C})$ is independent of $j_C$ at leading order under holographic measurements. One can also show 
\begin{align}
E^W(A:B)(\sigma_{AB}^{j_C}) \geq E^W(A:B)(\rho_{AB})
\end{align}
by observing that the minimal surface $\gamma_C$ for $\sigma_{ABC}^{j_C}$ does not extend beyond the minimal surface $\gamma_C$ for $\rho_{ABC}$ since holographic measurements on $C_0$ can place EoW brane-like objects only inside $\mathcal{E}_{C}$ for $\rho_{ABC}$. 
Hence, we obtain Eq.~\eqref{eq:1WAY_claim}.

The inequality (Eq.~\eqref{eq:1WAY_claim}) may be interpreted as a version of the monotonicity relation of $J^W(A|C)$ under holographic measurement on $C$. 
Note, however, that conventional monotonicity relations for locally accessible information $J(A|C)$ hold only under LOs acting on $A$ and $C$ separately~\cite{Henderson:2001wrr}. 
By contrast, the monotonicity relation for $J^W(A|C)$ applies to holographic measurements with CCs as well.   

\subsection{Optimality from $E_F\approx E^W$ proposal}
In the previous subsection, we proved the optimality of the distillation protocol when POVMs are restricted to holographic measurements. 
The optimality of the protocol can be proven, without the need of restricting to holographic measurements, by relying on the holographic relations for $E_F$ and $J(A|C)$:
\begin{align}
E_{F}(A:B) \approx E^W(A:B), \qquad J(A|C) \approx J^W(A|C)
\end{align}
where the second relation follows from the Koashi-Winter relation. 
Assuming this proposal and using claim~\ref{claim_JCA}, we arrive at the following result.

\begin{claim}
Assume the holographic proposal of $E_{F}(A:B) \approx E^W(A:B)$. We then have 
\begin{equation}
\begin{split}
&E_{D}^{[\mathrm{1WAY \ LOCC}](A\leftarrow C )}(A:C) \approx J^W(A|C) \\
&E_{D}^{[\mathrm{1WAY \ LOCC}]}(A:C) \approx J^W(A:C).
\end{split}
\end{equation}
That is, the aforementioned 1WAY LOCC protocol is optimal at leading order in $1/G_{N}$.
\end{claim}

Hence, the optimality of the protocol is reduced to the validity of the holographic proposals of $E_{F}(A:B) \approx E^W(A:B)$ and $J(A|C) \approx J^W(A|C)$. 
In the next two sections, we present additional pieces of evidence for $E_{F}(A:B) \approx E^W(A:B)$ and $J(A|C) \approx J^W(A|C)$. 

\section{On $E_F\approx E^W$ proposal }\label{sec:EF}

In this section, we discuss the holographic proposal of $E_F\approx E^W$.

In~\cite{Umemoto:2019jlz}, Umemoto presented a no-go argument, suggesting $E_{F}\not=E^W$ in general. 
Below, we reproduce the argument as we understand it, and explain how our holographic proposal can avoid the apparent contradiction. 

Recall that entanglement of formation satisfies the upper bound $E_{F}(A:B) \leq \min( S_{A},S_{B} )$.
Also, recall the Araki-Lieb inequality 
\begin{align}
S_{A} + S_{AB} - S_{B} \geq 0.
\end{align}
These two bounds are connected by the equivalence
\begin{align}
E_{F}(A:B) = S_{A} \ \Leftrightarrow \ S_{A} + S_{AB} - S_{B} = 0. \label{eq:iff}
\end{align}
Suppose $E_{F}(A:B) = E^W(A:B)$, to derive a contradiction. 
Umemoto pointed out that there are cases in holography where 
\begin{align}
S_{A} + S_{AB} - S_{B} = O(1/G_{N})\quad \text{but} \quad E^W(A:B) = S_{A}. \label{eq:disconnected}
\end{align}
This contradicts Eq.~\eqref{eq:iff}, suggesting that $E_{F}(A:B) \not= E^W(A:B)$ in general.

However, this argument can be circumvented if the relation ``$E_{F}=E^W$'' is only valid at leading order. That is, Umemoto's argument does not preclude the following situation:
\begin{align}
S_{A} + S_{AB} - S_{B} = O(1/G_{N}), \qquad E_F(A:B) \approx E^W(A:B) \approx S_{A}. \label{eq:Umemoto}
\end{align}
The key point is that the equivalence in Eq.~\eqref{eq:iff} is not robust under small corrections and does not hold approximately.
Such a situation can indeed occur in holography when $ABC$ is in a pure state.

In fact, an analogous situation arises for Haar random states as pointed out in section~\ref{sec:LOCC_Haar}. 
For $n_C< n_A + n_B$ and $n_A < n_B$, we have $E_{F}(A:B) \approx E^W(A:B) \approx n_A$ and $S_{A}+S_{AB} - S_{B} = O(n)$.
In this case, the Araki-Lieb inequality reduces to the non-negativity of mutual information, since $S_{A} + S_{AB} - S_{B} = I(A:C)$.  
Then the first of Eq.~\eqref{eq:Umemoto} implies that $A,C$ have a connected wedge, while the second implies that $J(A|C) \approx 0$ at leading order. 
This corresponds to the Regime II in Fig.~\ref{fig_claim} in pure AdS$_3$.


\section{On $J(A|C)\approx J^W(A|C)$ proposal }\label{sec:JW}

In this section, we discuss the holographic proposal of $J(A|C)\approx J^W(A|C)$. 



\subsection{Generalized RT formula}

\begin{figure}
\centering
\includegraphics[width=0.22\textwidth]{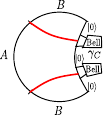}
\caption{
Bell measurements across the minimal cross section $\Sigma_{A:B}$.
}
\label{fig_EoW}
\end{figure}

In the next three subsections, we examine this problem from the perspective of the proposal $J(A|C)\approx J^W(A|C)$. 
According to this proposal, local projective measurements on $\gamma_C$ achieve the maximal entropy drop $\Delta S_{A}$. 
That is, placing EoW brane-like objects on $\gamma_C$ leads to
\begin{align}
S_{A}^{\text{before}} = \frac{1}{4G_{N}} \figbox{1.5}{fig_SA}\,  , \qquad S_{A}^{\text{after}} = \frac{1}{4G_{N}} \figbox{1.5}{fig_EWAB} \ , \label{eq:post-measurement}
\end{align}
yielding
\begin{align}
\Delta S_{A} = S_{A}^{\text{before}} - S_{A}^{\text{after}} \approx J^W(A|C). 
\end{align}
The central question is therefore whether holographic measurements achieve the maximal entropy reduction at leading order.

One piece of supporting evidence comes from the observation that projections onto entangled states tend to increase $S_{A}^{\text{after}}$.
For instance, consider performing Bell basis measurements across the minimal cross section $\Sigma_{A:B}$, as shown in Fig.~\ref{fig_EoW}, instead of local product basis measurements.
Such Bell projections introduce additional entanglement between $A$ and $B$, resulting in higher post-measurement entropy $S_A^{\text{after}}$.

This phenomenon can be understood naturally via the generalized RT formula. 
The minimal surface $\gamma_C$ may be interpreted as bulk DOFs that are isometrically encoded into boundary regions $AB$.
Let $|\psi_C\rangle$ denote the post-measurement state on $\gamma_C$.
Then, the generalized RT formula~\cite{Faulkner:2013ana} gives 
\begin{align} 
S_A^{\text{after}} \approx E^W(A:B) + S_{C_A}(|\psi_C\rangle), 
\end{align}
where $C_A$ is the portion of $\gamma_C$ that lies within the new entanglement wedge of $A$.
Here, the second term is analogous to the bulk entropy contribution.
To maximize the entropy drop $\Delta S_A$, one should minimize this bulk term. This suggests that local projective measurements on $\gamma_C$, which produce low-entropy pure states, are optimal.

A caveat to this argument is that the generalized RT formula is only valid when the bulk entropy is sub-leading. 
Indeed, known violations occur when the bulk entropy contributes at leading order~\cite{ Akers:2020pmf, Akers:2019wxj}. 
However, these examples involve bulk quantum states that are statistical mixtures of semiclassically distinct geometries.
In contrast, we are considering projections onto pure states.
Moreover, placing a mixed state in the bulk generically increases the entropy, whereas our goal is to achieve a maximal entropy drop.
For these reasons, we expect that our argument remains valid and is not susceptible to the kinds of counterexamples constructed in~\cite{Akers:2020pmf, Akers:2019wxj}.

\subsection{Violation of holographic proposals for R\'{e}nyi entropies}

In the previous subsection, we argued that the generalized RT formula supports the proposal $J(A|C) \approx J^W(A|C)$.
However, it remains difficult to fully exclude the possibility of fine-tuned POVMs that could achieve an entropy reduction larger than $J^W(A|C)$.
As a nontrivial test of this proposal, we examine whether applying the Petz recovery map, followed by projective measurements, can lead to a larger entropy drop.
The intuition is that the Petz map, acting on $C$, distills DOFs that are strongly entangled with $A$.
One might then expect that projectively measuring these distilled DOFs could enhance the entropy reduction.

In Appendices~\ref{app:Haar} and~\ref{app:TN}, we perform this analysis in both Haar random states and holographic setups.
The result is negative: the Petz map approach does not yield a larger entropy reduction when we focus on the von Neumann entropy.
However, there is an interesting twist: this approach does lead to larger entropy reduction when considering R\'{e}nyi entropies.
Specifically, we find that in certain regimes, 
\begin{align} 
J^{(m)}(A|C) > J^W(A|C) \qquad \text{for } m \geq 2, 
\end{align} 
or equivalently, 
\begin{align} E_F^{(m)}(A:B) < E^W(A:B) \qquad \text{for } m \geq 2, 
\end{align} 
at leading order.
Here, $E_F^{(m)}$ denotes the R\'{e}nyi entanglement of formation, where the usual von Neumann entropy in the convex sum is replaced by the R\'{e}nyi-$m$ entropy.
The R\'{e}nyi locally accessible information $J^{(m)}(A|C)$ is defined analogously.
The decomposition demonstrating this inequality is obtained by applying the Petz map to prepare a double-copy state, followed by projective measurements on the second copy.

This deviation can be understood from a quantum information perspective.
As discussed in Section~\ref{sec:instability}, even exponentially small perturbations to a separable state $\rho_A \otimes \rho_{A'}$ can produce leading-order discrepancies between von Neumann and R\'{e}nyi quantities.
The isotropic state exhibits precisely this behavior as we further discuss in section~\ref{sec:instability}.
From the holographic viewpoint, this discrepancy arises because the dominant saddle configurations contributing to $S_{A}^{(m)}$ depend on $m$.
In particular, these saddles become divergent as $m \rightarrow 1$, ensuring that the original proposal $E_F \approx E^W$ remains valid in the von Neumann limit.

In summary, we find that the holographic proposal $J(A|C) \approx J^W(A|C)$ can be violated at leading order for R\'{e}nyi entropies with $m > 1$.
However, this violation vanishes as $m \rightarrow 1$.
We interpret this subtle behavior as further evidence supporting the holographic proposal for $E_F \approx E^W$ and $J(A|C) \approx J^W(A|C)$ at $m=1$.

\subsection{Emergent bulk causality}

Another motivation for the holographic proposal $J(A|C) \approx J^W(A|C)$ comes from considerations of emergent bulk causality.
In holography, bulk DOFs in $\mathcal{E}_{AB}$ are spacelike separated from those in $\mathcal{E}_C$, implying that they commute and should remain causally independent.
Thus, local measurements on $C$ should not induce noticeable changes to bulk physics, including the geometry, within $\mathcal{E}_{AB}$.
Suppose, for contradiction, that some POVM on $C$ reduces the entropy $S_A$ to a value $S_A^{\text{after}} < E^W(A:B)$ at leading order.
This would imply that the minimal surface of $A$ becomes shorter than $E^W(A:B)$, suggesting that the geometry within $\mathcal{E}_{AB}$ has been significantly altered by backreaction from measurements on $C$.
However, such backreaction would be detectable by observers inside $\mathcal{E}_{AB}$, violating the bulk causality.
Therefore, we expect that $S_A$ can decrease only as far as $S_A^{\text{after}} \approx E^W(A:B)$ at leading order in $1/G_N$.

It is important to note, however, that there may exist certain projectors $\Pi_C$ that do induce drastic geometric changes within $\mathcal{E}_{AB}$.
A concrete example occurs in the thermofield double (TFD) state, where $AB$ is identified with one boundary and $C$ with the other side of the eternal black hole.
Projecting $C$ onto a low-energy subspace leads to a correspondingly lower-energy state on $AB$, thereby modifying the geometry substantially.
Nevertheless, the amplitude of such projections is exponentially suppressed in $1/G_N$.
See~\cite{Miyaji:2021ktr} for details.

\section{Bound entanglement in holography}\label{sec:bound_entanglement}

\subsection{NPT bound entanglement}

In quantum information theory, entangled states that are not distillable are known as \emph{bound entangled states}, and have been extensively studied~\cite{Horodecki:1998kf}; see Section III of~\cite{horodecki2020openproblemsquantuminformation} for a recent review.
One surprising aspect of our result ($E_{D}^{[\text{1WAY LOCC}]}(A:C) \approx J^W(A:C)$) is that certain holographic states with connected entanglement wedges may provide a version of bound entanglement in a 1WAY and one-shot setting, satisfying 
\begin{align} 
0 \approx E_{D}^{[\text{1WAY LOCC}]}(A:C)  < \frac{1}{2} I(A:C) < E_{F}(A:C) \approx E^W(A:C). \label{eq:bound-proposal} 
\end{align} 
See also Eq.~\eqref{eq:holo_ineq}.
Such a regime typically arises shortly after the entanglement wedge becomes connected while $J^W(A|C), J^W(C|A)$ remain zero.  

We now discuss how these ideas relate to the problem of negative partial transpose (NPT) bound entanglement.
Recall that a mixed state $\rho_{AC}$ is said to be NPT if its partial transpose $\rho_{AC}^{T_C}$ has a negative eigenvalue; otherwise, it is a positive partial transpose (PPT) state.
The Peres-Horodecki criterion establishes that 
\begin{align}
\text{$\rho_{AC}$ is separable} \quad \Rightarrow \quad \text{$\rho_{AC}$ is a PPT state}.
\end{align}
However, the converse does not hold.
There exist PPT states, known as PPT bound entangled states, that are entangled but non-distillable. 
(Note: PPT states are always non-distillable as  $E_D(A:C) \leq E_N(A:C) = 0$.)

A major open problem in quantum information theory is whether NPT bound entangled states (i.e., entangled NPT states that are nonetheless non-distillable) exist~\cite{DiVincenzo_2000,PhysRevA.61.062313}.
In this context, our work offers an intriguing new perspective.
In random tensor networks and fixed-area states, the logarithmic negativity $E_N(A:C)$ satisfies~\cite{Dong:2021clv} 
\begin{align} 
E_N(A:C) \approx \frac{1}{2} I(A:C).
\end{align} 
Thus, holographic states with connected wedges are NPT states at leading order.
Furthermore, we have shown that holographic states are not distillable under 1WAY LOCC at leading order when $J^W(A:C) \approx 0$.
This implies the existence of a regime where 
\begin{align} 
E_N(A:C) \sim O(1/G_N), \qquad E_D^{[\text{1WAY LOCC}]}(A:C) \approx 0, 
\end{align} 
thus suggesting that certain holographic states realize a form of NPT bound entanglement in the one-shot 1WAY LOCC setting at $G_N \rightarrow 0$. 
Meanwhile, as observed in Section~\ref{sec:LOCC_Haar}, Haar random states provides a closely related example:
in certain regimes, 
\begin{align} E_N(A:C) \sim O(n), \qquad E_D^{[\text{1WAY LOCC}]}(A:C) \approx 0. \end{align} 
Thus, Haar random states also serve as concrete examples of NPT bound entanglement at leading order, again in the one-shot, 1WAY LOCC setting.

Let us reiterate that, in conventional discussions, bound entanglement is considered in the asymptotic setting.
An essential difference is that, by preparing many copies of a given state, one may concentrate its entanglement and distill nearly perfect EPR pairs even from weakly entangled states.
In this context, the distillable entanglement $E_D$ is defined as the asymptotic rate at which EPR pairs can be distilled per copy. 
A state is considered distillable whenever $E_D \neq 0$, while it is considered NPT whenever $E_{N} \neq 0$.
On the contrary, in holographic states, we have $E_D^{[\text{1WAY LOCC}]}\rightarrow 0$ only at the limit of $G_N\rightarrow 0$.

\subsection{Remark on previous proposal}

Holographic bound entanglement has previously been discussed in the literature. 
In particular,~\cite{Vardhan:2021mdy,Vardhan:2021npf} considered certain finite temperature holographic states and observed the following
\begin{align}
E_N(A:B) \sim O(n), \qquad I(A:B) \sim o(n)
\end{align}
where $I(A:B)$ is sub-extensive in terms of the total number of qubits $n$.

Based on this observation, the authors proposed that a large (extensive) logarithmic negativity implies nontrivial quantum entanglement in the sense of entanglement cost $E_C$.
Recall that $E_{C}(A:B)$ quantifies the number of EPR pairs per copy needed to prepare $\rho_{AB}$ with an error vanishing in the asymptotic limit. 
While $E_C$ is difficult to compute in general, a stronger version—called the exact PPT entanglement cost, $E_C^{[\mathrm{ppt,exact}]}$—admits analytic control in certain settings.
This quantity is defined as the number of EPR pairs per copy needed to exactly prepare the target state using PPT-preserving operations, before taking the asymptotic limit.
Importantly, it satisfies the bound $E_C^{[\mathrm{ppt,exact}]} \geq E_N$~\cite{Audenaert_2003}.

From this,~\cite{Vardhan:2021mdy,Vardhan:2021npf} proposed that the finite-temperature states they examined may be examples of bound entangled states:
\begin{align}
E_C^{[\mathrm{ppt,exact}]}(A:B) \geq E_N(A:B) \sim O(n), \qquad E_{D}(A:B) \leq \frac{1}{2}I(A:B) \sim o(n).
\end{align}
At first glance, this proposal may seem to contradict our result.
We have suggested that bound entangled states can arise in holography when $I(A:C) \sim O(1/G_N)$, whereas~\cite{Vardhan:2021mdy,Vardhan:2021npf} consider a regime where $I(A:C) \sim o(1/G_N)$.

There are two possible resolutions to this apparent tension.
First, our analysis is guided by intuition from random tensor networks and fixed-area states which exhibit nearly flat entanglement spectra.
In contrast, full holographic states generally have non-flat spectra, with R\'{e}nyi entropies $S_A^{(m)}$ depending strongly on $m$.
This opens the possibility that the non-flatness of the spectrum plays a role in the emergence of bound entanglement in examples of~\cite{Vardhan:2021mdy,Vardhan:2021npf}.

Second, and more crucially, the distinction between $E_C$ and $E_C^{[\mathrm{ppt,exact}]}$ may be substantial.
The latter requires \emph{exact} state preparation prior to taking the asymptotic limit, and is in fact sensitive to exponentially small perturbations to the state.
Similarly, exponentially small amount of entanglement can lead to extensitive $E_N$.
This distinction becomes particularly significant when the entanglement spectrum is non-flat, with a large deviation between $S_A$ and $S_A^{(m)}$.
In section~\ref{sec:instability}, we will illustrate this difference by analyzing the entanglement structure of an isotropic state.

\addcontentsline{toc}{section}{Part III: Relevant topics}

\section*{Part III: Relevant topics}

\section{Subleading effects}\label{sec:subleading}

In this section, we discuss possible subleading contributions to $E_{D}^{[\text{LO}]}$ and $E_{D}^{[\text{LOCC}]}$. 
Specifically, we identify three potential physical mechanisms for subleading effects.

\begin{enumerate}[i)]
\item \emph{Traversable wormhole}: A 1WAY LOCC version of traversable wormhole protocol distills EPR pairs from $\rho_{AC}$ when $A,C$ are sufficiently large subsystems on two boundaries. This may allow subleading distillation even when $J^W(A:C) \approx 0$.
\item \emph{Holographic scattering}: The fact that bulk scattering processes require a connected entanglement wedge suggests the possibility of LOCC distillability from $\rho_{AC}$ at subleading order even when $J^W(A:C) \approx 0$.
\item \emph{Planck-scale effect}: Significant corrections to $E_{D}^{[\text{LO}]}(A:C)$ must arise when $\gamma_A,\gamma_C$ become Planck-scale close.
\end{enumerate}

Before beginning, it is worth noting that Umemoto's argument from section~\ref{sec:EF} also points to the possible 
subleading corrections to the holographic proposal $E_{F}\approx E^W$. 
Also, bulk matter field contributions to $E_D(A:C)$ are expected to be negligible as correlations in massive matter fields between two entanglement wedges $\mathcal{E}_A$ and $\mathcal{E}_C$ decay exponentially with spatial separation.

\subsection{Traversable wormhole}\label{sec:traversable}

Traversable wormholes are phenomena where quantum information, thrown from one side of a two-sided AdS black hole with inverse temperature $\beta$ and Bekenstein-Hawking entropy $S_{\mathrm{BH}}$ at $t=0$, can reach the other side by introducing a special interaction that couples two boundaries:
\begin{align}
U_{int}=\exp\qty( - i \theta \sum_{j} O_{j}\otimes O_{j}^* ) \label{eq:traversable_coupling}
\end{align}
where $O_{j}$ are simple operators such as few-body Pauli or Majorana operators~\cite{Gao:2016bin, Maldacena:2017axo, Gao:2019nyj, Brown:2019hmk, Nezami:2021yaq, Schuster:2021uvg}. 
This interaction, with a phase $\theta(\tau)$, must be applied at $t_{L} \approx - t_{R}= \tau$ with $\tau > 0$ satisfying $t_{th} \lesssim \tau \lesssim t_{scr}$ where $t_{th}\sim\beta$ and $t_{scr}\sim \beta\log S_{\mathrm{BH}}$ are the thermal and scrambling time respectively. 

\begin{figure}
\centering
\raisebox{\height}{a)\hspace{10pt}}\raisebox{-0.85\height}{\includegraphics[width=0.3\textwidth]{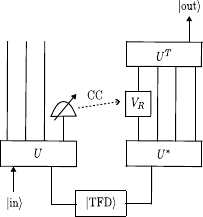}}
\hspace{10pt}
\raisebox{\height}{b)\hspace{5pt}}\raisebox{-1.2\height}{\includegraphics[width=0.28\textwidth]{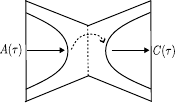}}
\caption{
a) Quantum circuit diagram for the LOCC traversable wormhole. Starting from the thermofield double at $t=0$, the system evolves to $t_L=\tau$ and $t_R=-\tau$ via $U\otimes U^*$, followed by measurements on the left and feedback operations on the right.
The system then evolves to $t_R = 0$ via $I \otimes U^T$, transmitting the input state $|\psi_{\text{in}}\rangle$ to the right.
b) Trajectory of the input particle on the static slice. The entanglement wedge $\mathcal{E}(A(t))$ is drawn to include the particle at its front edge. Near the horizon in AdS$_3$, the boundary region $A(t)$ expands at the speed of light.
}
\label{fig_traversable_1}
\end{figure}

At first glance, the ability to transmit information between directly coupled sides may not seem surprising. 
What is truly surprising is that it utilizes pre-shared quantum entanglement between two sides in order to transmit information.
This becomes clear when the same effect is reproduced by a quantum teleportation-like protocol where the unitary coupling of Eq.~\eqref{eq:traversable_coupling} is replaced with 1WAY LOCC. 
Concretely, suppose that $O_{j}$'s are mutually commuting single-body Pauli operators. 
One can send a signal through the wormhole by projectively measuring $O_{j}$ on the left and then applying the following on the right:
\begin{align}
U_{R} = \exp\qty( - i \theta \sum_{j} m_{j} O_{j}^* ) \label{eq:traversable_LOCC}
\end{align}
where $m_j = \pm1$ denotes the measurement outcome of $O_{j}$.
The corresponding quantum circuit is shown in Fig.~\ref{fig_traversable_1}(a). 
This process constitutes an LOCC protocol, involving local measurements on one side and classical communications to the other.
That quantum information can be sent by an LOCC implies that the traversable wormhole utilizes pre-existing entanglement between two parties. 

To connect this with entanglement distillation, we need a few additional ingredients. 
First, traversable wormhole phenomena remain possible even when only subsystems of the boundary Hilbert spaces are accessed.
Let us track the motion of an infalling signal by the growth of an entanglement wedge on the static slice as shown in Fig.~\ref{fig_traversable_1}(b).\footnote{
In the absence of the input state $|\psi_{in}\rangle$, the time evolution by $U \otimes U^*$ leaves $|\text{TFD}\rangle$ invariant  as in Fig.~\ref{fig_traversable_1}(a).
Thus, ignoring backreaction, the infalling signal's motion can be tracked entirely on the static slice.
} 
When the coupling (or LOCC) is applied, the signal jumps across the horizon to the right which can be understood as a result of the backreaction from the coupling.\footnote{Some readers may question applying backreaction effects on a static slice. While earlier works interpret traversable wormholes as resulting from negative energy shockwaves that cause the particle to cross to the other side, this view conflicts with the manifest left-right symmetry of the boundary time evolution (especially under Eq.~\eqref{eq:traversable_coupling}).
Our interpretation aligns with \cite{Maldacena:2017axo}, where in JT gravity, the coupling instantaneously shifts the horizon locations, effectively inducing a symmetric static-slice jump of the particle. 
}
Here, we define boundary subsystems $A(t)$ on the left such that $\mathcal{E}(A(t))$ is just large enough to contain the infalling signal. 
We take $A = A(\tau)$ on the left and also define $C = C(\tau)$ analogously on the right. 
Since the infalling/outgoing signals are recoverable from $A$ and $C$ via the entanglement wedge reconstruction, the traversable wormhole protocol can be implemented using only the mixed state $\rho_{AC}$, without access to complementary regions.\footnote{
Stanford and Mezei~\cite{Mezei:2016wfz} found that, for infalling massless signals near the horizon, the size of $A(t)$ grows at the speed of $\tilde{v}_{B} = \sqrt{\frac{d}{2(d-1)}}$ where $d$ is the boundary spacetime dimension.
For $d=2$, it is $\tilde{v}_{B}=1$, matching the speed of light. 
They also observed that this speed $\tilde{v}_{B}$ agrees with the butterfly velocity $v_{B}$, which characterizes the spread of local perturbations via out-of-time ordered correlation functions. 
Thus, it suffices to apply the coupling (Eq.~\eqref{eq:traversable_coupling}) or LOCC (Eq.~\eqref{eq:traversable_LOCC}) only on $A = A(\tau)$ and $C=C(\tau)$.
}

We now explain how EPR pairs can be distilled from $\rho_{AC}$ using this LOCC protocol.
Rather than sending a signal through the wormhole, we prepare an EPR pair on the left, retain one half, and send the other half through the wormhole via LOCC.
This creates an EPR pair shared between $A$ and $C$, thereby distilling entanglement.
The number of distillable pairs in this setting is limited.
A natural upper bound is set by the black hole entropy $S_{\text{BH}}$, but in practice, distillable entanglement is expected to remain subleading, as signals are transmitted through bulk matter fields.
Indeed, if a signal with entropy $O(1/G_N)$ is sent, its backreaction becomes significant, and entanglement distillation is expected to fail.

\begin{figure}
\centering
\raisebox{\height}{a)\hspace{10pt}}\raisebox{-0.85\height}{\includegraphics[width=0.24\textwidth]{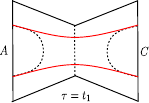}}
\hspace{10pt}
\raisebox{\height}{b)\hspace{10pt}}\raisebox{-0.85\height}{\includegraphics[width=0.24\textwidth]{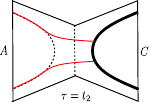}}
\caption{
a) The connected entanglement wedge at $\tau = t_1$, where the minimal surface of $AC$ (shown in red) becomes comparable to those of $A$ and $C$.
b) The locally accessible information satisfies $J^W(A:C) > 0$ for $\tau > t_2 > t_1$, where the minimal surface $\gamma_A$ becomes comparable to the entanglement wedge cross section $\Sigma_{A:B}$.
}
\label{fig_traversable_2}
\end{figure}

Finally, we discuss the possibility of subleading corrections to our proposal that $E_{D}^{[\text{1WAY LOCC}]} \approx J^W(A:C)$.
The key question is whether this LOCC protocol works in a regime where $I(A:C)=O(1/G_{N})$, but $J^W(A:C) = 0$. 
Here, it is convenient to identify two characteristic time scales $\tau_1$ and $\tau_2$ as shown in Fig.~\ref{fig_traversable_2}.
Namely, for $\tau > t_1 > 0$, the boundary regions $A$ and $C$ have a connected wedge with $I(A:C)>0$, and for $\tau > t_2 > t_1$, we have $J^W(A:C)>0$ at leading order.
We are particularly interested in whether the traversable wormhole phenomena can occur in the intermediate regime $t_{2}> \tau > t_{1}$, which would imply LOCC entanglement distillation at subleading order even when $J^W(A:C)= 0$.  
In Appendix~\ref{app:metric}, we explicitly compute $t_1,t_2$ in AdS$_{3}$ and find that $t_1,t_2 \sim t_{th}$.
Analyses from~\cite{Schuster:2021uvg} indicate that the traversable wormhole phenomena indeed become possible for $\tau \gtrsim t_{th}$, suggesting the possibility of such processes in this regime.
However, it remains unclear from the results of~\cite{Schuster:2021uvg} exactly how large $\tau$ must be for a traversable wormhole to occur. 
It would be interesting to ask whether a connected wedge ($\tau \geq t_1$) is necessary and/or sufficient for enabling a traversable wormhole.
Answering this question requires more detailed analyses, which we leave for future work.

\subsection{Holographic scattering}

\begin{figure}
\centering
\raisebox{\height}{a)\hspace{10pt}}\raisebox{-0.85\height}{\includegraphics[width=0.21\textwidth]{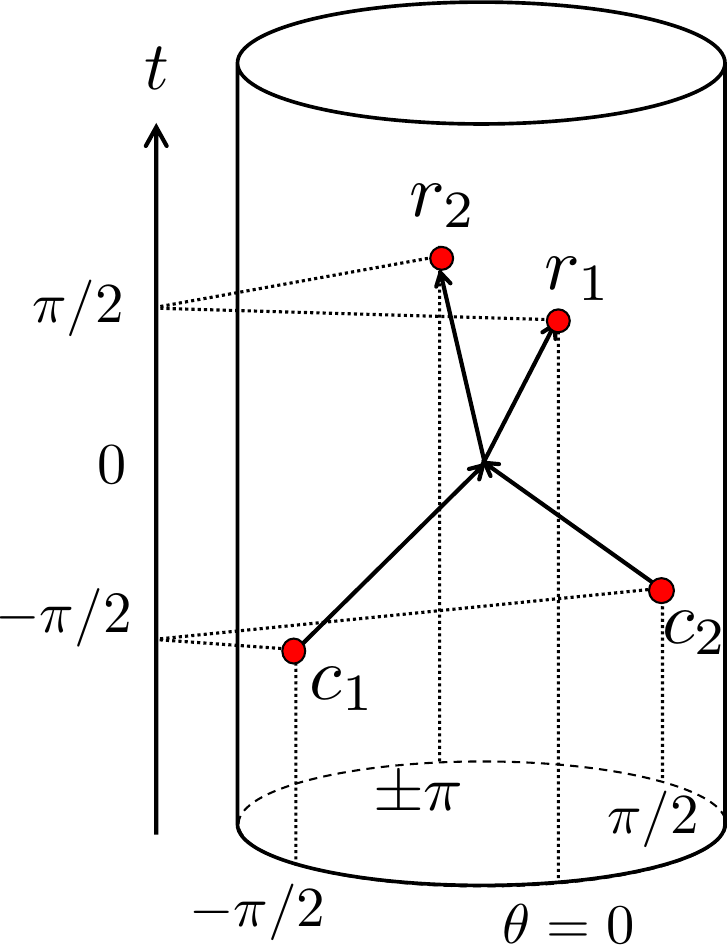}}
\hspace{10pt}
\raisebox{\height}{b)\hspace{10pt}}\raisebox{-0.85\height}{\includegraphics[width=0.3\textwidth]{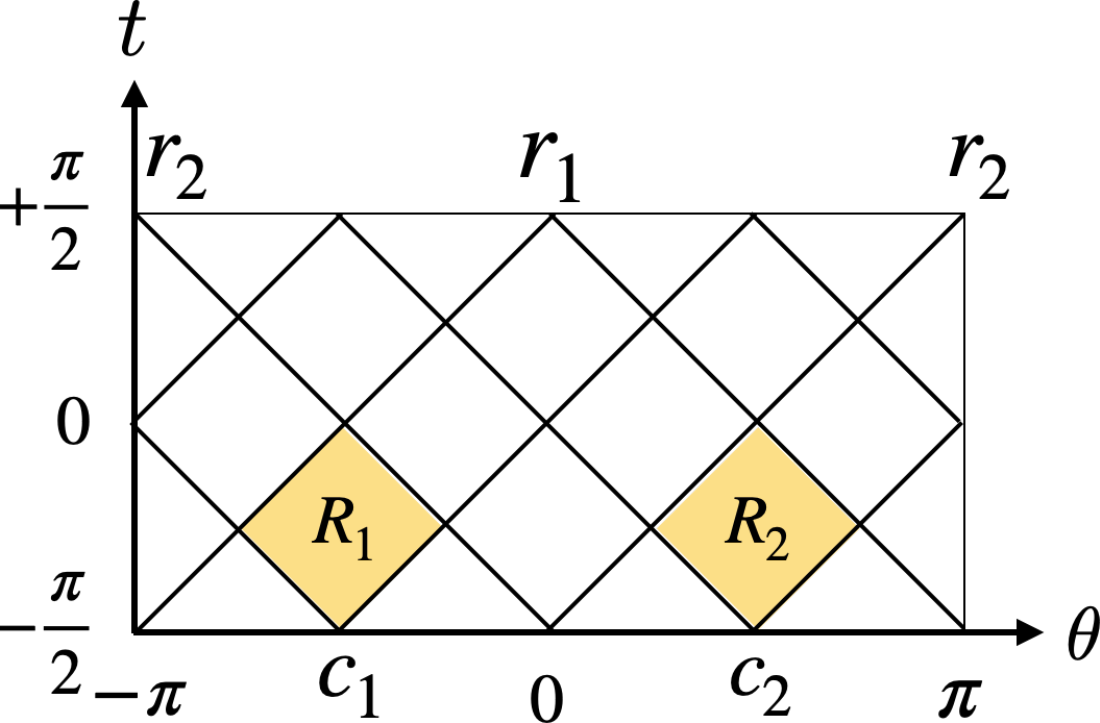}}
\caption{
a) Bulk scattering between two signals in AdS$_3$ that cannot be realized by direct boundary interaction due to causal separation.
b) Boundary causal structure. Interactions may be induced by utilizing pre-existing entanglement between $R_{1}$ and $R_{2}$ due to the connected wedge. 
}
\label{fig_holographic_scattering}
\end{figure}

Imagine two signals traveling to the center of pure AdS$_3$, interacting with each other, and then scattering back to the boundary (Fig.\ref{fig_holographic_scattering}(a)).
Specifically, we arrange the input points $c_1, c_2$ at $(\theta,t)=(- \pi/2,0),(\pi/2,0)$ and the output points $r_1, r_2$ at $(\theta,t)=(0,\pi/2),( \pi, \pi/2)$, respectively as shown in Fig.~\ref{fig_holographic_scattering}(b).
In this setup, while there is enough time for bulk scattering, there is not enough time for direct boundary interaction.
This raises the question: how does interaction between the two particles emerge on the boundary?

This apparent puzzle can be resolved by finding boundary spacetime regions 
accessible to each input point~\cite{May:2019yxi}. 
Specifically, let us define 
\begin{align}
R_{1} \equiv J_{+}(c_{1})  \cap J_{-}(r_{1}) \cap J_{-}(r_{2}), \qquad 
R_{2} \equiv J_{+}(c_{2})  \cap J_{-}(r_{1}) \cap J_{-}(r_{2})
\end{align}
where $J_{+}$ and $J_{-}$ represent the future and past light cones respectively. 
These regions can be signaled from one of the inputs and can influence both outputs.
The crucial observation is that $R_1$ and $R_2$ have a connected entanglement wedge, suggesting that interaction between the two particles may be mediated by pre-existing entanglement.
This idea has been strengthened by the ``connected wedge theorem''~\cite{May:2019odp, May:2022clu}, which shows that bulk scattering implies a connected entanglement wedge between $R_1$ and $R_2$.

In a series of works, May and collaborators proposed several quantum information protocols that model such interactions using pre-shared entanglement~\cite{May:2019yxi, May:2019odp, May:2022clu}.
Although these protocols differ in detail, they all resemble quantum teleportation, assuming access to clean EPR pairs, which could hypothetically be distilled from $R_1$ and $R_2$ soon after 
their wedge becomes connected.
(Note that, in this regime, $J^W(R_1:R_2) = 0$.)
Importantly, these protocols generally fail without access to clean EPR pairs, at least in their original formulations.
It remains unclear whether distillability of such EPR pairs is necessary for emergent interaction in holography.
If so, this would imply the existence of a subleading mechanism for LOCC entanglement distillation even when $J^W(A:C)=0$.

Nevertheless, such processes are expected to contribute only at subleading order.
This is because scattering with entropy $O(1/G_N)$ would induce significant geometric backreaction.
In particular, the connected wedge theorem has been explored in conical defect and BTZ black hole geometries~\cite{Caminiti:2024ctd}, where bulk scattering becomes more constrained due to backreaction.

\subsection{Planck-scale effect}

Finally, let us return to the original setup in Fig.~\ref{fig_connected}, where $A$ and $C$ are separated by $B$ in pure AdS$_{3}$. 
For large $B$, the minimal surfaces $\gamma_A, \gamma_C$ of $A,C$ are macroscopically separated on the AdS scale even if $A$ and $C$ have a connected wedge. 
In this regime, we proposed that $E_{D}^{[\text{LO}]}(A:C)\approx0$ at leading order. 
However, as $B$ becomes smaller and eventually vanishes, we will have $E_{D}^{[\text{LO}]}(A:C)\approx S_{A}$. 
This suggests that as $\gamma_A$ and $\gamma_{C}$ approach each other, subleading corrections to $E_{D}^{[\text{LO}]}$ begin to emerge and eventually become dominant.

We expect this subleading effect to become significant when $\gamma_A$ and $\gamma_{C}$ are separated by a Planck-scale distance. 
Specifically, in random tensor network models, the reduced state $\rho_{AA'}$ of the double-copy state can be approximated as a maximally mixed state supported on the minimal surface of $AA'$ or $BB'$, while subleading corrections to $\rho_{AA'}$ are exponentially suppressed in $1/G_N$.
When $\gamma_A$ and $\gamma_C$ are Planck-scale close, these subleading terms can become dominant, leading to distillable entanglement between $A$ and $C$.
This point is elaborated in Appendix~\ref{app:TN}.

\section{Instability of entanglement measures}\label{sec:instability}

In this section, we study the entanglement properties of isotropic states.
In particular, we demonstrate that certain entanglement measures, such as the logarithmic negativity $E_N$ and the exact entanglement cost $E_C^{[\mathrm{exact}]}$, are highly sensitive to exponentially small perturbations.
Our goal is to highlight this subtlety and caution against the use of these quantities in characterizing entanglement in many-body quantum systems.

\subsection{Instability of logarithmic negativity}

Recall that an isotropic state is defined by
\begin{align}
\rho_{AA'} = \frac{1-F}{{d_A}^2-1} (I - |\text{EPR}\rangle \langle \text{EPR} |)  +  F |\text{EPR}\rangle \langle \text{EPR} |,\quad 0\le F\le 1. 
\end{align}
As discussed in Section~\ref{sec:LO_Petz}, an isotropic state arises by applying the Petz map to a tripartite Haar random state $|\psi_{ABC}\rangle$ in the regime $n_A < n_B$, with EPR fidelity 
\begin{align} 
F \approx 2^{-\Delta} = \frac{d_C}{d_A d_B}. 
\end{align}
We focus on the regime $F \ll 1$, where the isotropic state is exponentially close to the maximally mixed state: 
\begin{align} 
\left\Vert\rho_{AA'} - \frac{1}{d_{A}^2}I_A\otimes I_{A'} \right\Vert_{1} \approx F. 
\end{align}

That $\rho_{AA'}$ is nearly disentangled can be also seen by computing its mutual information.
Since $S_A = S_{A'} = \log d_A$, and using a lower bound on $S_{AA'}$, \begin{align} 
S_{AA'} \geq (1-F) \left( 2 \log d_A - \frac{1}{d_A^2-1}\right), 
\end{align} 
we find 
\begin{align} I(A:A') \approx 2 n_A F \approx o(1), 
\end{align} 
which is exponentially small in $n$, consistent with $\rho_{AA'}$ being nearly decoupled.

Next, consider the logarithmic negativity, defined as 
\begin{align} 
E_N(A:A') \equiv \log \Big( \sum_j |\lambda_j| \Big), 
\end{align} 
where $\lambda_j$ are eigenvalues of $\rho_{AA'}^{T_{A'}}$. For small $F$, we find \begin{align} 
\rho_{AA'}^{T_{A'}} \approx \frac{1}{{d_A}^2} I + F \frac{1}{d_A} \text{SWAP}.
\end{align} 

Since the SWAP operator has eigenvalues $\pm1$ corresponding to symmetric and antisymmetric sectors, the eigenvalues of $\rho_{AA'}^{T_{A'}}$ are approximately: \begin{align} 
\text{symmetric:} \quad \frac{1}{d_A^2}\left(1 + F d_A\right), \qquad \text{antisymmetric:} \quad \frac{1}{d_A^2}\left(1 - F d_A\right). 
\end{align} 
In the regime $F d_A \gg 1$ (equivalently, $n_C > n_B$ in the original tripartite state), these eigenvalues can be approximated as $\pm \frac{F}{d_A}$.
Thus, $\sum_j |\lambda_j| \approx F d_A$, and we obtain 
\begin{align} 
E_N(A:A') \approx \max\left( n_A + \log_2 F, 0 \right) \approx \max\left( n_C -  n_B, 0 \right) .
\end{align} 
Here $E_N$ can be extensive, $O(n)$, even though $\rho_{AA'}$ is almost separable with exponentially small $F$.

A related instability appears in R\'{e}nyi entropies.
The R\'{e}nyi entropy $S_{AA'}^{(m)}$ deviates significantly from the von Neumann entropy $S_{AA'}^{(1)}$, as shown in Appendix~\ref{app:Haar}: 
\begin{align} 
S_{AA'}^{(m)} = \frac{1}{1-m}\Tr(\rho_{AA'}^m) \approx \frac{1}{m-1} \min( m\Delta, (m-1)2n_A), 
\end{align} 
where $\Delta = n_A + n_B - n_C$.
In particular, for $m=2$, 
\begin{align} S_{AA'}^{(2)} = 2(n_A + n_B - n_C) \qquad \text{(for $n_C>n_B$)}, \end{align} 
leading to a (na\"{i}ve un-sandwitched) R\'{e}nyi-2 mutual information \begin{align} 
\frac{1}{2}I^{(2)}(A:A') \approx E_{N}(A:A') \approx n_C - n_B. 
\end{align}
The match between $I^{(2)}(A:A')$ and $E_N(A:A')$ is not accidental: as emphasized in~\cite{Dong:2021clv, Dong:2024gud}, the computation of $E_N$ is primarily controlled by the even R\'{e}nyi entropies.

We have seen that the logarithmic negativity $E_N$ is highly sensitive to exponentially small perturbations of the maximally mixed state.
This instability suggests that $E_N$ may not be a reliable measure of mixed-state entanglement, especially in many-body quantum systems.\footnote{
Another possible perspective is that a statistical mixture of two distinct quantum phases is intrinsically unstable as the mixing parameter $F$ may not be continuously changed by geometrically local operations. 
We thank Ryohei Kobayashi for discussion on this. 
}

\subsection{Exact vs. approximate entanglement cost}

We now turn to an important distinction between exact and approximate entanglement cost.
In particular, we will see that there exists a regime in an isotropic state where $E_C^{[\mathrm{exact}]} \sim O(n)$ while $E_C \sim o(1)$.

Recall that $E_C^{[\mathrm{exact}]}$ is defined as the number of EPR pairs per copy required to \emph{exactly} create the target state before taking the asymptotic limit, whereas $E_C$ only requires \emph{approximate} creation with asymptotically vanishing error.
Since $E_C^{[\mathrm{exact}]} \geq E_N$, we immediately obtain \begin{align} E_C^{[\mathrm{exact}]} \sim O(n) \qquad (Fd_A \gg 1). \end{align}
In contrast, recall that analytical expressions for $E_F(A:A')$ were obtained in~\cite{terhal2000entanglement, wang2016entanglement}:
\[
E_{F}(A:A') =
\begin{cases} 
0, &  F \in [0, 1/d_A ], \\
-(1-F) \frac{d_A}{d_A-2}\log (d_A-1) + \log d_A, &  F \in \left[\frac{4(d_A-1)}{d_A^2}, 1 \right],
\end{cases}
\]
where $F = \frac{d_C}{d_A d_B}$.
For small $F$, we find 
\begin{align} 
E_F(A:A') \lesssim F n_A \sim o(1). \label{eq:EF_bound} 
\end{align} 
Since $E_C$ is defined as the asymptotic limit of $E_F$, we have 
\begin{align} 
E_C(A:A') \leq E_F(A:A') \sim o(1), \label{eq:LN-iso} 
\end{align} 
revealing a dramatic separation between $E_C^{[\mathrm{exact}]}$ and $E_C$.

Moreover, we can explicitly verify that 
\begin{align} 
E_C^{[\mathrm{exact}]} \approx E_{N} \approx n_C - n_B. \label{eq:EC_exact} 
\end{align} 
To see this, it suffices to construct an exact preparation of $\rho_{AA'}$ using $n_C - n_B$ EPR pairs.
The construction proceeds as follows.
Prepare the initial state on $AA'$: 
\begin{align} |0\rangle^{\otimes (n_A + n_B - n_C)} \otimes |\text{EPR}\rangle^{\otimes (n_C - n_B)} \otimes |0\rangle^{\otimes (n_A + n_B - n_C)}, 
\end{align} 
where $n_C - n_B$ EPR pairs are shared between $A$ and $A'$.
Applying the two-fold Haar twirl (see Eq.\eqref{eq:twirl}) symmetrizes this state, effectively depolarizing everything orthogonal to $|\text{EPR}\rangle_{AA'}$, and yields $\rho_{AA'}$.
This establishes Eq.\eqref{eq:EC_exact}.

On the other hand, the \emph{approximate} entanglement cost $E_C$ can be much smaller, because it only requires approximate preparation with asymptotically vanishing error.
Specifically, since $\rho_{AA'}$ contains $|\text{EPR}\rangle_{AA'}$ with probability $F$, it suffices to prepare $\sim m F$ copies of $|\text{EPR}\rangle_{AA'}$ and randomly distribute them across $m$ instances.
Note that this procedure resembles Shannon's source coding applied to EPR pairs.
This method approximates $\rho_{AA'}^{\otimes m}$ with vanishing error as $m \to \infty$, by the law of large numbers.
Thus, the approximate scheme requires only $\sim n_A F$ EPR pairs per copy, leading to the bound\footnote{While the exact value of $E_C$ for isotropic states is not known, it suffices here to establish an upper bound.} 
\begin{align} 
E_C \lesssim F n_A. \label{eq:EC_bound} 
\end{align} 
This also follows directly from Eq.\eqref{eq:EF_bound}, given that $E_C \leq E_F$ by definition.





\section{Outlook}\label{sec:outlook}

Our main results can be summarized as follows
\begin{align}
&E_{D}^{[\text{LO}]}(A:C) \approx 0  \qquad (\text{when $\gamma_A,\gamma_C$ do not overlap}) \label{eq:result_LO} \\
&E_{D}^{[\text{1WAY LOCC}]}(A\leftarrow C) \approx J^W(A|C) \label{eq:result_LOCC}
\end{align}
at leading order in $1/G_N$ where $J^W(A|C) \equiv S_{A} - E^W(A:B)$.
Also, we argued 
\begin{align}
E_F(A:C)\approx E^W(A:C), \qquad J(A|C) \approx J^W(A|C).  
\end{align}
Furthermore, we obtained the Koashi-Winter monogamy relation in the holographic context: 
\begin{align}
E_{D}^{[\text{1WAY LOCC}]}(A\leftarrow C) \approx S_{A} - E_{F}(A:B)
\end{align}
The hierarchy of holographic entanglement measures is summarized below: 
\begin{align}
\underbrace{E_{D}^{[\text{LO}]}}_{\approx 0} \leq \text{hash}(A:C)\leq \underbrace{E_{D}^{[\text{1WAY LOCC}]}}_{\approx J^W(A:C)} \leq \underbrace{E_{sq}}_{\approx \frac{1}{2}I(A:C)} \leq \underbrace{E_{F}, E_{P}}_{\approx E^{W}(A:C)} \leq \min(S_A,S_C).
\end{align}

For Haar random states, these results were obtained rigorously at leading order in $n$ via counting arguments and concentration of measure. 
For holography, we proved a weaker bound on $E_{D}^{[\text{LO}]}(A:C)$ using the Petz map.
As for $E_{D}^{[\text{1WAY LOCC}]}(A:C)$, this was shown from the holographic proposal $E_{F}\approx E^W$.
Furthermore, we presented an optimal LOCC protocol that implements holographic measurements placing EoW brane-like objects.

There are several key limitations in our work. 
First, we have mostly focused on one-shot settings for $E_{D}^{[\text{LO}]}$ and one-shot 1WAY settings for $E_{D}^{[\text{1WAY LOCC}]}$. 
Namely, whether our result applies to 2WAY LOCC scenarios remains open. 
Second, much of our studies focused on random tensor networks and fixed-area states, ignoring subleading fluctuations of area operators.
This limitation could potentially be overcome by constructing an explicit distilling operation in the dual conformal field theories, taking into account their nontrivial entanglement spectrum and subleading corrections.
Third, our proposal on LOCC-distillable entanglement crucially relies on the assumption concerning the existence of disentangled basis states.  
We hope to address these limitations in the future work.


Even though a connected wedge does not guarantee distillable EPR pairs under our proposal, it \emph{does not} imply any discontinuity of the spacetime within the entanglement wedge.
Rather, we expect that a connected wedge corresponds to a smooth spacetime region where subleading matter fields can move and interact.
Indeed, we proposed that traversable wormholes and holographic scattering are possible manifestations emerging from such connectivity.
It would be interesting to verify these ideas through quantitative analysis.

One promising future direction is the application of our analysis on $E_{D}^{[\text{LO}]}$ to many-body quantum systems, such as random unitary quantum circuits and related toy models of scrambling dynamics.
Namely, it will be interesting to study how $E_{D}^{[\text{LO}]}$ changes by introducing $T$ gates to random Clifford circuits or by adding interactions to otherwise free Hamiltonian dynamics. 
Furthermore, our proposals on $E_{D}^{[\text{1WAY LOCC}]}$ may shed light on the entanglement structure in monitored (hybrid) quantum circuits and their connection to black hole physics, where local projective measurements can be incorporated into many-body and gravitational dynamics~\cite{Yoshida:2022srg}.

\subsection*{Acknowledgment}

We thank Hideo Furugori, Jonah Kudler-Flam, Ryohei Kobayashi, Zhi Li, Alex May, Yoshifumi Nakata, Tadashi Takayanagi, Kotaro Tamaoka, and Shreya Vardhan for helpful discussions.
Research at Perimeter Institute is supported in part by the Government of Canada through the Department of Innovation, Science and Economic Development and by the Province of Ontario through the Ministry of Colleges and Universities. This work was supported by JSPS KAKENHI Grant Number 23KJ1154, 24K17047.

\appendix

\addcontentsline{toc}{section}{Appendix}

\section{Entanglement wedge cross section for AdS${}_3$/CFT${}_{2}$}
\label{app:hol-calc}

\begin{figure}
    \centering
    \includegraphics[width=0.27\linewidth]{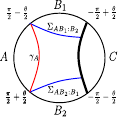}
    \caption{Two candidate configurations for $E^W(A:B)$.}
    \label{fig:vacAdS-ewcs}
\end{figure}

In this appendix, we present calculations of $E^W(A:C)$ and $E^W(A:B)$ in pure AdS$_3$. 
For simplicity, we set the AdS radius to unity and the circumference to $2\pi$. Let us consider a symmetric configuration on the boundary $S^1$ where $A$ and $C$ are disjoint intervals.
Namely, we take $A:[\frac{\pi}{2}-\frac{\theta}{2},\frac{\pi}{2}+\frac{\theta}{2}]$, $C:[-\frac{\pi}{2}+\frac{\theta}{2},-\frac{\pi}{2}-\frac{\theta}{2}]$ and $B$ to be their complement. 
We will call two symmetric complementary subsystems as $B_1$ and $B_2$ as shown in Fig.~\ref{fig:vacAdS-ewcs}.

We begin by computing $E^W(A:C)$.
When the entanglement wedge of $AC$ is connected (i.e., $\theta > \frac{\pi}{2}$), the area of the cross section $\Sigma_{A:C}$ is given by a function of the conformal cross ratio $z$~\cite{Takayanagi:2017knl}:
\begin{align}
\text{Area}(\Sigma_{A:C}) = \text{Cross}(z_{A:C}), 
\end{align}
where
\begin{equation}
    \text{Cross}(z) \equiv \log(1+2z+2\sqrt{z(z+1)}). 
\end{equation}
When $z\gg 1$, $\text{Cross}(z)\rightarrow \frac{1}{4G_N}\log(4z)$.
The cross ratio $z_{A:C}$ is given by 
\begin{equation}
z_{A:C}=\frac{\displaystyle\sin\frac{\abs{A}}{2} \sin\frac{\abs{C}}{2}}{\displaystyle\sin\frac{\abs{B_1}}{2} \sin\frac{\abs{B_2}}{2}}=\tan^2\frac{\theta}{2}.
\end{equation}
Hence, we have 
\begin{equation}
\boxed{  \
\begin{split}
    E^W(A:C) = 
    \begin{cases}
        \displaystyle\frac{1}{4G_N} \log\qty(1+2\tan^2\frac{\theta}{2} + 2\frac{\tan\frac{\theta}{2}}{\cos\frac{\theta}{2}}) \qquad \qquad &\frac{\pi}{2}\le \theta \\
        0  \qquad \qquad   &\theta \le \frac{\pi}{2}. 
    \end{cases} 
\end{split} \
}
\end{equation}

Next, we compute  $E^W(A:B)$. 
As shown in Fig.~\ref{fig:vacAdS-ewcs}, there are two candidate surfaces for the minimal cross section.
The first is given by the minimal surface $\gamma_A$:
\begin{align}
\text{Area}(\gamma_A) = 2\log\frac{2\sin\frac{\theta}{2}}{\epsilon}
\end{align}
where $\epsilon$ is a UV cutoff. 
The second has two contributions:
\begin{align}
\text{Area}(\Sigma_{AB_2:B_1}) + \text{Area}(\Sigma_{AB_1:B_2}).
\end{align}
The contribution from $\Sigma_{AB_2:B_1}$ is
\begin{align}
\text{Area}(\Sigma_{AB_2:B_1}) = \text{Cross}(z_{AB_2:B_1}),
\end{align}
where the cross ratio is
\begin{align}
z_{AB_2:B_1} 
=\frac{\displaystyle\sin\frac{\abs{AB_1}}{2} \sin\frac{\abs{B_2}}{2}}{\displaystyle \sin\frac{\abs{C}}{2} \sin{\epsilon}} \approx \frac{\cot\theta}{\epsilon}\gg 1.
\end{align}
Here we use the same cutoff $\epsilon$ as for $\gamma_A$ so that the area difference remains finite:
\begin{align}
\text{Area}(\gamma_A) - (\text{Area}(\Sigma_{AB_2:B_1}) + \text{Area}(\Sigma_{AB_1:B_2})).
\end{align}
When the second candidate surface dominates, we have:
\begin{align}
E^W(A:B) = \frac{2}{4G_N} \log ( \frac{4 \cot\theta}{\epsilon} ).
\end{align}
and thus
\begin{equation}
\boxed{  \
\begin{split}
J^W(A:C) \equiv S_A - E^W(A:B)  =
\begin{cases}
    \displaystyle\frac{1}{2G_N} \log\frac{\sin\frac{\theta}{2}\tan\frac{\theta}{2}}{2}, \qquad &\theta\ge \theta_\ast \\
    0 \qquad  &\theta\le \theta_\ast
\end{cases}
\end{split}
\ } 
\end{equation}
which is UV finite. 
Finally, the critical value $\theta=\theta_\ast$ is given by
\begin{equation}
    \text{Area}(\gamma_A)=\text{Area}(\Sigma_{A:B}) 
    \Rightarrow \theta_\ast = 4 \mathrm{arctan}\sqrt{\sqrt{2}-1}\approx 2.287.
\end{equation}

\begin{figure}
    \centering
    \includegraphics[width=1\linewidth]{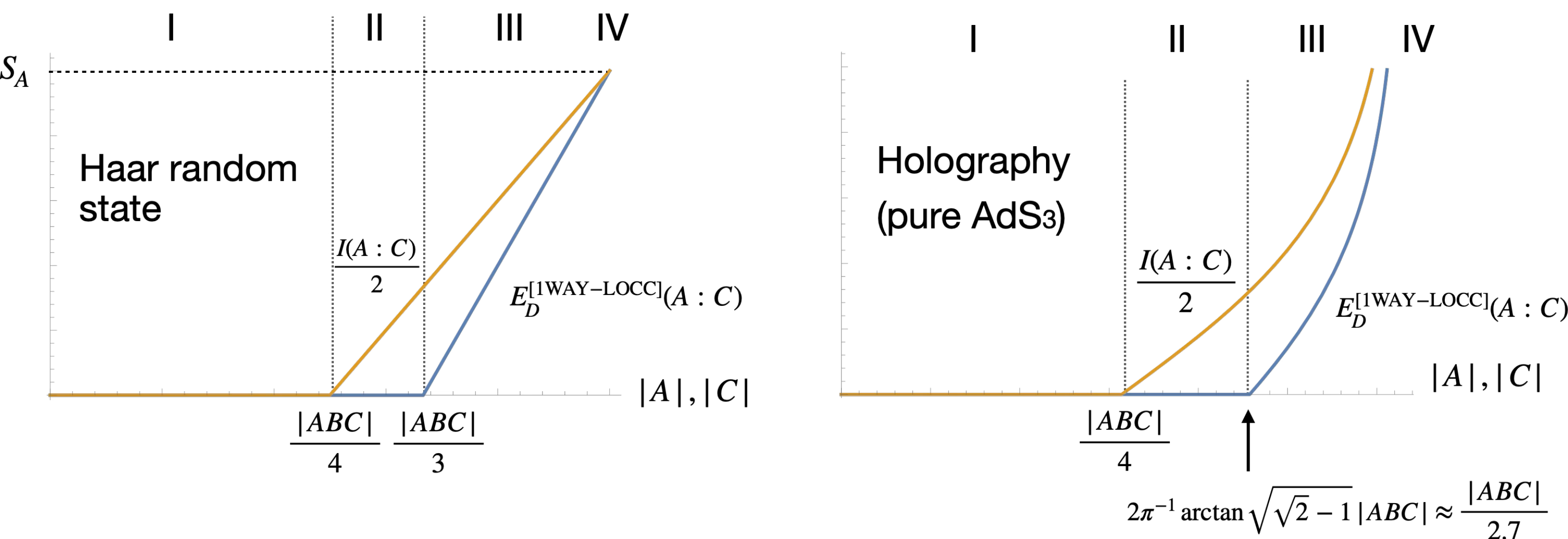}
    \caption{The leading-order behavior of $I(A:C)/2$ and $E_D^{[\text{1WAY-LOCC}]}(A:C)$ of a Haar random state and the holographic vacuum state dual to pure AdS$_3$. $\abs{A}=n_A,\abs{C}=n_C$ for the Haar random state and $\abs{A}=\abs{C}=\theta,\abs{ABC}=2\pi$ for the vacuum state.}
    \label{fig_claim2}
\end{figure}

The result is summarized in Fig.~\ref{fig_claim2}. A leading-order result for Haar random states with $n_A=n_C$ is also presented for a comparison.

\section{Entanglement wedge transitions in a planar BTZ black hole}\label{app:metric}

In this appendix, we compute the transition times $t_1$ and $t_2$ associated with $I(A:C)$ and $J^W(A|C)$ in the BTZ black hole (see Fig.~\ref{fig_traversable_2}). 
These transitions were discussed in the context of traversable wormholes in Section\ref{sec:traversable}.

Consider the thermofield double (TFD) state of a $(1+1)$-dimensional holographic CFT at the inverse temperature $\beta$. 
The subsystems $A$ and $C$, located on the left and right boundaries respectively, are taken to be symmetric and of equal size, growing linearly with one-sided time: $\abs{A}=\abs{C}=2\tau$ after time $\tau$.
In the case of the AdS$_3$/CFT$_2$, the gravity dual of the TFD state is described by the BTZ black hole. When the boundary topology is $\mathbb{R}^{1,1}$, the dual bulk metric is given by the planar BTZ black hole,
\begin{equation}
    ds^2=\frac{1}{z^2}\qty(-f(z)dt^2+dx^2+\frac{dz^2}{f(z)}),\quad f(z)=1-\frac{z^2}{z_H^2},
    \label{eq:btz-planar-metric}
\end{equation}
where we have set the AdS radius to unity. The horizon radius is related to the inverse temperature by $z_H=\frac{\beta}{2\pi}$.

To compute geodesic lengths in this geometry, we introduce embedding coordinates $(X_0, X_1, X_2, X_3)$:
\begin{equation}
    \begin{split}
        X_0 & = \frac{z_H}{z}\cosh\frac{x}{z_H},\\
        X_3 & = \sqrt{\qty(\frac{z_H}{z})^2-1}\sinh\frac{t}{z_H},\\
        X_1 & = \frac{z_H}{z}\sinh\frac{x}{z_H},\\
        X_2 & = \sqrt{\qty(\frac{z_H}{z})^2-1}\cosh\frac{t}{z_H}.
    \end{split}
\end{equation}
Plugging them into
\begin{equation}
    ds^2= -dX_0^2-dX_3^2+dX_1^2+dX_2^2, 
\end{equation}
one recovers the the BTZ metric (Eq.~\eqref{eq:btz-planar-metric}). In these coordinates, the geodesic distance between two bulk points $X_A$ and $X_B$ is given by
\begin{equation}
    \mathrm{arccosh}\qty(-X_A\cdot X_B),
    \label{eq:geo-emb}
\end{equation}
where the inner product is defined as $X\cdot Y = -X_0 Y_0 - X_3 Y_3 + X_1 Y_1 + X_2 Y_2$.

\begin{figure}
    \centering
    \includegraphics[width=0.3\linewidth]{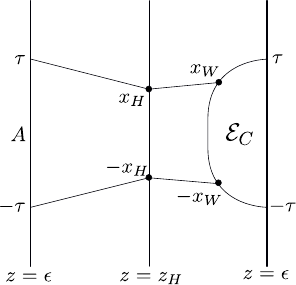}
    \caption{The cross-section $\gamma_{A:B}$ before minimization. $x_H,x_W$ are the $x$ coordinates on the horizon $z=z_H$ or the edge of the entanglement wedge of $C$, respectively.}
    \label{fig:WH-geo}
\end{figure}

To compute $J^W(A|C) \equiv S_A - E^W(A:B)$, we evaluate two candidate surfaces. 
The first candidate $\frac{\text{Area}(\gamma_A)}{4G_N}$ corresponds to the minimal surface of $A$. 
The second candidate $\frac{\text{Area}(\gamma_{A:B})}{4G_N}$ corresponds to the cross section with respect to $A,B$ where $B=(AC)^c$. 
When the subsystem size is $2\tau$, the lengths of two candidate geodesics are
\begin{align}
    \text{Area}(\gamma_A(\tau)) &= 2\log\frac{z_H}{\epsilon} + \log\qty(2\cosh\frac{2\tau}{z_H}-2),\\
    \text{Area}(\gamma_{A:B}(\tau)) &= \min_{x_H,x_W}\qty[2\log\qty(\frac{2z_H}{\epsilon}\cosh\frac{\tau-x_H}{z_H})+2\mathrm{arccosh}\qty(\frac{\cosh\frac{\tau}{z_H}}{\sqrt{\cosh^2\frac{\tau}{z_H}-\cosh^2\frac{x_W}{z_H}}} \cosh\frac{x_H-x_W}{z_H})],
\end{align}
where $\epsilon (\ll 1)$ is the UV cutoff. 
The second expression involves a minimization over $x_H$ and $x_W$, which are the $x$ coordinates at the intersection of the geodesic with the horizon or the edge of the complement entanglement wedge respectively, as illustrated in Fig.~\ref{fig:WH-geo}.
Hence, 
\begin{equation}
    \begin{split}
        J^W(A|C)&= \frac{1}{4G_N}\big[\text{Area}(\gamma_A(\tau)) - \min[\text{Area}(\gamma_A(\tau)),\text{Area}(\gamma_{A:B}(\tau))]\big]\\
        &=\frac{1}{4G_N}\max\qty[0,\text{Area}(\gamma_A(\tau))-\text{Area}(\gamma_{A:B}(\tau))].
    \end{split}
\end{equation}
In the early time, $\text{Area}(\gamma_A(\tau))<\text{Area}(\gamma_{A:B}(\tau))$ and in the late time $\text{Area}(\gamma_A(\tau))>\text{Area}(\gamma_{A:B}(\tau))$. The transition happens when $\tau=t_2$ such that $\text{Area}(\gamma_A(t_2))=\text{Area}(\gamma_{A:B}(t_2))$. Since $\tau$ only appears with $z_H$, the transition time is $\tau \sim z_H$ (multiplied by the AdS radius). A numerical solution gives 
\begin{equation}
    t_2\approx 0.21 \beta.
\end{equation}

As for $I(A:C)$, the length of the geodesics connecting $A$ and $C$ is 
\begin{equation}
    \text{Area}(\gamma_{AC}(\tau))=4\int_\epsilon^{z_H} \frac{1}{z}\qty(1-\frac{z^2}{z_H^2})^{-1/2}
    =4\log\qty(\frac{2z_H}{\epsilon}).
    \label{eq:dis-MI-WH}
\end{equation}
Note that this does not depend on $\tau$. The entanglement wedge between $A$ and $C$ has a transition at time $\tau=t_1$ such that $2\text{Area}(\gamma_A (t_1))=\text{Area}(\gamma_{AC})$. This again happens of order $\beta$ because
\begin{equation}
    \cosh\frac{2t_1}{z_H}=\frac{3}{2}\Leftrightarrow t_1 =\frac{z_H}{2}\mathrm{arccosh}\frac{3}{2}\approx 0.08\beta.
\end{equation}

\section{Haar random double-copy state}\label{app:Haar}

In this appendix, we study the entanglement properties of the double-copy state constructed from a Haar random state.
Consider an $n$-qubit Haar random state $|\psi\rangle$ with a tripartition into $A,B,C$ such that $n_R < \frac{n}{2}$ for each $R=A,B,C$. 
Viewing $\rho_{AC}$ as a quantum channel from $A$ to $C$ and applying the Petz recovery map on $C$, we obtain the double-copy state:
\begin{align}
|\Phi^{(\text{double})}_{ABA'B'}\rangle \approx \sqrt{d_C} \cdot \figbox{1.7}{fig_Haar_double} \ . 
\end{align}
Our main claim is that
\begin{align}
S_{AA'} \approx \min\qty(\figbox{0.17}{fig_SASAdouble-Haar}\, , \figbox{0.17}{fig_SAAdouble-Haar}) = 2 \min (n_A, n_B)
\end{align}
which implies $\frac{1}{2}I(A:A') \approx \max (0, n_A - n_B)$.
We also find that $\rho_{AA'}$ is given by
\begin{align}
\rho_{AA'} \approx 2^{-\Delta}|\text{EPR}\rangle \langle \text{EPR}|_{AA'}  + (1-2^{-\Delta})\mu_{\text{max}}, \qquad \Delta = n_A + n_B - n_C > 0
 \label{eq:mixture}
\end{align}
where $\mu_{\text{max}}$ denotes a maximally mixed state on $AA'$ for $n_{A} < n_{B}$, and a maximally mixed state on a $d_{B}^2$-dimensional subspace of $AA'$ for $n_{B} < n_{A}$.  

\subsection{Entanglement in double-copy state}

Let us derive Eq.~\eqref{eq:mixture} by evaluating $\Tr\big[ (\rho_{AA'})^m \big]$ using standard Haar calculus (see~\cite{Hayden:2016cfa} for a review). 
We can express $\Tr\big[ (\rho_{AA'})^m \big]$ as 
\begin{align}
\Tr\big[ (\rho_{AA'})^m \big] \approx \ \frac{1}{{d_C}^m} \figbox{0.25}{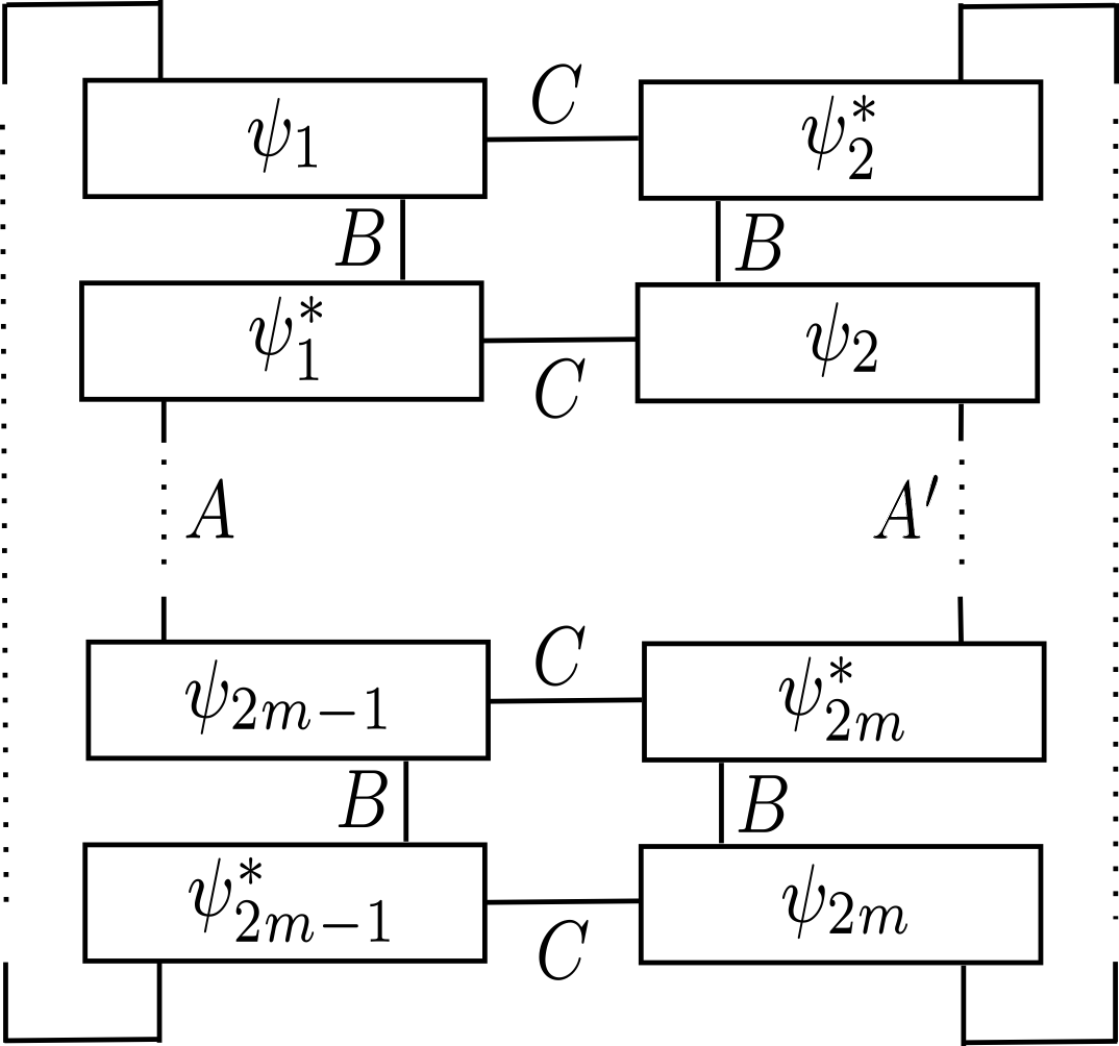}
\end{align}
Recall that the $m$-fold random projector can be written as a uniform sum over permutation operators:
\begin{align}
\int d\psi \ |\psi\rangle \langle \psi|^{\otimes m} \propto \sum_{V \in S_{m}} V.
\end{align}
Since $|\Phi^{(\text{double})}_{ABA'B'}\rangle$ involves both $|\psi\rangle$ and $|\psi^*\rangle$, the calculation requires $2m$ copies of $|\psi\rangle$, involving the $S_{2m}$ permutation group. 
The boundary conditions on $AA',BB',CC'$ are given by
\begin{align}
\Tr\big[ (\rho_{AA'})^m \big] : \ \figbox{1.7}{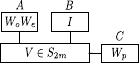}.
\end{align}
Here, $W_{o}$ is a cyclic permutation on $(1,3,\cdots, 2m-1)$, $W_{e}$ is a reverse cyclic permutation on $(2,4,\cdots, 2m)^{-1}$, and $W_{p}$ is the pairwise swap, namely $W_{p}=(1,2)(3,4)\cdots (2m-1,2m)$. 

The dominant contribution comes from the choice of $V\in S_{2m}$ that minimizes the number of cycles (equivalently, the ``energy'' of the $S_{2m}$ ferromagnetic Hamiltonian). 
Among all possible $V$, the following three permutations are candidates for the leading contributions, $V = W_{o}W_{e}, I, W_{p}$:
\begin{align}
&\figbox{1.7}{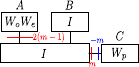}  &\figbox{1.7}{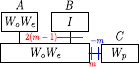} \qquad  &\figbox{1.7}{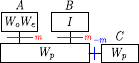} \\
&F_{\text{RT-$A$}}^{(m)}  = 2(m-1)n_A 
&F_{\text{RT-$B$}}^{(m)}  = 2(m-1)n_B \qquad \quad
&F_{\text{tri}}^{(m)}  = m(n_A+n_B - n_C). \notag
\end{align}
We have subtracted $mn_{c}$ from each expression to account for the normalization of $|\Phi^{(\text{double})}_{ABA^\prime B^\prime}\rangle$.
Red cuts represent contributions with respect to the specified boundary conditions, while blue cuts subtract the normalization.
Hence, we obtain
\begin{align}
\Tr\big[ (\rho_{AA'})^m \big] = \frac{1}{2^{2(m-1)n_{A}}} + \frac{1}{2^{2(m-1)n_{B}}} + \frac{1}{2^{m(n_A+n_{B}-n_C)}} + \cdots.
\end{align}
The first two terms correspond to the RT contributions (minimal surfaces homologous to $AA'$), while the third is a tripartite term.  
In the regime of $n_A,n_B,n_C < \frac{n}{2}$, the third term dominates for large $m$:
\begin{align}
- \log \Tr\big[ (\rho_{AA'})^m \big] \approx m(n_A+n_{B}-n_C) \qquad \mbox{for large $m$}. 
\end{align}
Hence, 
\begin{align}
S_{AA'}^{(m)} \approx n_A+n_{B}-n_C  \qquad \mbox{for large $m$}, 
\end{align}
exhibiting a leading-order deviation from the RT formula for R\'{e}nyi entropy at large $m$. 

Now consider the von Neumann entropy $S_{AA'}$. 
Using the R\'{e}nyi-$m$ definition:
\begin{align}
S_{AA'}^{(m)} = - \frac{1}{m-1} \log \Tr\big[ (\rho_{AA'})^m \big]
\end{align}
we examine the following contributions as $m\rightarrow 1$: 
\begin{align}
\lim_{m\rightarrow 1} \frac{F_{\text{RT-$A$}}^{(m)}}{m-1}, \frac{F_{\text{RT-$B$}}^{(m)}}{m-1},\frac{F_{\text{tri}}^{(m)}}{m-1}.
\end{align}
Because the tripartite term diverges like $1/(m-1)$, it becomes subleading near $m=1$, so the RT contributions dominate:
\begin{align}
S_{AA'} \approx \min\qty(\figbox{0.17}{fig_SASAdouble-Haar}\, , \figbox{0.17}{fig_SAAdouble-Haar}) = 2 \min (n_A, n_B).
\end{align}

We now infer the spectrum of $\rho_{AA'}$ from these moments.
By treating $\log \Tr\big[ (\rho_{AA'})^m \big]$ as a moment-generating function and applying inverse Laplace transformation, the linear dependence on $m$ suggests a delta peak in the spectrum while the $(m-1)$ scaling suggests a flat spectrum.
This leads to the ansatz:
\begin{align}
\rho_{AA'} \approx 2^{- \Delta} |\Psi_{\text{max}}\rangle \langle \Psi_{\text{max}}| + (1- 2^{-\Delta}) \sigma, \qquad \Delta = n_{A} + n_{B} - n_C
\label{eq:isotropic}
\end{align}
with $|\Psi_{\text{max}}\rangle$ a pure state and $\sigma$ a mixed state.
One can further deduce the structure of $\sigma$ using the next-to-leading contribution.
When $n_A < n_B$, we have
\begin{align}
\log \Tr\big[ (\rho_{AA'})^m \big] \approx m(n_A+n_B - n_C) + 2(m-1)n_A
\end{align}
indicating that $\sigma$ is the maximally mixed state on $AA'$.
On the contrary, if $n_B < n_A$,
\begin{align}
\log \Tr\big[ (\rho_{AA'})^m \big] \approx m(n_A+n_B - n_C) + 2(m-1)n_B,
\end{align}
so $\sigma$ is a maximally mixed state over a $2^{2n_B}$-dimensional subspace of $AA'$.

The remaining task is to determine the form of $|\Psi_{\text{max}}\rangle$.
Our claim that $|\Psi_{\text{max}}\rangle \approx |\text{EPR}\rangle_{AA'}$ is supported by evaluating the overlap
\begin{align}
\Tr\big[ (\rho_{AA'})^m|\text{EPR}\rangle\langle\text{EPR}| \big] \approx 2^{-m \Delta}. \label{eq:fidelity_calculation}
\end{align} 
The boundary conditions required for this calculation are given by 
\begin{align}
\Tr\big[ \rho_{AA'}^m|\text{EPR}\rangle\langle\text{EPR}| \big]  : \figbox{1.7}{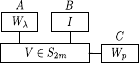}
\end{align}
with an additional contribution of $- m n_C - n_A$ from normalizations. 
(Here, the $-n_A$ term comes from the normalization of the EPR projector.)
The permutation $W_\lambda$ on $A$ is a cyclic permutation, as illustrated below:
\begin{align}
\Tr\big[ (\rho_{AA'})^m \dyad{\text{EPR}}\big]
\approx \ \frac{1}{{d_C}^m d_A} \figbox{1.7}{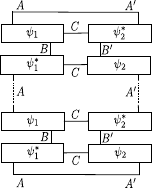}
\end{align}
Several key contributions arise from different choices of $V$ in $S_{2m}$:
\begin{align}
&\figbox{1.7}{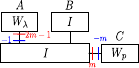} 
&\figbox{1.7}{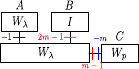} \\
&2(m-1)n_A  & -n_A + (2m-1)n_B - n_C \notag  \\ 
&\figbox{1.7}{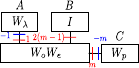} 
&\figbox{1.7}{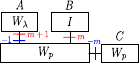}  \\
& 2(m-1)n_B 
& m(n_A + n_B - n_C). \notag
\end{align}
Among these, the contribution from $V = W_p$ dominates in the large $m$ limit, confirming Eq.~\eqref{eq:fidelity_calculation}. 

\subsection{Entanglement in post-measurement state}

Next, we evaluate the entropy drop $\Delta S_{A}$ resulting from a measurement on $A'$ in the double-copy state. 
Namely, we demonstrate that our proposal $E_F(A:C) \approx E^W(A:C)$ can be violated for R\'{e}nyi-$m$ entropies for $m\geq 2$. 

Letting $\rho_{A}^{\text{after}}$ denote the post-measurement state on $A$, we begin by computing $\Tr\big[(\rho_{A}^{\text{after}})^m\big]$.
Recall that the $m$-fold random projector can be written as a uniform sum over permutation operators. 
This suggests that, in the Haar calculus, random projections impose open boundary conditions. 
As a result, the boundary conditions for computing $\Tr\ [(\rho_{A}^{\text{after}})^m]$ are given by
\begin{align}
\Tr\ [(\rho_{A}^{\text{after}})^m] \ : \ \figbox{1.7}{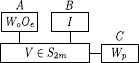}
\end{align}
where an open boundary condition $O_e$ is imposed at $A'$ (corresponding to even copies), allowing an arbitrary permutation element that minimizes the domain wall energy. 

The key contributions are
\begin{align}
\figbox{1.7}{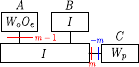} 
\qquad \figbox{1.7}{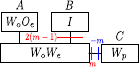} 
\qquad \figbox{1.7}{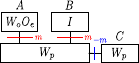}\ \ \ \\
F_{\text{RT-$A$}}^{(m)}  = (m-1)n_A  
\qquad F_{\text{RT-$B$}}^{(m)}  = 2(m-1)n_B
\ \ \qquad F_{\text{tri}}^{(m)}  = m(n_A+n_B - n_C) \notag
\end{align}
The first two correspond to RT-like contributions while the third arises from a tripartite configuration. 
When $n_C > n_B$, the tripartite contribution becomes dominant at large $m$. 
Namely, we find
\begin{equation}
\begin{split}
S_{A}^{\text{after}(m)}
\approx n_A+n_{B}-n_C \qq{when} m>\frac{n_A}{n_C-n_B}>0
\end{split} 
\end{equation}
Since $n_A+n_B-n_C>0$ and $n_C>n_B$ imply $\frac{n_A}{n_C - n_B} > 1$, the tripartite contribution dominates for some $m > 1$.
However, as $m \to 1$, the RT-like terms dominate, and we find:
\begin{align}
S_{A}^{\text{after}} \approx \min(n_{A},2n_B).
\end{align}
Recalling the R\'{e}nyi analog of the Koashi-Winter relation~\cite{debarba2017koashi}, we obtain the bound
\begin{align}
E_{F}^{(m)}(A:B) \lesssim S_{A}^{\text{after}(m)}. 
\end{align}
Therefore, for $m\ge 2$, we have  
\begin{align}
\boxed{ \
E_{F}^{(m)}(A:B) \lesssim \frac{m}{m-1} (n_A + n_B - n_C)\le 2(n_A+n_B-n_C)
\ } \\
    \phantom{E_{F}^{(m)}(A:B)}\underset{m = \infty}{\longrightarrow} n_A + n_B - n_C. \phantom{\le 2(n_A+n_B-n_C)\qquad} \notag
\end{align}
This upper bound can be smaller than $E^W(A:B) = \min(n_A, 2n_B)$ at leading order.

\begin{figure}
    \centering
    \raisebox{\height}{a)}\raisebox{-0.75\height}{\includegraphics[width=0.3\linewidth]{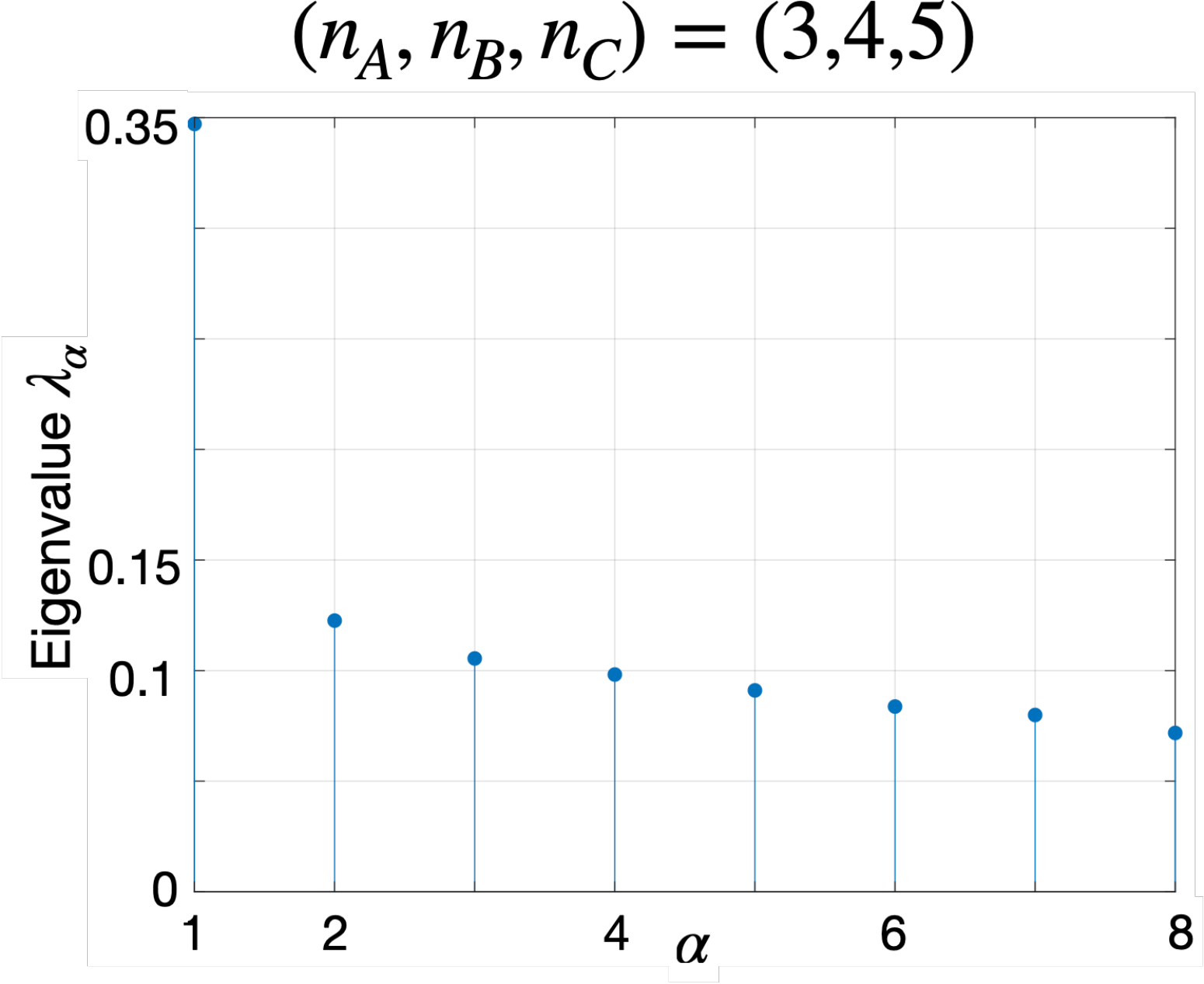}}
    \hspace{2pt}
    \raisebox{\height}{b)}\raisebox{-0.75\height}{\includegraphics[width=0.3\linewidth]{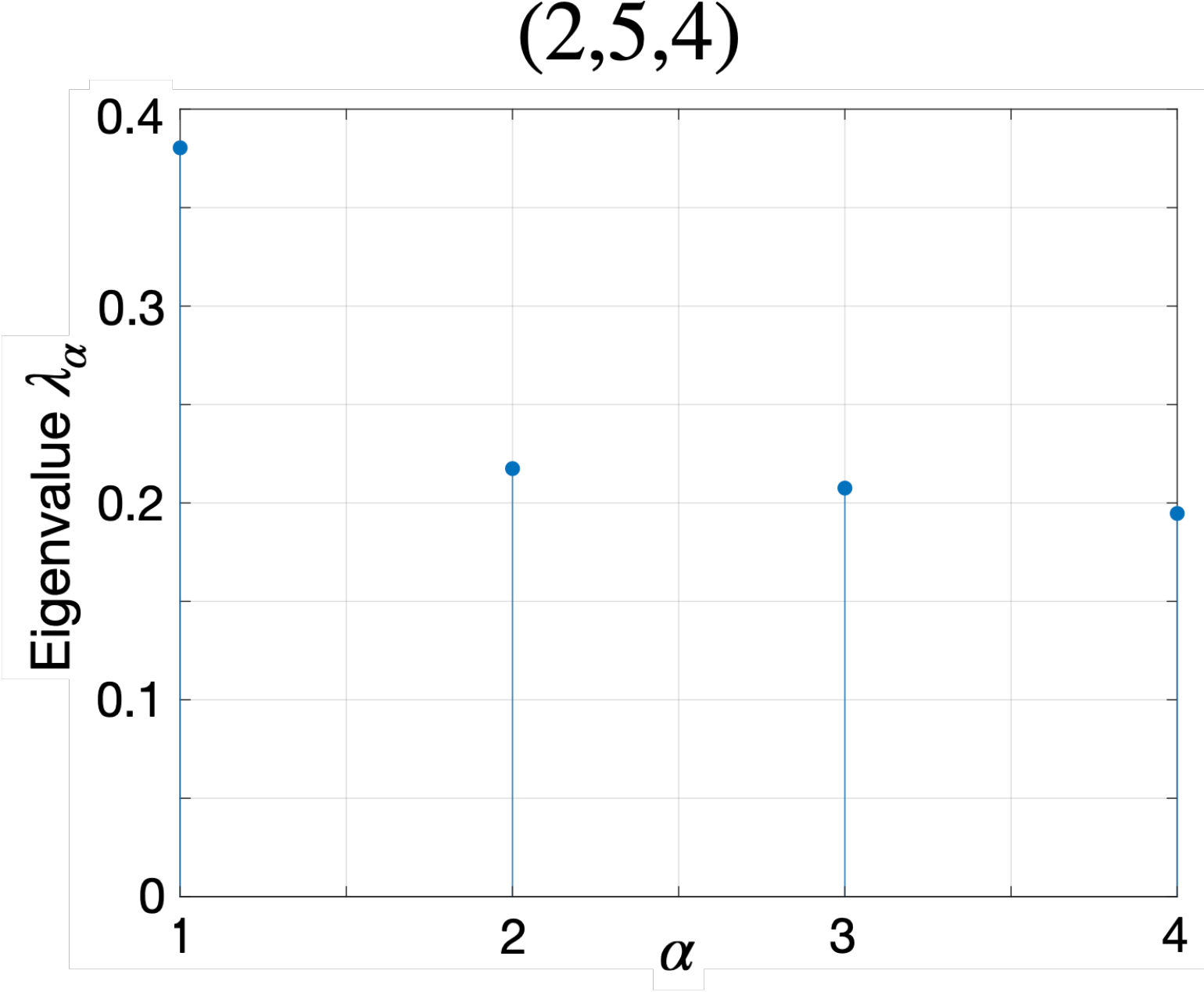}}
    \hspace{2pt}
    \raisebox{\height}{c)}\raisebox{-0.75\height}{\includegraphics[width=0.3\linewidth]{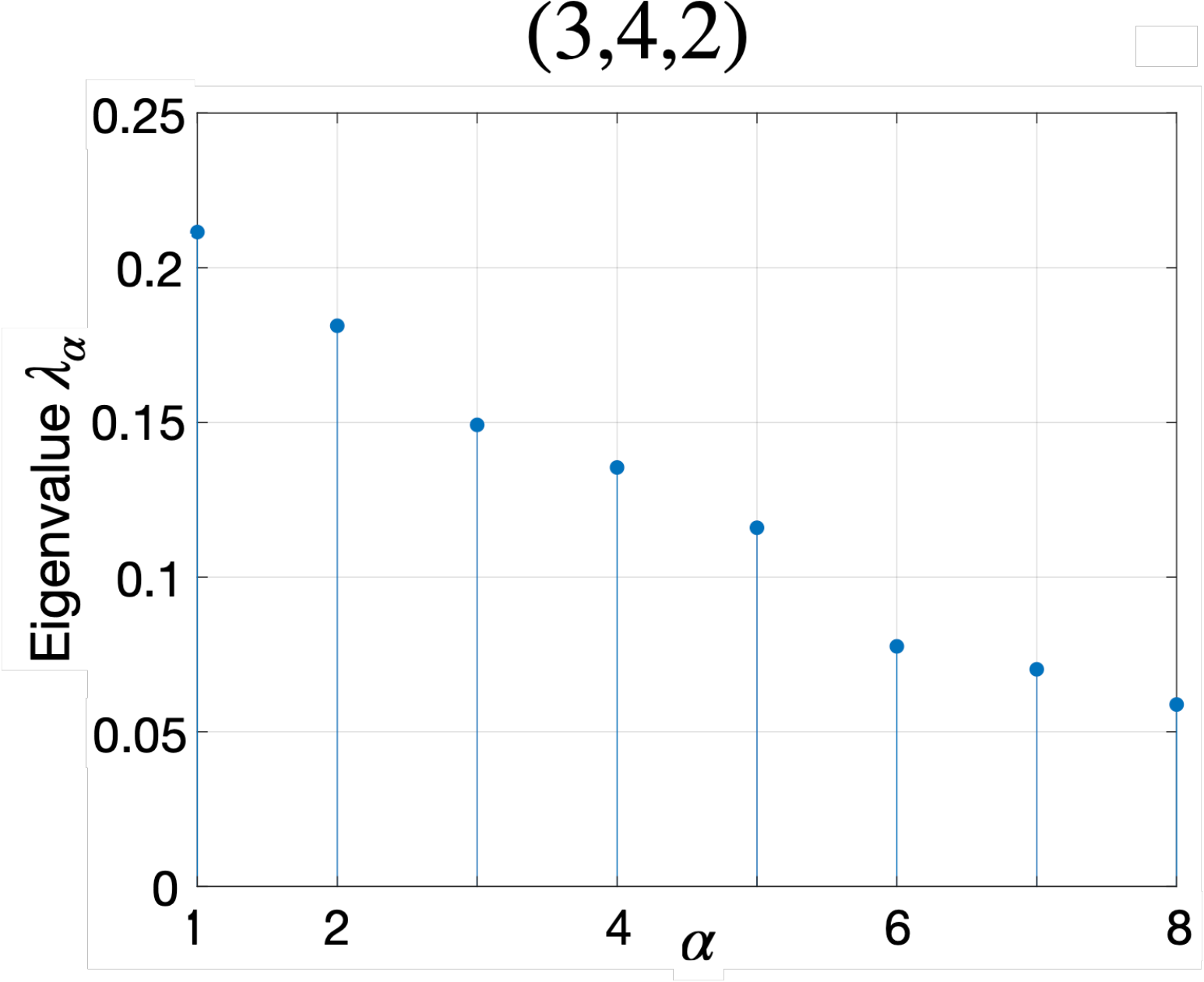}}
    \caption{Spectra of the reduced density matrix $\rho_{A}^{\text{after}}$ of the double-copy state after a random measurement on $A'$.
    Note that each state corresponds to the respective state in Fig.~\ref{fig:numeric} with the same label.
    }
    \label{fig:numeric2}
\end{figure}

Let us conclude with some numerical evidence.
Since the original state $\rho_{AA'}$ is ab isotropic state (for $n_A < n_B$), we expect the post-measurement spectrum to have a peaked structure over a flat background.
Fig.~\ref{fig:numeric2} shows numerically computed eigenvalue spectra of $\rho_{A}^{\text{after}}$.
Each post-measurement state was obtained by computing $\Tr_{A'}(\rho_{AA'} \dyad{\psi}_{A'})$ (up to normalization), where $\ket{\psi}$ is sampled from the Haar measure.
Fig.\ref{fig:numeric2}(a) shows the case $(n_A, n_B, n_C) = (3,4,5)$ with a clear peak plus flat background.
Fig.\ref{fig:numeric2}(b) shows a similar structure but with a smaller peak amplitude.
Finally, Fig.~\ref{fig:numeric2}(c) displays a decaying spectrum, where no dominant peak is visible.
These results are consistent with expectations based on projecting an isotropic state.

\section{Holographic double-copy state}\label{app:TN}

In this appendix, we study the double-copy state for random tensor networks. We focus on the following setup, where two copies of the tensor network, the original $|\psi_{ABC}\rangle$ and the complex conjugated $|\psi_{ABC}^*\rangle$, are glued (contracted) at the minimal surface $\gamma_C$ of $C$:
\begin{align}
|\psi_{ABC}\rangle=\figbox{1.7}{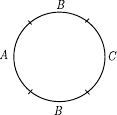}, \qquad
|\Phi_{ABA'B'}^{(\text{double})}\rangle =\figbox{1.7}{fig_TN_glue}\ .
\end{align}

\subsection{Entanglement in double-copy state}

We begin by evaluating $\Tr\big[ (\rho_{AA'})^m \big]$. 
The boundary conditions are the same as in the case of Haar random states. 
Here, we assume a tiling of Haar random tensors down to length scales smaller than the AdS scale, but larger than the Planck scale. 

As before, we employ the folded geometry with $2m$ copies of $|\psi\rangle$ ($m$ copies of $|\psi\rangle$ and $|\psi^*\rangle$).
Key contributions are listed below: 
\begin{align}
F_{\text{RT-$A$}}^{(m)} = \figbox{1.7}{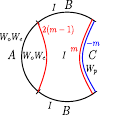}, \ 
F_{\text{RT-$B$}}^{(m)} = \figbox{1.7}{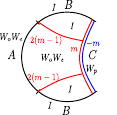}, \  F_{\text{tri}}^{(m)} =\figbox{1.7}{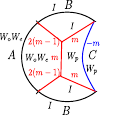}
\label{eq:hol-tri-dc}
\end{align}
where we used $C$ to denote $\gamma_C$ for notational symplicity. 
Note that $-m$ contribution on $C$ results from normalization. 
These diagrams result from placing generalized spins $V=W_o W_e, I, W_p \in S_{2m}$ in the bulk where their domain walls make energetic contributions to the effective ferromagnet Hamiltonian. 
The above diagrams make the following contributions:
\begin{align}
\Tr\big[ (\rho_{AA'})^m \big] = 2^{-F_{\text{RT-$A$}}^{(m)}} + 2^{-F_{\text{RT-$B$}}^{(m)}} + 2^{-F_{\text{tri}}^{(m)}} \cdots.
\end{align}
The first and second terms represent RT-like contributions evaluating the minimal surfaces of $AA'$ in the glued geometry. For large $m$, the RT contribution asymptotes to geodesic lengths:
\begin{align}
\frac{F_{\text{RT-$A$}}^{(m)}}{m-1} \underset{m=\infty}{\longrightarrow}\   \figbox{1.7}{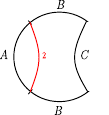} = 
\figbox{1.7}{fig_TN_glue_RT_A_geo} \label{eq:FRT-A}\\
\frac{F_{\text{RT-$B$}}^{(m)}}{m-1} \underset{m=\infty}{\longrightarrow}\   \figbox{1.7}{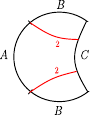} = 
\figbox{1.7}{fig_TN_glue_RT_B_geo} \label{eq:FRT-B}
\end{align}
where we illustrated the geodesics in both folded and unfolded glued geometries. 
The second diagram becomes dominant over the first one for $J^W(A|C) > 0$ which is the regime when 1WAY LOCC distillation becomes possible by measuring $C$.  

The third term, the tripartite contribution $F_{\text{tri}}^{(m)}$, arises by choosing two interior vertices and constructing a tripartite diagram. 
Note that two interior vertices may be placed on minimal surfaces $\gamma_A, \gamma_B, \gamma_C$, or on the asymptotic boundary. It is also possible to bring two vertices together. 
The diagram for $F_{\text{tri}}^{(m)}$ in Eq.~\eqref{eq:hol-tri-dc} represents one specific example of tripartite contributions. 
Also, note that the second RT diagram can be regarded as a special case of the tripartite contributions where inner vertices are placed on $\gamma_C$.

For large $m$, it asymptotes to
\begin{align}
\frac{F_{\text{tri}}^{(m)}}{m-1} \ \underset{m=\infty}{\longrightarrow} \ \figbox{1.7}{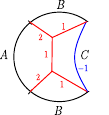} \ .
\end{align}
In this diagram, we have not yet optimized the locations of two interior vertices for minimization. 
We find that the minimal configuration at large $m$ can be constructed by moving the interior vertices to the boundary, as schematically shown below
\begin{align}
\figbox{1.7}{fig_TN_RT_tri_geo}  \ \geq \ \figbox{1.7}{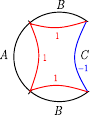} ,
\end{align}
which follows from the geodesic extremality.
Thus, as $m\rightarrow \infty$, the tripartite contribution asymptotes to
\begin{align}
\frac{F_{\text{tri}}^{(m)}}{m-1} \underset{m=\infty}{\longrightarrow}\ \figbox{1.7}{fig_TN_RT_tri_geo_min} \ .
\end{align}

We now show that the tripartite contribution always dominates at large $m$ whenever $A$ and $C$ have connected wedges. 
Let us first consider the regime where the first RT contribution dominates the second RT contribution. 
We have
\begin{align}
\figbox{1.7}{fig_TN_RT_A_geo}\ -\ \figbox{1.7}{fig_TN_RT_tri_geo_min} \ = \figbox{1.7}{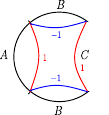}\ = I(A:C)  > 0
\end{align}
where the inequality follows from the connected wedge, with $I(A:C) = O(1/G_{N})$. 
Next, consider the regime where the second RT contribution dominates:
\begin{equation}
\begin{split}
\figbox{1.7}{fig_TN_RT_B_geo}\ -\ \figbox{1.7}{fig_TN_RT_tri_geo_min} \ &= 
I(A:C) - 2J^W(A|C) \geq 0.
\end{split}
\label{eq:I-2Jw}
\end{equation}
This inequality can be derived in two ways. 
First, it can be derived geometrically since Eq.~\eqref{eq:I-2Jw} is equivalent to $E^W(A:B)\ge \frac{1}{2}I(A:B)$~\cite{Takayanagi:2017knl}. 
Second, since one can LOCC distill at least $J^W(A|C)$ EPR pairs from $\rho_{AC}$, we must have $\frac{1}{2}I(A:C)\geq J^W(A|C)$. 

To summarize, when $A,C$ have connected wedge with $I(A:C)\sim O(1/G_{N})$, we find 
\begin{align}
S_{AA'}^{(m)} \underset{m=\infty}{\longrightarrow} \figbox{1.7}{fig_TN_RT_tri_geo_min}  ,\qquad \text{when $I(A:C)\sim O(1/G_N)$ }\ . 
\end{align}
This demonstrates a leading-order deviation from the RT formula at large $m$. 

A similar violation of the RT formula occurs for R\'{e}nyi-$2$ entropy. 
We focus on the regime where the first RT contribution dominates, namely $J^W(A|C) = 0$. 
We demonstrate that a leading-order violation for $m=2$ occurs when $J^W(A|C) = 0$, but $J^W(C|A) > 0$. 
For $m=2$, while we were unable to identify the global minimum for the tripartite contribution, we found the a candidate configuration:
\begin{align}
\figbox{1.7}{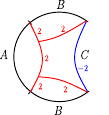} \geq \min F_{\text{tri}}^{(2)}
\end{align}
where two interior vertices are placed on $\gamma_A$. 
We now demonstrate that this diagram dominates over the first RT contribution in a specific regime.
Observe:
\begin{align}
\figbox{1.7}{fig_TN_RT_A_geo} - \figbox{1.7}{fig_TN_RT_tri_geo_m2_min}  
= 2(S_C - E^W(B:C) )= 2J^W(C|A)
\end{align}
suggesting that, as long as $J^W(C|A)>0$, the tripartite contribution dominates. 
This gives an upper bound on the R\'{e}nyi-$2$ entropy:
\begin{align}
\boxed{ \  
S_{AA'}^{(2)} \approx \min F_{\text{tri}}^{(2)} \lesssim 2 S_A - 2 J^W(C|A), \qquad \frac{1}{2}I^{(2)}(A:A') \gtrsim J^W(C|A) \ } \label{eq:Renyi-2-TN}
\end{align}
where $I^{(2)}(A:A') \equiv S_A^{(2)} + S_{A'}^{(2)} - S_{AA'}^{(2)}$ is the na\"{i}ve R\'{e}nyi-$2$ mutual information. 
Here, $S_{AA'}^{(2)}$ deviates from the RT formula by at least $2J^W(C|A)$. 

An interesting observation is that the (a)symmetry between $A$ and $C$ is closely related to the R\'enyi-$2$ entropy $S_{AA'}^{(2)}$.
Recall that we are considering the regime $J^W(A|C)=0$.
Eq.~\eqref{eq:Renyi-2-TN} suggests that when $S_{AA^\prime}^{(2)}\approx 2S_A$, we must also have $J^W(C|A)=0$. This implies that 1WAY LOCC distillable entanglement vanishes at leading order, i.e., $J^W(A:C) = 0$.

As for the von Neumann entropy $S_{AA'}$, recalling the formula for R\'{e}nyi-$m$ entropy,
\begin{align}
S_{AA'}^{(m)} = - \frac{1}{m-1} \log \Tr\big[ (\rho_{AA'})^m \big]
\end{align}
we will need to evaluate the following in the $m\rightarrow 1$ limit: 
\begin{align}
\lim_{m\rightarrow 1} \frac{F_{\text{RT}}^{(m)}}{m-1}, \frac{F_{\text{tri}}^{(m)}}{m-1}.
\end{align}
We find that the tripartite diagram yields larger contributions due to the $1/(m-1)$ factor as $m\rightarrow 1$ unless the interior vertices exactly lie on $\gamma_C$, in which case the minimization reproduces the second RT contribution. Thus, the minimal configuration is always given by the RT contributions. Hence, we find
\begin{align}
S_{AA'} \approx \min\qty( \figbox{1.7}{fig_TN_RT_A_geo}, \ \figbox{1.7}{fig_TN_RT_B_geo}  ).
\end{align}

Finally, we comment on subleading contributions to $\rho_{AA'}$.
For the double-copy state constructed from a Haar random state, $\rho_{AA'}$ exhibits a single peak on $|\text{EPR}\rangle_{AA'}$ with a flat background.
That is, $\rho_{AA'}$ approximates a maximally mixed state on $AA'$ (for $n_A < n_B$) with exponentially small corrections of order $2^{-\Delta}$ from $|\text{EPR}\rangle\langle \text{EPR}|_{AA'}$.
For random tensor networks, $\rho_{AA'}$ is similarly approximated by a maximally mixed state over the minimal surface of $AA'$ or $BB'$, with small corrections from entangled states.
While solving for the full spectrum is challenging, 
partial spectral information can be inferred from the tripartite diagrams.
Here we focus on contributions that result from the following diagram:
\begin{align}
\figbox{1.7}{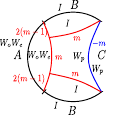} 
\end{align}
where two inner vertices are placed on $\gamma_A$. 
Looking at the scaling with respect to $m$, one can deduce that the following state contributes to $\rho_{AA'}$: 
\begin{align}
2^{-\Delta}\cdot \figbox{2.0}{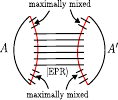} \ ,  \qquad 
\Delta = \frac{1}{4G_N}\figbox{1.7}{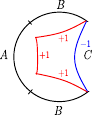} 
\end{align}
where $\Delta$ corresponds to the length difference between the red and blue curves, and the probability amplitude is exponentially suppressed with respect to $1/G_N$. 
When $\gamma_A,\gamma_C$ approach at the Planck scale, the length difference $\Delta$ will also be order of the Planck length. 
In such a regime, the correction terms become dominant. 

\subsection{Entanglement in post-measurement state} 

Next, we study $J(A|C)$ and $E_F(A:B)$ by evaluating the entropy drop resulting from measuring $A'$.  
Namely, we demonstrate that the holographic proposal $E_F(A:B) \approx E^W(A:B)$ can be violated for R\'{e}nyi-$m$ entropy, namely $E_F^{(m)}(A:B) \not\approx E^W(A:B)$ for large $m$ in general. 

Let us discuss the effect of measuring the minimal surface of $A'$ in the double-copy state $|\Phi\rangle$. 
Measuring $\gamma_A'$ in a random basis places open boundary conditions along $\gamma_A'$ (even copies of $A$). Important contributions are given by
\begin{align}
F_{\text{RT-$A$}}^{\text{after}(m)} = \figbox{1.7}{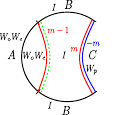}  ,\
F_{\text{RT-$B$}}^{\text{after}(m)} = \figbox{1.7}{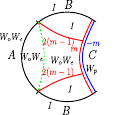} ,  \
F_{\text{tri}}^{\text{after}(m)} =\figbox{1.7}{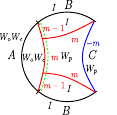}
\label{eq:RT-tri-hol-A} \notag
\end{align}
where open boundary conditions are shown in dotted green lines. We emphasize that the boundary spins $O_e$ along these lines cannot be placed beyond $\mathcal{E}_{A^\prime}$ since we perform measurements on $A^\prime$.
For large $m$, these contributions asymptote to 
\begin{align}
\frac{F_{\text{RT-$A$}}^{\text{after}(m)}}{m-1} \rightarrow    \figbox{1.7}{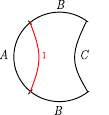}\!, \ 
\frac{F_{\text{RT-$B$}}^{\text{after}(m)}}{m-1} \rightarrow   \figbox{1.7}{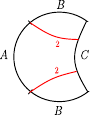} \!, \
\frac{F_{\text{tri}}^{\text{after}(m)}}{m-1} \rightarrow    \figbox{1.7}{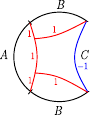} \! . \notag
\end{align}
We focus on a regime where the first diagram dominates over the second.
In this regime, we find that the tripartite contribution becomes dominant over others at large $m$ when
\begin{align}
J^W(C|A) > 0.
\end{align}
To see this, observe 
\begin{align}
\figbox{1.7}{fig_TN_RT_A_geo_after} -  
\figbox{1.7}{fig_TN_tri_geo_after} = J^W(C|A). 
\end{align}
This suggests 
\begin{align}
E_{F}^{(m)}(A:B) < S_{A} - J^W(C|A) \qq{for large $m$.}
\end{align}
Recalling that $E^W(A:B) = S_{A} - J^W(A|C)$, we find that 
\begin{align}
E_{F}^{(m)}(A:B) < E^W(A:B) \qq{for large $m$} \quad \mbox{if $J^W(C|A) > J^W(A|C)$}.
\end{align}
Hence, we find that there exist regimes where $E_{F}^{(m)}(A:B) \not\approx E^W(A:B)$ for large $m$.

Finally, we discuss the effect of measuring both $A'$ and $B'$ in the double-copy state. In this case, one can in principle measure DOFs behind the minimal surface $\gamma_{A'}$. 
We focus on the following measurement pattern and the resulting tripartite contribution:
\begin{align}
F_{\text{tri}}^{\text{after}(m)} \ = \ \figbox{1.7}{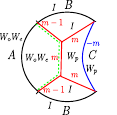}  , \qquad
\frac{F_{\text{tri}}^{\text{after}(m)}}{m-1} \rightarrow  \figbox{1.7}{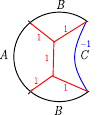}
\end{align}
where the measurement surface is shown in a dotted green line. 
Here the measurement surface can be placed beyond $\gamma_{A^\prime}$ as one has access to both $A^\prime B^\prime$.
This leaves a possibility of optimizing the locations of interior vertices and potentially further reducing $S_{A}^{\text{after}(m)}$ at large $m$.\footnote{
This diagram is similar to the one studied in~\cite{Penington:2022dhr}. 
Whether such tripartite contributions play significant roles in characterizing entanglement in $\rho_{AC}$ remains unclear at this moment. 
For one thing, tripartite contributions do not seem to survive at the $m\rightarrow 1$ limit for the same logic presented in the previous subsection, suggesting that these would make negligible (exponentially suppressed in $1/G_N$) perturbations only. 
In the main text, we highlighted a similar observation by studying the entanglement properties of an isotropic state.
It should be, however, noted that there is a special way of taking the $m\rightarrow 1$ limit for a certain family of tripartite diagrams for which a smooth limit appears to exist~\cite{Penington:2022dhr}. 
Whether this particular procedure of taking the $m\rightarrow 1$ limit can be observed in robust entanglement measures or not remains to be seen. 
}

\mciteSetMidEndSepPunct{}{\ifmciteBstWouldAddEndPunct.\else\fi}{\relax}
\bibliographystyle{JHEP}
\bibliography{ref.bib}
\end{document}